\documentclass[journal]{IEEEtran}
\usepackage[dvips]{graphicx,color}
\usepackage{latexsym}
\usepackage{array}
\begin{document}

%
%
\arraycolsep 0.5mm

\newcommand{\bfig}{\begin{figure}[t]}
\newcommand{\efig}{\end{figure}}
\setcounter{page}{1}
\newenvironment{indention}[1]{\par
\addtolength{\leftskip}{#1}\begingroup}{\endgroup\par}
%
\newcommand{\namelistlabel}[1]{\mbox{#1}\hfill} 
\newenvironment{namelist}[1]{%
\begin{list}{}
{\let\makelabel\namelistlabel
\settowidth{\labelwidth}{#1}
\setlength{\leftmargin}{1.1\labelwidth}}
}{%
\end{list}}
\newcommand{\bc}{\begin{center}}  %
\newcommand{\ec}{\end{center}}
\newcommand{\befi}{\begin{figure}[h]}  %
\newcommand{\enfi}{\end{figure}}
\newcommand{\bsb}{\begin{shadebox}\begin{center}}   %
\newcommand{\esb}{\end{center}\end{shadebox}}
\newcommand{\bs}{\begin{screen}}     %
\newcommand{\es}{\end{screen}}
\newcommand{\bib}{\begin{itembox}}   %
\newcommand{\eib}{\end{itembox}}
\newcommand{\bit}{\begin{itemize}}   %
\newcommand{\eit}{\end{itemize}}
\newcommand{\defeq}{\stackrel{\triangle}{=}}
\newcommand{\qed}{\hbox{\rule[-2pt]{3pt}{6pt}}}
\newcommand{\beq}{\begin{equation}}
\newcommand{\eeq}{\end{equation}}
\newcommand{\beqa}{\begin{eqnarray}}
\newcommand{\eeqa}{\end{eqnarray}}
\newcommand{\beqno}{\begin{eqnarray*}}
\newcommand{\eeqno}{\end{eqnarray*}}
\newcommand{\ba}{\begin{array}}
\newcommand{\ea}{\end{array}}
\newcommand{\vc}[1]{\mbox{\boldmath $#1$}}
\newcommand{\lvc}[1]{\mbox{\scriptsize\boldmath $#1$}}
\newcommand{\svc}[1]{\mbox{\scriptsize \boldmath$#1$}}
\newcommand{\wh}{\widehat}
\newcommand{\wt}{\widetilde}
\newcommand{\ts}{\textstyle}
\newcommand{\ds}{\displaystyle}
\newcommand{\scs}{\scriptstyle}
\newcommand{\vep}{\varepsilon}
\newcommand{\rhp}{\rightharpoonup}
\newcommand{\cl}{\circ\!\!\!\!\!-}
\newcommand{\bcs}{\dot{\,}.\dot{\,}}
\newcommand{\eqv}{\Leftrightarrow}
\newcommand{\leqv}{\Longleftrightarrow}
\newcommand{\lan}{\langle}
\newcommand{\ran}{\rangle}
\newcommand{\beqal}{\begin{eqnalp}}
\newcommand{\eeqal}{\end{eqnalp}}
\newcommand{\beqala}{\begin{eqnalp2}}
\newcommand{\eeqala}{\end{eqnalp2}}

\newfont{\bg}{cmr10 scaled \magstep4}
\newcommand{\bigzerol}{\smash{\hbox{\bg 0}}}
\newcommand{\bigzerou}{\smash{\lower1.7ex\hbox{\bg 0}}}
\newcommand{\nbn}{\frac{1}{n}}
\newcommand{\ra}{\rightarrow}
\newcommand{\la}{\leftarrow}
\newcommand{\ldo}{\ldots}
\newcommand{\ep}{\epsilon }
\newcommand{\typi}{A_{\epsilon }^{n}}
\newcommand{\bx}{\hspace*{\fill}$\Box$}
\newcommand{\pa}{\vert}
\newcommand{\ignore}[1]{}
\newtheorem{co}{Corollary} 
\newtheorem{lm}{Lemma} 
\newtheorem{Ex}{Example} 
\newtheorem{Th}{Theorem}
\newtheorem{df}{Definition} 
\newtheorem{pr}{Property} 
\newtheorem{pro}{Proposition} 
\newtheorem{rem}{Remark} 

\newcommand{\lcv}{convex } 

\newcommand{\hugel}{{\arraycolsep 0mm
                    \left\{\ba{l}{\,}\\{\,}\ea\right.\!\!}}
\newcommand{\Hugel}{{\arraycolsep 0mm
                    \left\{\ba{l}{\,}\\{\,}\\{\,}\ea\right.\!\!}}
\newcommand{\HUgel}{{\arraycolsep 0mm
                    \left\{\ba{l}{\,}\\{\,}\\{\,}\vspace{-1mm}
                    \\{\,}\ea\right.\!\!}}
\newcommand{\huger}{{\arraycolsep 0mm
                    \left.\ba{l}{\,}\\{\,}\ea\!\!\right\}}}
\newcommand{\Huger}{{\arraycolsep 0mm
                    \left.\ba{l}{\,}\\{\,}\\{\,}\ea\!\!\right\}}}
\newcommand{\HUger}{{\arraycolsep 0mm
                    \left.\ba{l}{\,}\\{\,}\\{\,}\vspace{-1mm}
                    \\{\,}\ea\!\!\right\}}}

\newcommand{\hugebl}{{\arraycolsep 0mm
                    \left[\ba{l}{\,}\\{\,}\ea\right.\!\!}}
\newcommand{\Hugebl}{{\arraycolsep 0mm
                    \left[\ba{l}{\,}\\{\,}\\{\,}\ea\right.\!\!}}
\newcommand{\HUgebl}{{\arraycolsep 0mm
                    \left[\ba{l}{\,}\\{\,}\\{\,}\vspace{-1mm}
                    \\{\,}\ea\right.\!\!}}
\newcommand{\hugebr}{{\arraycolsep 0mm
                    \left.\ba{l}{\,}\\{\,}\ea\!\!\right]}}
\newcommand{\Hugebr}{{\arraycolsep 0mm
                    \left.\ba{l}{\,}\\{\,}\\{\,}\ea\!\!\right]}}
\newcommand{\HUgebr}{{\arraycolsep 0mm
                    \left.\ba{l}{\,}\\{\,}\\{\,}\vspace{-1mm}
                    \\{\,}\ea\!\!\right]}}

\newcommand{\hugecl}{{\arraycolsep 0mm
                    \left(\ba{l}{\,}\\{\,}\ea\right.\!\!}}
\newcommand{\Hugecl}{{\arraycolsep 0mm
                    \left(\ba{l}{\,}\\{\,}\\{\,}\ea\right.\!\!}}
\newcommand{\hugecr}{{\arraycolsep 0mm
                    \left.\ba{l}{\,}\\{\,}\ea\!\!\right)}}
\newcommand{\Hugecr}{{\arraycolsep 0mm
                    \left.\ba{l}{\,}\\{\,}\\{\,}\ea\!\!\right)}}

\newcommand{\hugepl}{{\arraycolsep 0mm
                    \left|\ba{l}{\,}\\{\,}\ea\right.\!\!}}
\newcommand{\Hugepl}{{\arraycolsep 0mm
                    \left|\ba{l}{\,}\\{\,}\\{\,}\ea\right.\!\!}}
\newcommand{\hugepr}{{\arraycolsep 0mm
                    \left.\ba{l}{\,}\\{\,}\ea\!\!\right|}}
\newcommand{\Hugepr}{{\arraycolsep 0mm
                    \left.\ba{l}{\,}\\{\,}\\{\,}\ea\!\!\right|}}

\newenvironment{jenumerate}
	{\begin{enumerate}\itemsep=-0.25em \parindent=1zw}{\end{enumerate}}
\newenvironment{jdescription}
	{\begin{description}\itemsep=-0.25em \parindent=1zw}{\end{description}}
\newenvironment{jitemize}
	{\begin{itemize}\itemsep=-0.25em \parindent=1zw}{\end{itemize}}
\renewcommand{\labelitemii}{$\cdot$}

\newcommand{\iro}[2]{{\color[named]{#1}#2\normalcolor}}
\newcommand{\irr}[1]{{\color[named]{Red}#1\normalcolor}}
\newcommand{\Irr}[1]{{\color[named]{Black}#1\normalcolor}}

\newcommand{\irg}[1]{{\color[named]{Green}#1\normalcolor}}
\newcommand{\irb}[1]{{\color[named]{Blue}#1\normalcolor}}
\newcommand{\irBl}[1]{{\color[named]{Black}#1\normalcolor}}
\newcommand{\irWh}[1]{{\color[named]{White}#1\normalcolor}}

\newcommand{\irY}[1]{{\color[named]{Yellow}#1\normalcolor}}
\newcommand{\irO}[1]{{\color[named]{Orange}#1\normalcolor}}
\newcommand{\irBr}[1]{{\color[named]{Purple}#1\normalcolor}}
\newcommand{\IrBr}[1]{{\color[named]{Purple}#1\normalcolor}}
\newcommand{\irBw}[1]{{\color[named]{Brown}#1\normalcolor}}
\newcommand{\irPk}[1]{{\color[named]{Magenta}#1\normalcolor}}
\newcommand{\irCb}[1]{{\color[named]{CadetBlue}#1\normalcolor}}

\newcommand{\irMho}[1]{{\color[named]{Mahogany}#1\normalcolor}}
\newcommand{\irOlg}[1]{{\color[named]{OliveGreen}#1\normalcolor}}
\newcommand{\irBg}[1]{{\color[named]{BlueGreen}#1\normalcolor}}
\newcommand{\irCy}[1]{{\color[named]{Cyan}#1\normalcolor}}
\newcommand{\irRyp }[1]{{\color[named]{RoyalPurple}#1\normalcolor}}

\newcommand{\irAqm}[1]{{\color[named]{Aquamarine}#1\normalcolor}}
\newcommand{\irRyb}[1]{{\color[named]{RoyalBule}#1\normalcolor}}
\newcommand{\irNvb}[1]{{\color[named]{NavyBlue}#1\normalcolor}}
\newcommand{\irSkb}[1]{{\color[named]{SkyBlue}#1\normalcolor}}
\newcommand{\irTeb}[1]{{\color[named]{TeaBlue}#1\normalcolor}}
\newcommand{\irSep}[1]{{\color[named]{Sepia}#1\normalcolor}}
\newcommand{\irReo}[1]{{\color[named]{RedOrange}#1\normalcolor}}
\newcommand{\irRur}[1]{{\color[named]{RubineRed}#1\normalcolor}}
\newcommand{\irSa }[1]{{\color[named]{Salmon}#1\normalcolor}}
\newcommand{\irAp}[1]{{\color[named]{Apricot}#1\normalcolor}}

%
%

\newcommand{\Ch}{{\Gamma}}
\newcommand{\Rw}{{W}}

\newcommand{\Cd}{{\cal R}_{\rm d}(\Ch)}
\newcommand{\Cdi}{{\cal R}_{\rm d}^{\rm (in)}(\Ch)}
\newcommand{\Cdo}{{\cal R}_{\rm d}^{\rm (out)}(\Ch)}

\newcommand{\tCdi}{\tilde{\cal R}_{\rm d}^{\rm (in)}(\Ch)}
\newcommand{\tCdo}{\tilde{\cal R}_{\rm d}^{\rm (out)}(\Ch)}
\newcommand{\hCdo}{  \hat{\cal R}_{\rm d}^{\rm (out)}(\Ch)}

\newcommand{\Cs}{{\cal R}_{\rm s}(\Ch)}
\newcommand{\Csi}{{\cal R}_{\rm s}^{\rm (in)}(\Ch)}
\newcommand{\Cso}{{\cal R}_{\rm s}^{\rm (out)}(\Ch)}
\newcommand{\tCsi}{\tilde{\cal R}_{\rm s}^{\rm (in)}(\Ch)}
\newcommand{\tCso}{\tilde{\cal R}_{\rm s}^{\rm (out)}(\Ch)}
\newcommand{\cCsi}{\check{\cal R}_{\rm s}^{\rm (in)}(\Ch)}
\newcommand{\Cds}{{\cal C}_{\rm ds}(\Ch)}
\newcommand{\Cdsi}{{\cal C}_{\rm ds}^{\rm (in)}(\Ch)}
\newcommand{\Cdso}{{\cal C}_{\rm ds}^{\rm (out)}(\Ch)}
\newcommand{\tCdsi}{\tilde{\cal C}_{\rm ds}^{\rm (in)}(\Ch)}
\newcommand{\tCdso}{\tilde{\cal C}_{\rm ds}^{\rm (out)}(\Ch)}
\newcommand{\hCdso}{\hat{\cal C}_{\rm ds}^{\rm (out)}(\Ch)}
\newcommand{\Css}{{\cal C}_{\rm ss}(\Ch)}
\newcommand{\Cssi}{{\cal C}_{\rm ss}^{\rm (in)}(\Ch)}
\newcommand{\Csso}{{\cal C}_{\rm ss}^{\rm (out)}(\Ch)}
\newcommand{\tCssi}{\tilde{\cal C}_{\rm ss}^{\rm (in)}(\Ch)}
\newcommand{\tCsso}{\tilde{\cal C}_{\rm ss}^{\rm (out)}(\Ch)}
\newcommand{\Cde}{{\cal R}_{\rm d1e}(\Ch)}
\newcommand{\Cdei}{{\cal R}_{\rm d1e}^{\rm (in)}(\Ch)}
\newcommand{\Cdeo}{{\cal R}_{\rm d1e}^{\rm (out)}(\Ch)}
\newcommand{\tCdei}{\tilde{\cal R}_{\rm d1e}^{\rm (in)}(\Ch)}
\newcommand{\tCdeo}{\tilde{\cal R}_{\rm d1e}^{\rm (out)}(\Ch)}
\newcommand{\hCdeo}{  \hat{\cal R}_{\rm d1e}^{\rm (out)}(\Ch)} 
\newcommand{\Cse}{{\cal R}_{\rm s1e}(\Ch)}
\newcommand{\Csei}{{\cal R}_{\rm s1e}^{\rm (in)}(\Ch)}
\newcommand{\Cseo}{{\cal R}_{\rm s1e}^{\rm (out)}(\Ch)}
\newcommand{\tCsei}{\tilde{\cal R}_{\rm s1e}^{\rm (in)}(\Ch)}
\newcommand{\tCseo}{\tilde{\cal R}_{\rm s1e}^{\rm (out)}(\Ch)}

\newcommand{\Capa}{C}
\newcommand{\tCapa}{\tilde{C}}

\newcommand{\MEq}[1]{\stackrel{
{\rm (#1)}}{=}}

\newcommand{\MLeq}[1]{\stackrel{
{\rm (#1)}}{\leq}}

\newcommand{\MGeq}[1]{\stackrel{
{\rm (#1)}}{\geq}}

\newcommand{\ZeTa}{\zeta(S;Y,Z|U)}
\newcommand{\ZeTaI}{\zeta(S_i;Y_i,Z_i|U_i)}

\newcommand{\MarkOh}[1]{{\color[named]{Black}#1\normalcolor}}
\newcommand{\Rev}[1]{{\color[named]{Black}#1\normalcolor}}
\newcommand{\RevP}[1]{{\color[named]{Black}#1\normalcolor}}
\newcommand{\CEreg}{rate\MarkOh{-equivocation} }
\newcommand{\Cls}{class NL}
\newcommand{\vSpa}{\vspace{0.3mm}}
%
%
%
%
%
\title{
Capacity Results 
for Relay Channels with Confidential Messages
}
%
%
\author{Yasutada~Oohama 
        and Shun~Watanabe
\thanks{
}%
\thanks{Y. Oohama and S. Watanabe 
        are with the Department of Information Science 
        and Intelligent Systems,
        University of Tokushima, 
        2-1 Minami Josanjima-Cho, 
        Tokushima 770-8506, Japan.}
}
\markboth{
}
{
}
%

%
%
\maketitle
\begin{abstract}
We consider a {communication system} where a relay helps 
transmission of messages from {a} sender to {a} receiver. 
The relay is considered not only as a helper but as a wire-tapper who 
can obtain some knowledge about transmitted messages. In this paper 
{we study a relay channel with confidential messages(RCC), where 
a sender attempts to transmit common information to both a receiver and 
a relay and also has private information intended for the receiver and 
confidential to the relay. The level of secrecy of private
information confidential to the relay is measured by the 
equivocation rate, i.e., the entropy rate of private information 
conditioned on channel outputs at the relay. 
The performance measure of interest for the RCC is the rate triple 
that includes the common rate, the private rate, and the equivocation rate 
as components.} The \CEreg region is defined by the set {that 
consists of all these achievable rate triples.} In this 
paper we give two definition{s} of the \CEreg region. We first 
define the \CEreg region in the case of deterministic encoder and call 
it the deterministic \CEreg region. Next, we define 
the \CEreg region in the case 
of stochastic encoder and call it the stochastic \CEreg region. 
We derive explicit inner and outer bounds for the above two 
\CEreg regions. On the deterministic/stochastic \CEreg region we present 
two classes of relay channels where inner and outer bounds match. 
We also evaluate the deterministic and stochastic \CEreg regions of the 
Gaussian RCC.
\end{abstract}

%
%
\IEEEpeerreviewmaketitle

\section{Introduction}

{Security of communications can be studied 
from information 
theoretical viewpoint by regarding them as a 
communication system in which some messages transmitted 
through channel should be confidential to anyone except 
for authorized receivers.}

Information theoretical approach to security problem 
in communications was first attempted by Shannon \cite{sh}.   
He discussed a theoretical model of 
cryptosystems using the framework of classical one way noiseless 
channels and derived some conditions for secure communication. 
Yamamoto \cite{yama1},\cite{yama2} investigated 
some extensions of Shannon's cipher system. 

Various types of multi-terminal communication systems have been 
investigated so far in the field of multi-user information theory. 
In those systems we can consider the case where a confidentiality of 
transmitted messages is required from standpoint of security. In this 
case it is of importance to analyze security of communications from 
viewpoint of multi-user information theory. 

The security of communication for the broadcast channel was studied by
Wyner \cite{Wyn1} and Csisz{\' a}r and K{\"o}rner \cite{CsiKor1}.
Yamamoto \cite{yama3}-\cite{yama7} studied several secure
communication systems under the framework of multi-terminal source or
channel coding systems.  Maurer \cite{Maurer}, 
Ahlswede and Csisz{\'a}r \cite{ac1}, Csisz\'ar and Narayan
\cite{cn}-{\cite{cn3}}, studied public key agreements 
under the framework of multi-user information theory. 
Oohama \cite{ohrcc} discussed the
security of communication for the relay channel. He posed and
investigated the relay channel with confidential messages, where the
relay acts as both a helper and a wire-tapper.  
Subsequently, the above security problem in relay communication was studied 
in detail by Oohama \cite{ohrcc07} and He and Yener \cite{heye}, \cite{heye3}. 
Liang and Poor
\cite{lp} discussed the security of communication for the multiple
access channel. They formulated and studied the multiple access
channel with confidential messages.  Liu {et al.} \cite{liu}
considered interference and broadcast channels with confidential
messages.  Tekin and Yener \cite{ty} studied general Gaussian multiple
access and two way wire tap channels.  Lai and El Gamal \cite{lg}
investigated the security of relay channels in a problem set up
different form \cite{ohrcc}.

In this paper we discuss the security of communication for the relay 
channel under the framework that Oohama introduced in \cite{ohrcc}. 
In the relay channel {a} relay is considered not only as a sender 
who helps transmission of messages but as a wire-tapper who wish to know 
something about the transmitted messages. Coding theorem for the relay 
channel was first established by Cover and El~Gamal \cite{cg}. By 
carefully checking their coding scheme used for the proof of the direct 
coding theorem, we can see that in their scheme the relay helps 
transmission of messages by learning all of them. Hence, this coding 
scheme is not adequate when some messages should be confidential to the 
relay.

Oohama \cite{ohrcc} studied the security of communication for the 
relay channel under the situation that some of transmitted messages 
should be confidential to the relay. For analysis of this 
situation Oohama posed the communication system called 
the relay channel with confidential messages or briefly said the RCC. 
In the RCC, a sender wishes to transmit two types of message{s}. 
One is a message called {\it {a} common message} which is sent 
to {a legitimate} receiver and {a} relay. 
The other is a message called {\it {a} private message} which is  
sent only to the legitimate receiver and should be 
confidential to the relay as much as possible.
The level of secrecy of private information 
confidential to the relay can be measured by the equivocation rate, i.e., 
the entropy rate of private messages conditioned on channel 
outputs at the relay. 
The performance measure of interest is the rate triple 
that includes the transmission rates of common and private messages 
and the equivocation rate as components. We 
refer to the set that consists of all achievable rate triples 
as the \CEreg region. Oohama \cite{ohrcc} derived an 
inner bound of the \CEreg region. 

In this paper we study the coding problem of the RCC. In general two
cases of encoding can be considered in the problem of channel coding.
One is a case where deterministic encoders are used for transmission
of messages and the other is a case where stochastic encoders are
used. It is well known that for problems involving secrecy,
randomization of encoding enhances the security of communication.
From this reason, stochastic encoding was always assumed in the
previous works treating security problems in communication. However,
in those works it is not clear how much advantage stochastic encoding
can offer in secure communication. To know a merit of stochastic
encoding precisely we must also know a fundamental theoretical limit
of secure communication when encoding is restricted to be {\it
  deterministic}. In this paper we discuss security problems in the
RCC in two cases.  One is a case of deterministic encoder, where the
sender must use a deterministic encoder. The other is a case of
stochastic encoder, where the sender is allowed to use a stochastic
encoder. The former case
models an {\it insecure} communication scheme and the latter case
models a {\it secure} communication scheme. We define two \CEreg
regions. One is a \CEreg region in the case of deterministic encoder
and call it the deterministic \CEreg region. The other is a \CEreg
region in the case of stochastic encoder and call it the stochastic
\CEreg region. 

In this paper, we derive several results on the deterministic and
stochastic \CEreg regions.
\footnote{The same determination problems of the two \CEreg regions 
were investigated 
by Oohama \cite {ohrcc07}. However, his results on the deterministic 
\CEreg region contain some mistakes. The results on the deterministic \CEreg 
we derive in this paper correct those mistakes.
} Cover and El~Gamal \cite{cg} determined
the capacity for two classes of relay channels. One is a degraded
relay channel and the other is a reversely degraded channel. In the
degraded relay channel, channel outputs obtained by the relay are less
noisy than those obtained by the receiver.  Conversely, in the
reversely degraded relay channel, channel outputs obtained by the
relay are more noisy than those obtained by the receiver.  Our
capacity results have a close connection with the above two classes of
relay channels.

On the deterministic \CEreg 
region, we derive two pairs of inner and outer bounds.  On the first 
pair of inner and outer bounds we show that they match for the class of 
reversely degraded relay channels. Furthermore, we show that if the 
relay channel is degraded, no security is guaranteed for transmission of 
private messages. On the second pair of inner and outer bounds we show 
that they match for the class of relay channels having some 
deterministic component in their stochastic matrix. We further derive an 
explicit outer bound effective for a class of relay channels where 
channel outputs obtained by the relay depend only on channel inputs from 
the sender. 

On the stochastic \CEreg region, we derive two pairs of inner and
outer bounds. On the first pair, inner and outer bound match for the
class of reversely degraded channels. On the second one, inner and 
outer bounds match for the class of semi deterministic relay channels.
We show that when the relay channel is degraded, no security is
guaranteed for transmission of private messages even if we use
stochastic encoders.

We compare the deterministic \CEreg region with the stochastic \CEreg
region to show that the former is strictly smaller than the latter.
It is obvious that the maximum secrecy rate
attained by the deterministic encoding does not exceed that of
stochastic encoding. We demonstrate that for the reversely degraded
relay channel the former is equal to the latter.

When the relay is kept completely ignorant of private message in the
RCC, we say that the prefect secrecy is established. We show that the
prefect secrecy can hardly be attained by the deterministic
encoder. In the case of stochastic encoder the secrecy capacity is
defined by the maximum transmission rate of private message under the
condition of prefect secrecy.  From the results on the stochastic
\CEreg regions, we can obtain inner and outer bounds of the stochastic
secrecy capacities. In particular, when the relay channel is reversely
degraded or semi deterministic, we determine the stochastic secrecy
capacity.

We also study the Gaussian RCC, where transmissions are corrupted by 
additive Gaussian noise. We evaluate the deterministic and stochastic 
\CEreg regions of the Gaussian RCC. For each \CEreg we derive a pair of 
explicit inner and outer bounds to show that those bounds match for the 
class of reversely degraded relay channels. 

On our results on the inner bounds of the rate-equivocation region we 
give their rigorous proofs. The method Csisz\'ar and K\"orner 
\cite{CsiKor1} used for computation of the equivocation is a 
combinatorial method based on the type of sequences \cite{ckB}. Their 
method has a problem that it is not directly applicable to the Gaussian 
case. To overcome this problem we introduce a new unified way of 
estimating error probabilities and equivocation rate for both discrete 
and Gaussian cases. Our method is based on the information spectrum 
method introduced and developed by Han \cite{han}. Our derivation of the 
inner bounds is simple and straightforward without using any particular 
property on the sets of jointly typical sequences. 

\Rev{In the RCC, the relay also act as a receiver with respect to the
common message. This implies that when there is no security
requirement in the RCC, its communication scheme is equal to that of
a special case of cooperative relay broadcast channels(RBCs) posed
and investigated by Liang and Veeravalli \cite{lv} and Liang and
Kramer \cite{lk}. }  Cooperation and security are two important
features in communication networks. Coding problems for the RCC
provide an interesting interplay between cooperation and security.


\section{Relay Channels with Confidential Messages} 

Let ${\cal X},{\cal S},{\cal Y},$ ${\cal Z}$ be finite sets.
The relay channel dealt with in this paper is defined 
by a discrete memoryless channel specified with the following 
stochastic matrix:
\beq
{\Ch} \defeq \{ {\Ch}(y,z\mid x,s)\}_{
(x,s,y,z) 
\in    {\cal X}
\times {\cal S}
\times {\cal Y} 
\times {\cal Z}}\,.
\eeq
Let $X$ be a random variable taking values in ${\cal X}$ and
$X^n=X_{1}X_{2}$ $\cdots X_{n}$ be a random vector taking 
values in ${\cal X}^n$. We write an element of ${\cal X}^n$ as   
${\vc x}=x_{1}x_{2}$ 
$\cdots x_{n}.$
Similar notations are adopted for  
$S,Y,$ and  $Z$.

In the RCC, we consider the following scenario of communication.  
Let $K_n$ and  $M_n$ be uniformly distributed random 
variables taking values in message sets ${\cal K}_n $ and ${\cal M}_n$, 
respectively. The random variable $M_n$ is a common message 
sent to a relay and a legitimate receiver. 
The random variable $K_n$ is a private message sent only to the 
receiver and contains an information confidential to the relay. A sender 
transforms $K_n$ and $M_n$ into a transmitted sequence $X^n$ using an 
encoder function $f_n$ and sends it to the relay 
and the legitimate receiver. 
For the encoder function $f_n$, we consider two cases; one is the case 
where $f_n$ is {\it deterministic} and the other is the case 
where $f_n$ is {\it stochastic}. In the former case $f_n$ 
is a one to one mapping 
from ${\cal K}_n\times {\cal M}_n$ to ${\cal X}^n$. 
In the latter case  
$f_n: {\cal K}_n\times {\cal M}_n$ 
$\to {\cal X}^n$ is a stochastic matrix 
defined by 
$$
f_n(k,m)
=\{f_n({\vc x}|k,m)\}_{{\svc x}\in {\cal X}^n },
(k,m)\in{\cal K}_n \times {\cal M}_n\,. 
$$
Here, $f_n({\vc x}|k,m)$ is a probability that the encoder 
$f_n$ generates a channel input ${\vc x}$ from the message 
pair $(k,m)$.
\begin{figure}[t]
\bc
\includegraphics[width=3.7cm]{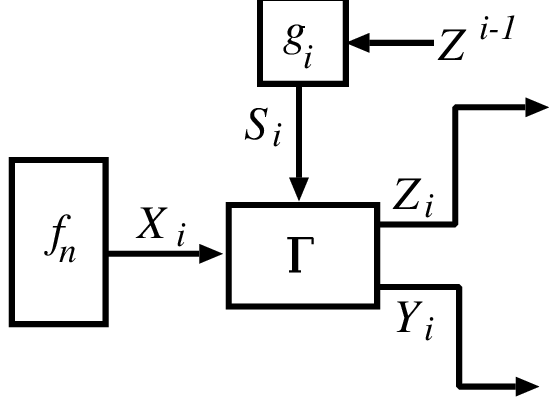}
\caption{Channel inputs and outputs at the $i$th transmission.}
\label{fig:rccblockn} 
\ec
\vspace*{-4mm}
\end{figure}
Channel inputs and outputs at the $i$th transmission is shown 
in Fig. \ref{fig:rccblockn}. At the $i$th transmission, the relay 
observes the random sequence $Z^{i-1}\defeq (Z_1,$ $Z_{2},$ 
$\cdots, Z_{i-1})$ transmitted by the sender through noisy 
channel, encodes them into the random variable $S_{i}$ and 
sends it to the receiver. 

\begin{figure}[t]
\bc
\includegraphics[width=8.7cm]{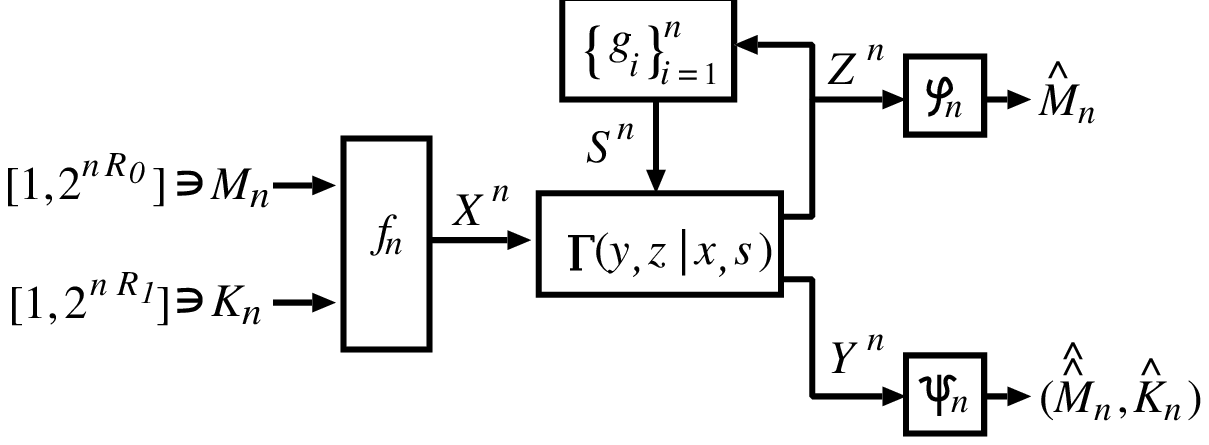}
\caption{Transmission of messages via relay channel 
using $(f_n,\{g_i\}_{i=1}^n,$ $\psi_n,\varphi_n)$.}
\label{fig:blockn} 
\ec
\vspace*{-4mm}
\end{figure}

The relay also wishes to decode the common message from observed 
channel outputs. The encoder function at the relay 
is defined by the sequence of 
functions $\{g_i \}_{i=1}^{n}$. Each $g_i$ is defined by $g_i: 
{\cal Z}^{i-1}\to {\cal S}$. Note that the channel input $S_{i}$ 
that the relay sends at the $i$th transmission depends solely 
on the output random sequence $Z^{i-1}$ that the relay previously 
obtained as channel outputs. The decoding functions at 
the {legitimate} receiver and the relay are 
denoted by ${\psi}_n$ and ${\varphi}_n$, respectively. 
Those functions are formally defined by
$
{\psi}_n: {\cal Y}^{n}   \to {\cal K}_n \times {\cal M}_n\,,
{\varphi}_n: {\cal Z}^{n} \to {\cal M}_n\,.
$
Transmission of messages via relay channel using 
$(f_n,$ $\{g_i \}_{i=1}^{n}${,} $\psi_n,\varphi_n)$ 
is shown in Fig. \ref{fig:blockn}.   
When $f_n$ is a deterministic encoder, 
the joint probability mass function on 
       ${\cal K}_n\times {\cal M}_n$
$\times {\cal Y}^n \times {\cal Z}^n$ is given by
\beqa
& &\Pr\{(K_n,M_n,Y^n,Z^n)=(k,m,{\vc y},{\vc z})\}
\nonumber\\
&=&
\frac{1}{\pa{\cal K}_n\pa\pa{\cal M}_n\pa}
\prod_{i=1}^n {\Ch}\hspace*{-1mm}
\left(y_i,z_i\left|x_i(k,m),g_i(z^{i-1})\right.\right),
\nonumber
\eeqa
where $x_i(k,m)$ is the $i$th component of ${\vc x}=f_n(k,m)$ 
and $\pa {\cal K}_n \pa$ is a cardinality 
of the set ${\cal K}_n$. 
When $f_n$ is a stochastic encoder, 
the joint probability mass function on 
${\cal K}_n\times {\cal M}_n$
$\times {\cal X}^n$
$\times {\cal Y}^n \times {\cal Z}^n$ is given by
\beqa
& &\Pr\{(K_n,M_n,X^n,Y^n,Z^n)=(k,m,{\vc x},{\vc y},{\vc z})\}
\nonumber\\
&=&
\frac{ f_n({\vc x}|k,m) }{\pa{\cal K}_n\pa\pa{\cal M}_n\pa}
\prod_{i=1}^n {\Ch}\hspace*{-1mm}
\left(y_i,z_i\left|x_i(k,m),g_i(z^{i-1})\right.\right).
\nonumber
\eeqa
Error probabilities of decoding at the receiver and the relay 
are defined by 
\beqno
{\lambda}_{1}^{(n)}& \defeq & \Pr\{\psi_{n}(Y^n)\neq (K_n,M_n)\}
\mbox{ and } 
\\
{\lambda}_{2}^{(n)}&\defeq &\Pr\{\varphi_{n}(Z^n)\neq M_n\},
\eeqno
respectively.
\newcommand{\wid}{\hspace*{-1mm}}

In the RCC, the relay act as a {\it wire-tapper} with respect to the 
private message $K_n$. 
The level of ignorance of the relay with respect to $K_n$ is measured by 
the equivocation rate, i.e., the entropy rate 
$\frac{1}{n}H(K_n|Z^n)$ conditioned on the channel output 
$Z^n$ at the relay. Throughout the paper, the logarithmic function 
is to the base 2. The equivocation rate should be 
greater than or equal to a prescribed positive level. 

A triple $(R_0,R_1,R_{\rm e})$ is {\it achievable}
if there exists a sequence of quadruples 
$\{(f_n, \{g_i\}_{i=1}^n,$ ${\psi }_n ,{\varphi }_n)\}_{n=1}^{\infty}$ 
such that 
\beqa 
\lim_{n\to\infty}{\lambda}_{1}^{(n)}
&=&\lim_{n\to\infty}{\lambda}_{2}^{(n)}=0, 
\nonumber\\
\lim_{n\to\infty} \nbn \log \pa {\cal M}_n \pa & = & R_0,
\lim_{n\to\infty} \
\nbn \log \pa {\cal K}_n \pa = R_1,
\nonumber\\
\lim_{n\to\infty} \nbn H(K_n|Z^{n}) &\geq & R_{\rm e}\,.
\nonumber
\eeqa
When $f_n$ and $\{g_i\}_{n=1}$ are restricted to be deterministic, 
the set that consists of all achievable rate triple is denoted by 
$\Cd$, which is called the deterministic \CEreg region of the RCC. 
When $f_n$ is allowed to be stochastic and $\{g_i\}_{n=1}$ 
is restricted to be deterministic, the set that consists 
of all achievable rate triple is denoted by $\Cs$, which 
is called the stochastic \CEreg region. 
Main results on $\Cd$ and $\Cs$ will 
be described in the next section. 

In the above problem set up the relay encoder 
$\{g_i\}_{i=1}^n$ is a {\it deterministic encoder}.
We can also consider the case where we may use a 
{\it stochastic encoder } as $\{g_i\}_{i=1}^n$.  
In this case the relay function 
$g_i(z^{i-1})$ $\in {\cal S}$, 
$z^{i-1}\in {\cal Z}^{i-1}$ is a stochastic matrix given by 
$$
g_i(z^{i-1})=\left\{g_i(s|z^{i-1})\right\}_{s\in\cal S}.
$$ 
Here $g_i(s|z^{i-1})$ is a conditional probability 
of $S_i=s$ conditioned on $Z^{i-1}=$ $z^{i-1}$.   
When $f_n$ is deterministic and $\{g_i\}_{i=1}^n$ 
is stochastic, the joint probability mass function on 
       ${\cal K}_n\times {\cal M}_n$
$\times {\cal S}^n$
$\times {\cal Y}^n \times {\cal Z}^n$ is given by
\beqno
& &\Pr\{(K_n,M_n,S^n,X^n,Y^n,Z^n)=(k,m,{\vc s},{\vc y},{\vc z})\}
\\
&=&
\frac{1}{\pa{\cal K}_n\pa\pa{\cal M}_n\pa}
\prod_{i=1}^n {\Ch}\hspace*{-1mm}
\left(y_i,z_i\left|x_i(k,m),s_i\right.\right)
g_i(s_i|z^{i-1}).
\eeqno
When $f_n$ and $\{g_i\}_{i=1}^n$ 
are stochastic, the joint probability mass function on 
       ${\cal K}_n\times {\cal M}_n$
$\times {\cal S}^n\times {\cal X}^n$
$\times {\cal Y}^n \times {\cal Z}^n$ is given by
\beqno
& &\Pr\{(K_n,M_n,S^n,X^n,Y^n,Z^n)=(k,m,{\vc s},{\vc x},{\vc y},{\vc z})\}
\\
&=&
\frac{ f_n({\vc x}|k,m) }{\pa{\cal K}_n\pa\pa{\cal M}_n\pa}
\prod_{i=1}^n {\Ch}\hspace*{-1mm}
\left(y_i,z_i\left|x_i(k,m),s_i\right.\right)
g_i(s_i|z^{i-1}).
\eeqno
Capacity results in the case of stochastic relay 
encoder will be stated in Section III-C.     

In the remaining part of this section, we state relations 
between the RCC and previous works. When $|{\cal S}|=1$,
$\Ch$ becomes a broadcast channel, and the coding scheme 
of the RCC coincides with that of the broadcast channel 
with confidential messages(the BCC)
investigated by Csisz\'ar and K\"orner \cite{CsiKor1}. They 
determined the stochastic \CEreg region for the BCC. 

Liang and Veeravalli \cite{lv} and Liang and Krammer \cite{lk} 
posed and investigated a new theoretical model of cooperative 
communication network called the partially/fully cooperative relay 
broadcast channel(RBC). The RCC can be regarded as a communication 
system where a security requirement is imposed on the RBC. 
In fact, setting 
$$
{\cal C}_{\rm rbc
}(\Ch) \defeq
\Cd \cap 
\{(R_0,R_1,R_{\rm e}): R_{\rm e}=0 \}, 
$$
${\cal C}_{\rm rbc
}(\Ch)$ defines the capacity region of 
a special case of the partially 
cooperative RBC. Liang and Veeravalli \cite{lv} 
and Liang and Krammer \cite{lk} 
considered the determination problem of 
${\cal C}_{\rm rbc
}(\Ch)$ and determined it for some class of relay channels. 
The determination problem of ${\cal C}_{\rm rbc
}(\Ch)$ for general $\Ch$ still remains  open. 

\section{Main Results}

In this section we state our main results. Proofs of the results 
are stated in Sections VI and VII.
\subsection{Deterministic \Rev{Coding} Case}

In this subsection we state our results on inner 
and outer bounds of ${\Cd}$. Let $U$ be an auxiliary random 
variable taking values in finite 
set ${\cal U}$. Define the set of random triples $(U,$ $X,$ $S)$ 
$\in$ ${\cal U}$
$\times{\cal X}$
$\times{\cal S}$
by
\beqno
{\cal P}_1
&\defeq &
\{(U,X,S)
:
\ba[t]{l} 
\pa {\cal U} \pa 
\leq \pa {\cal X} \pa \pa {\cal S} \pa + 3\,,
\vSpa\\
\:\:U \ra XS \ra YZ \}\,,
\ea
\eeqno
where    
$U \ra XS \ra YZ$ means that random variables $U,(X,S)$ and 
$(Y,Z)$ form a Markov chain in this order. Set 
\beqno
\tilde{\cal R}_{\rm d}^{({\rm in})}(U,X,S|\Ch)
&\defeq &
\ba[t]{l}
\{(R_0,R_1,R_{\rm e}) : R_0,R_1,R_{\rm e} \geq 0\,,
\vSpa\\
\ba{rcl}
R_0 & \leq &\min \{I(US;Y), I(U;Z|S) \}\,,
\vSpa\\
R_1 & \leq & I(X;Y|US)\,,
\vSpa\\
R_{\rm e} & \leq &[R_1-I(X;Z|US)]^{+}\,.\}\,,
\ea
\ea
\\
\tilde{\cal R}_{\rm d}^{({\rm out})}(U,X,S|\Ch)
&\defeq &
\ba[t]{l}
\{(R_0,R_1,R_{\rm e}) : R_0,R_1,R_{\rm e} \geq 0\,,
\vSpa\\
\ba[t]{rcl}
R_0 &\leq & \min \{I(US;Y),I(U;Z|S)\}\,,
\vSpa\\
R_1 &\leq & I(X;YZ|US)\,,
\vSpa\\
R_0 &+    &R_1\leq I(XS;Y)\,,
\vSpa\\
R_{\rm e} &\leq &[R_1-I(X;Z|US)]^{+}\,.\}\,,
\ea
\ea
\eeqno
where $[a]^{+}=\max\{0,a\}$. Set
\beqno
\tCdi& \defeq &
\bigcup_{(U,X,S)\in {\cal P}_1}
\tilde{\cal R}_{\rm d}^{({\rm in})}(U,X,S|\Ch)\,,
\\
\tCdo& \defeq &
\bigcup_{(U,X,S)\in {\cal P}_1}
\tilde{\cal R}_{\rm d}^{({\rm in})}(U,X,S|\Ch)\,.
\eeqno
Then we have the following. 
\begin{Th}\label{th:ddirect}
{\rm For any relay channel $\Ch$, 
$$\tCdi \subseteq \Cd\subseteq \tCdo\,.$$
}\end{Th}

An essential difference between $\tCdi$ and $\tCdo$ 
is a gap $\Delta$ given by 
\beqno
\Delta &\defeq&I(X;Y|ZUS)-[I(X;Y|US)-I(X;Z|US)]
\nonumber\\
&=&I(X;ZY|US)-I(X;Y|US)
=I(X;Z|YUS)\,.
\eeqno
Observe that
\beqa
\Delta
&=&H(Z|YUS)-H(Z|YXUS)
\nonumber\\
&{\MEq{a}}&H(Z|YUS)-H(Z|YXS)
\nonumber\\
&\leq&H(Z|YS)-H(Z|YXS)=I(X;Z|YS){\,.}
\nonumber
\eeqa
{Equality {\rm (a)}} follows from the Markov condition 
$U \to XS$ $ \to YZ$. Hence, $\Delta$ vanishes if the relay 
channel ${\Ch}=\{{\Ch}(z,y|x,s)$ $\}_{(x,s,y,z)
\in {\cal X}\times{\cal S}\times{\cal Y}\times{\cal Z}}$ 
satisfies the following:
\beq
{\Ch}(z,y|x,s)={\Ch}(z|y,s){\Ch}(y|x,s).
\eeq
The above condition is equivalent to the condition that
$X,S,Y,Z$ form a Markov chain $X \to SY\to Z$ in 
this order. Cover and El. Gamal \cite{cg} called this 
relay channel the reversely degraded relay channel. 
On the other hand, we have   
\beqa
& &I(X;Y|ZUS)
=H(Y|ZUS)-H(Y|ZXUS)
\nonumber\\
&\leq &H(Y|ZS)-H(Y|ZXS)=I(X;Y|ZS)\,,
\label{eqn:delta2}
\eeqa
where the last inequality follows from the 
Markov condition $U \to $ $XSZ\to Y$. 
From (\ref{eqn:delta2}) we can see that 
the quantity $I(X;Y|ZUS)$ vanishes 
if the relay 
channel $\Ch$ satisfies the following:
\beq
{\Ch}(z,y|x,s)={\Ch}(y|z,s){\Ch}(z|x,s).
\label{eqn:degraded}
\eeq
Hence, if the relay channel $\Ch$ satisfies (\ref{eqn:degraded}), then, 
$R_{\rm e}$ should be zero. This implies that no security on the private 
messages is guaranteed. 
The condition (\ref{eqn:degraded}) is equivalent 
to the condition that $X,S,Y,Z$ form a Markov chain $X \to SZ\to Y$ 
in this order. Cover and El. Gamal \cite{cg} called 
this relay channel the degraded relay channel. Summarizing 
the above arguments, we obtain the following 
two corollaries. 
\begin{co}\label{co:Cor1}{\rm For the reversely degraded 
relay channel $\Ch$, 
$$
\tCdi=\Cd=\tCdo\,.
$$
}
\end{co}

\begin{co}\label{co:Cor2}{\rm In the deterministic \Rev{coding} case, 
if the relay channel $\Ch$ is degraded, then no security 
on the private messages is guaranteed.
}
\end{co}

Corollary \ref{co:Cor1} implies that the suggested 
strategy in Theorem \ref{th:ddirect} is optimal 
in the case of reversely degraded relay channels.
Corollary \ref{co:Cor2} meets our intuition 
in the sense that if the relay channel is degraded, the relay can 
do anything that the destination can. 

Next we define another pair of inner and outer bounds.   
Define a set of random triples $(U,$$X,$ $S)$ 
$\in$ ${\cal U}$
$\times{\cal X}$
$\times{\cal S}$
by
\beqno
{\cal P}_2
&\defeq &
\{(U,X,S): 
\ba[t]{l} 
\pa {\cal U} \pa 
\leq \pa {\cal Z} \pa \pa {\cal X} \pa \pa {\cal S} \pa + 3\,,
\vSpa\\
U \ra XSZ \ra Y \}\,.
\ea
\eeqno
It is obvious that ${\cal P}_1\subseteq{\cal P}_2$. 
Set
\beqno
& & 
{\cal R}_{\rm d}^{({\rm in})}(U,X,S|\Ch)
\\
&\defeq &
\ba[t]{l}
\{(R_0,R_1,R_{\rm e}): R_0,R_1,R_{\rm e}\geq 0\,,
\vSpa\\
\ba[t]{rcl}
R_0&\leq &\min \{I(US;Y), I(U;Z|S)\}\,,
\vSpa\\
R_0&+& R_1 \leq I(X;Y|US)
\vSpa\\
& &+\min \{I(U;Z|S),I(US;Y) \}\,,
\vSpa\\
R_{\rm e} &\leq & [R_1-I(X;Z|US)]^{+}\,, 
\vSpa\\
R_{\rm e} &\leq & [I(X;Y|US)-I(X;Z|US)]^{+}\,.\}\,,
\ea
\ea
\\
& &
{\cal R}_{\rm d}^{({\rm out})}(U,X,S|\Ch)
\\
&\defeq &
\ba[t]{l}
\{(R_0,R_1,R_{\rm e}) : R_0,R_1,R_{\rm e} \geq 0\,,
\vSpa\\
\ba[t]{rcl}
R_0&\leq &\min \{I(US;Y), I(U;Z|S)\}\,,
\vSpa\\
R_0&+& R_1 \leq I(X;Y|US)
\vSpa\\
& &+\min \{I(U;Z|S),I(US;Y) \}\,,
\vSpa\\
R_{\rm e} &\leq &[R_1-I(X;Z|US)+I(U;Z|XS)]^{+}\,,
\vSpa\\
R_{\rm e} &\leq & [I(X;Y|US)-I(X;Z|US)]^{+}\,.\}\,.
\ea
\ea
\eeqno
Furthermore, set
\beqno
\Cdi& \defeq &
\bigcup_{(U,X,S)\in {\cal P}_1}
{\cal R}_{\rm d}^{({\rm in})}(U,X,S|\Ch)\,,
\\
\Cdo& \defeq &
\bigcup_{(U,X,S)\in {\cal P}_2}
{\cal R}_{\rm d}^{({\rm out})}(U,X,S|\Ch)\,.
\eeqno
Then, we have the following theorem. 
\begin{Th}\label{th:main2}
{\rm For any relay channel $\Ch$,
$$ \tCdi \subseteq \Cdi \subseteq \Cd \subseteq \Cdo\,.$$
\label{th:dconv}
}\end{Th}

Here we consider a class of relay channels in which 
$Z$ is a function of $XS$. We call this class of relay 
channels the semi deterministic relay channel. If $\Ch$ 
is semi deterministic, 
$U\to XS\to Z$
for any $(U,X,S)\in{\cal P}_2$. On the other hand,  
we have
$U\to ZXS\to Y$
for any $(U,X,S)\in{\cal P}_2$. 
From those two Markov chains
we have $U\to XS \to YZ$, which implies that $\Cdi$$=\Cdo$.  
Summarizing the above argument we have the following.
\begin{co}If $\Ch$ belongs to the class of 
semi deterministic relay channels, 
$$\Cdo =\Cd=\Cdi\,.$$ 
\end{co}

Finally, we derive the third outer bound of $\Cd$ which 
is effective for a certain class of relay channels. We 
consider the case where the relay channel 
$\Ch$ satisfies 
\beq
{\Ch}(y,z|x,s)={\Ch}(y|z,x,s){\Ch}(z|x).
\label{eqn:indep}
\eeq
The above condition on $\Ch$ is equivalent to the 
condition that $X,S,Z$ satisfy the Markov chain 
$S\to X \to Z$. 
This condition corresponds to a situation where 
the outputs of the relay encoder does not directly 
affect the communication from the sender to the relay.
This situation can be regarded as a natural communication 
link in practical relay communication systems. 
In this sense we say that the relay channel $\Ch$ belongs 
to the class of natural communication link or briefly 
the class NL if it satisfies (\ref{eqn:indep}). 
%
%

For given 
$(U,X,S)$ 
$\in$ ${\cal U}$
$\times{\cal X}$
$\times{\cal S}$,
set
\beqno 
& &\hat{\cal R}_{\rm d}^{({\rm out})}(U,X,S|\Ch)
\\
&\defeq &
\ba[t]{l}
\{(R_0,R_1,R_{\rm e}): R_0,R_1,R_{\rm e}\geq 0\,,
\vSpa\\
\ba[t]{rcl}
R_0&\leq &\min \{I(U;Y),I(U;Z|S)\}\,,
\vSpa\\
R_0&+&R_1\leq I(X;Y|US)
\vSpa\\
& &+\min \{I(US;Y),I(U;Z|S)+\ZeTa\}\,,
\vSpa\\
R_{\rm e} &\leq & [R_1-I(X;Z|US)]^{+}\,, 
\vSpa\\
R_{\rm e}&\leq  & [I(X;Y|US)-I(X;Z|US)
+\ZeTa]^{+}\,.\}\,,
\ea
\ea
\eeqno
where we set 
\beqno
\ZeTa 
&\defeq&   
I(XS;Y|U)-I(XS;Z|U)
\nonumber\\
& &-[I(X;Y|US)-I(X;Z|US)]
\nonumber\\
&=&I(S;Y|U)-I(S;Z|U)
\nonumber\\
&=&H(S|ZU)-H(S|YU)\,.
\eeqno
The quantity $\ZeTa$ satisfies the following. 
\begin{pr} \label{pro:pro1} For any $(U,X,S)\in {\cal P}_2$,   
\beqno
\ZeTa & \leq & 
{\min\{H(S|Z),\irBl{I(XS;Y|Z)}\}}\,.
\label{eqn:zeta}
\eeqno
\end{pr}

{\it Proof:} 
{It is obvious that $\ZeTa \leq H(S|Z)$.
We prove $\ZeTa$ $\leq I(XS;Y|Z)$.}
We have the following chain of inequalities:
\beqa
& &\ZeTa
= H(S|ZU)-H(S|YU)
\nonumber\\
&\leq& H(S|ZU)-H(S|YZU)= I(S;Y|ZU)
\nonumber\\
& = & H(Y|ZU)-H(Y|ZUS)
\nonumber\\
&\leq & H(Y|Z)-H(Y|ZUS)
\nonumber\\
&\leq & H(Y|Z)-H(Y|ZXSU)
\nonumber\\
&= & H(Y|Z)-H(Y|ZXS)=I(XS;Y|Z),
\nonumber
\eeqa
where the last equality follows from 
the Markov condition $U\to$ 
$ZXS \to Y.$
\hfill\QED  

Set 
\beqno
\hCdo& \defeq &
\bigcup_{(U,X,S)\in {\cal P}_1}
\hat{\cal R}_{\rm d}^{(\rm out)}(U,X,S|\Ch)\,.
\eeqno
Our result is the following. 
\begin{Th}\label{th:thDI}{\rm If $\Ch$ belongs to the \Cls, 
we have 
$$
\Cd\subseteq \hCdo\,.
$$
}
\end{Th}

It is obvious that if $\ZeTa\leq 0$ for $(U,X,S)\in {\cal P}_1$, 
we have  
$$
\Cdi=\Cd=\hCdo.
$$
By Property \ref{pro:pro1}, the condition that 
\beq
\min\{H(S|Z),I(XS,Y|Z)\}=0
\mbox{ for any }(X,S)
\label{eqn:Cond}
\eeq
is a sufficient condition for $\ZeTa$ to be non positive 
on $(U,X,S)\in {\cal P}_1$. 
The condition (\ref{eqn:Cond}) on $\Ch$ 
is very severe. We do not have found so for any 
effective condition on $\Ch$ such that 
$\ZeTa\leq 0$ for any $(U,X,S)\in {\cal P}_1$. 
When $|{\cal S}|=1$, then by Property \ref{pro:pro1}, 
we have $\ZeTa\leq 0$. Hence $\hCdo$ coincides with $\Cdi$. 
In this case, the {\Cls} becomes a class of general 
broadcast channels with one output and two input. 
Thus, the coding strategy achieving $\Cdi$ in Theorem \ref{th:main2} is optimal 
in the case of BCC and deterministic coding.

\subsection{Stochastic Encoding Case}

In this subsection we state our results on inner and 
outer bounds of ${\Cs}$. Set 
\beqno
& &\tilde{\cal R}_{\rm s}^{({\rm in})}(U,X,S|\Ch)
\\
&\defeq &
\ba[t]{l}
\{(R_0,R_1,R_{\rm e}):0\leq R_0,0\leq R_{\rm e}\leq R_1,
\vSpa\\
\ba{rcl}
R_0 & \leq &\min \{I(US;Y), I(U;Z|S) \}\,,
\vSpa\\
R_1 & \leq & I(X;Y|US)\,,
\vSpa\\
R_{\rm e} & \leq & [I(X;Y|US)-I(X;Z|US)]^{+}
\}\,,
\ea
\ea
\\
& &\tilde{\cal R}_{\rm s}^{({\rm out})}(U,X,S|\Ch)
\\
&\defeq &
\ba[t]{l}
\{(R_0,R_1,R_{\rm e}) : 0\leq R_0,0\leq R_{\rm e}\leq R_1,
\vSpa\\
\ba[t]{rcl}
R_0 &\leq & \min \{I(US;Y),I(U;Z|S)\}\,,
\vSpa\\
R_1 &\leq & I(X;YZ|US)\,,
\vSpa\\
R_0 &+& R_1\leq I(XS;Y)\,,
\vSpa\\
R_{\rm e} & \leq & I(X;Y|ZUS)\,.\}\,.
\ea
\ea
\eeqno
Furthermore, set
\beqno
\tCsi& \defeq &
\bigcup_{(U,X,S)\in {\cal P}_1}
\tilde{\cal R}_{\rm s}^{({\rm in})}(U,X,S|\Ch)\,,
\\
\tCso& \defeq &
\bigcup_{(U,X,S)\in {\cal P}_1}
\tilde{\cal R}_{\rm s}^{({\rm out})}(U,X,S|\Ch)\,.
\eeqno
We further present another pair of inner and outer bounds of 
$\Cs$. To this end define sets of random quadruples 
$(U,$$V,$$X,$$S)$ 
$\in$ ${\cal U}$
$\times{\cal V}$
$\times{\cal X}$
$\times{\cal S}$
by
\beqno
{\cal Q}_1
&\defeq &
\ba[t]{l} 
\{(U,V,X,S):
\pa {\cal U} \pa 
\leq  \pa {\cal X}\pa\pa{\cal S}\pa + 3,
\vSpa\\
\:\:\pa {\cal V} \pa 
  \leq \left(\pa {\cal X} \pa\pa {\cal S}\pa\right)^2  
  +4\pa {\cal X}\pa\pa{\cal S}\pa + 3,
\vSpa\\
\:\:U \ra V \ra XS \ra YZ,
US \ra V \ra X.\},
\ea
\\
{\cal Q}_2
&\defeq &
\ba[t]{l} 
\{(U,V,X,S):
\pa {\cal U} \pa 
\leq \pa {\cal Z}\pa\pa{\cal X}\pa\pa{\cal S} \pa + 3,
\vSpa\\
\:\:\pa {\cal V} \pa 
  \leq \left(\pa{\cal Z}\pa\pa{\cal X}\pa\pa{\cal S}\pa\right)^2  
  +4\pa {\cal Z}\pa\pa{\cal X}\pa\pa{\cal S}\pa + 3,
\vSpa\\
\:\:U \ra V \ra XSZ \ra Y,
US \ra VX\ra Z,
\vSpa\\
\:\:US \ra V \ra X.\}.
\ea
\eeqno
It is obvious that ${\cal Q}_1\subseteq{\cal Q}_2$. For given 
$(U,V,X,S)$ 
$\in$ ${\cal U}$
$\times{\cal V}$
$\times{\cal X}$
$\times{\cal S}$,
set
\beqno
& & {\cal R}(U,V,X,S|\Ch)
\\
&\defeq &
\ba[t]{l}
\{(R_0, R_1, R_{\rm e}): 0\leq R_0,0\leq R_{\rm e}\leq R_1, 
\vSpa\\
\ba[t]{rcl}
R_0&\leq &\min \{I(US;Y),I(U;Z|S)\},
\vSpa\\
R_0&+& R_1 \leq I(V;Y|US)+\min \{I(US;Y),I(U;Z|S)\},
\vSpa\\
R_{\rm e}&\leq & [I(V;Y|US)-I(V;Z|US)]^{+}.\}.
\ea
\ea
\eeqno
Furthermore, set 
\beqno
\Csi&\defeq&
\bigcup_{(U,V,X,S)\in {\cal Q}_1}
{\cal R}(U,V,X,S|\Ch),
\\
\Cso&\defeq&
\bigcup_{(U,V,X,S)\in {\cal Q}_2}
{\cal R}(U,V,X,S|\Ch).
\eeqno
Our capacity results in the case of stochastic encoding are 
as follows. 
\begin{Th}\label{th:sreg1}
{\rm For any relay channel $\Ch$, 
$$\tCsi \subseteq \Cs\subseteq \tCso\,.$$
}\end{Th}

\begin{Th}\label{th:sreg2} {\rm For any relay channel $\Ch$, 
$$\tCsi \subseteq \Csi\subseteq \Cs \subseteq \Cso \,.$$
}
\end{Th}

The above two theorems together with arguments similar 
to those in the case of deterministic coding yield 
the following three corollaries.  
\begin{co}\label{co:Cor1z}{\rm If the the relay 
channel $\Ch$ is reversely degraded, 
$$
\tCsi=\Cs=\tCso\,.
$$
}
\end{co}

\begin{co}\label{co:Cor2z}{\rm If  
the relay channel $\Ch$ is semi deterministic, 
$$
\Csi=\Cs=\Cso\,.
$$
}
\end{co}

\begin{co}\label{co:Cor3z}{
If the relay channel $\Ch$ is degraded, then no security 
on the private messages is guaranteed 
even if $f_n$ is a stochastic encoder.    
}
\end{co}

When $|{\cal S}|=1$, the reversely degraded relay channel 
becomes the degraded broadcast channel. Wyner \cite{Wyn1} 
discussed the wire-tap channel in the case of 
degraded broadcast channels. Corollary \ref{co:Cor1z} 
can be regarded as an extension of his result to the case 
where wire-tapper may assist the transmission of common 
messages. Corollary \ref{co:Cor3z} meets our intuition 
in the sense that if the relay channel is degraded, the relay can 
do anything that the destination can. 

\subsection{Stochastic Relay Function} 

In this subsection we state our results 
in the case where the relay may use 
a stochastic encoder.  
Let ${\cal R}_{\rm d}^{\ast}(\Ch)$
and ${\cal R}_{\rm s}^{\ast}(\Ch)$ 
be denoted by the deterministic and stochastic 
rate equivocation regions\Irr{, respectively,} in 
the case where the stochastic relay encoder may be used. 
It is obvious that 
$\tCdi$ and $\Cdi$ still serve as inner bounds of 
${\cal R}_{\rm d}^{\ast}(\Ch)$. Similarly, 
$\tCsi$ and $\Csi$ serve as inner bounds of 
${\cal R}_{\rm s}^{\ast}(\Ch)$. 
Our capacity results on outer bounds 
in the case of stochastic relay encoder \Irr{are} described 
in the following theorem. 
\begin{Th}\label{th:sdRegssReg} \  
If $\Ch$ belongs to the \Cls, $\tCdo$, $\Cdo$, 
and $\hCdo$ still serve as outer bounds 
of ${\cal R}_{\rm d}^{\ast}(\Ch)$.
Similarly, if $\Ch$ belongs to the \Cls, 
$\tCso$ and $\Cso$ still serve as outer bounds of ${\cal R}_{\rm s}^{\ast}(\Ch)$.
\end{Th} 

\section{Secrecy Capacities of the RCC} 

In this section we derive explicit inner and outer bounds of the secrecy
capacity region by using the results in the previous section. 
We first consider the special case of no common message. Define  
\beqno
{\Cde}&\defeq &\{(R_1,R_{\rm e}):(0,R_1,R_{\rm e})\in \Cd \}\,,
\\
{\Cse}&\defeq & \{(R_1,R_{\rm e}):(0,R_1,R_{\rm e})\in \Cs \}\,.
\eeqno
To state a result on $\Cde$ and $\Cse$ set 
\beqno
\tilde{\cal R}_{\rm d1e}^{\rm (in)}(U,X,S|\Ch)
&\defeq &
\ba[t]{l}
\{(R_1,R_{\rm e}): R_1,R_{\rm e} \geq 0\,, 
\vSpa\\
\ba[t]{rcl}
R_1 &\leq &I(X;Y|US)\,,
\vSpa\\
R_{\rm e} &\leq & [R_1-I(X;Z|US)]^{+}\,.\}\,,
\ea
\ea
\\
\tilde{\cal R}_{\rm d1e}^{\rm (out)}(U,X,S|\Ch)
&\defeq &
\ba[t]{l}
\{(R_1,R_{\rm e}): R_1,R_{\rm e} \geq 0\,, 
\vSpa\\
\ba[t]{rcl}
R_1 &\leq &I(X;YZ|US)\,,
\vSpa\\
R_{\rm e} &\leq & [R_1-I(X;Z|US)]^{+}\,.\}\,,
\ea
\ea
\\
\tilde{\cal R}_{\rm s1e}^{\rm (in)}(U,X,S|\Ch)
&\defeq &
\ba[t]{l}
\{(R_1,R_{\rm e}): R_1,R_{\rm e} \geq 0\,, 
\vSpa\\
\ba[t]{rcl}
R_{\rm e} &\leq & R_1\leq I(X;Y|US)\,,
\vSpa\\
R_{\rm e} &\leq & [I(X;Y|US)
\vSpa\\
             &     & -I(X;Z|US)]^{+}\,.\}\,,
\ea
\ea
\\
\tilde{\cal R}_{\rm s1e}^{\rm (out)}(U,X,S|\Ch)
&\defeq &
\ba[t]{l}
\{(R_1,R_{\rm e}): R_1,R_{\rm e} \geq 0\,, 
\vSpa\\
\ba[t]{rcl}
R_{\rm e} &\leq & R_1\leq I(X;YZ|US)\,,
\vSpa\\
R_{\rm e} &\leq & I(X;Y|ZUS)\,.\}\,,
\ea
\ea
\\
\tCdei&\defeq&
\bigcup_{(U,X,S)\in {\cal P}_1}
\tilde{\cal R}_{\rm d1e}^{(\rm in)}(U,X,S|\Ch)\,,
\\
\tCdeo&\defeq&
\bigcup_{(U,X,S)\in {\cal P}_1}
\tilde{\cal R}_{\rm d1e}^{(\rm out)}(U,X,S|\Ch)\,,
\\
\tCsei&\defeq&
\bigcup_{(U,X,S)\in {\cal P}_1}
\tilde{\cal R}_{\rm s1e}^{(\rm in)}(U,X,S|\Ch)\,,
\\
\tCseo&\defeq&
\bigcup_{(U,X,S)\in {\cal P}_1}
\tilde{\cal R}_{\rm s1e}^{(\rm out)}(U,X,S|\Ch)\,.
\eeqno
From Theorems \ref{th:ddirect} and \ref{th:sreg1}, we have 
the following corollary. 
\begin{co}\label{co:aaa001} \ For any relay channel $\Ch$, 
\beqno
& & \tCdei \subseteq \Cde \subseteq \tCdeo\,, 
\\
& & \tCsei \subseteq \Cse \subseteq \tCseo\,. 
\eeqno
In particular, if $\Ch$ is reversely degraded, 
\beqno
&& \tCdei=\Cde = \tCdeo,
\\  
&& \tCsei=\Cse = \tCseo.  
\eeqno
\end{co}

\begin{figure}[t]
\bc
\includegraphics[width=6.0cm]{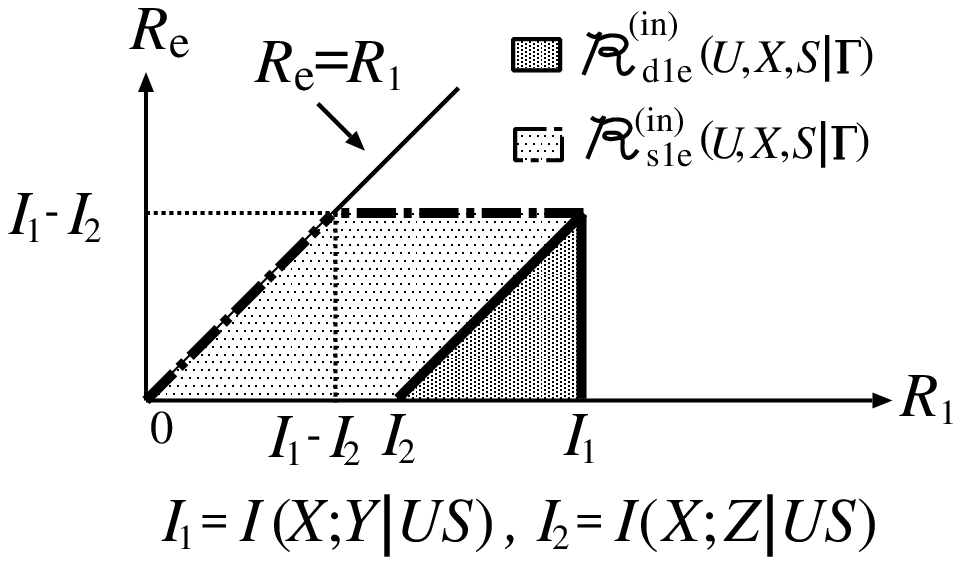}
\caption{ 
The regions 
$\tilde{\cal R}_{\rm d1e}^{(\rm in)}(U,X,S|\Ch)$
and 
$\tilde{\cal R}_{\rm s1e}^{(\rm in)}(U,X,S|\Ch)$.
} 
\label{fig:RegS}
\ec
\vspace*{-3mm}
\end{figure}

Now we consider the case where $\Ch$ is reversely degraded.
In this case we compare $\tCdei=\Cde$ and $\tCsei=\Cse$.  
The regions
$\tilde{\cal R}_{\rm d1e}^{(\rm in)}(U,X,S|\Ch)$
and 
$\tilde{\cal R}_{\rm s1e}^{(\rm in)}(U,X,S|\Ch)$
in this case are shown in Fig. \ref{fig:RegS}. 
It can be seen from this figure that 
the region $\tilde{\cal R}_{\rm d1e}^{(\rm in)}(U,X,S|\Ch)$ 
is strictly smaller than 
$\tilde{\cal R}_{\rm s1e}^{(\rm in)}(U,X,S|\Ch)$. 
In $\tilde{\cal R}_{\rm s1e}^{(\rm in)}(U,X,S|\Ch)$,  
the point $(R_1^*,R_{\rm e}^*)$ whose components are given by
\beq
R_1^{*}=R_{\rm e}^{*}=I(X;Y|US)-I(X;Z|US)
\eeq  
belongs to ${\cal R}_{\rm s1e}(\Ch)$. This implies that 
the relay is kept completely ignorant of the private 
message. In this case we say that 
the perfect secrecy on the private message 
is established. The stochastic secrecy capacity 
region ${\Css}$ and the secrecy capacity 
$C_{\rm ss}(\Ch)$ for the RCC are defined by 
\beqno
&&\Css\defeq\{(R_0,R_1):(R_0,R_1,R_1)\in \Cs \}\,,
\label{df:Css_region}
\\
&&C_{\rm ss}(\Ch)\defeq 
 \max_{(R_1,R_1)\in \Cse}R_1
=\max_{(0,R_1)\in \Css}R_1\,.
\eeqno
On the other hand, if we require the perfect secrecy in the 
case of deterministic encoding, we must have  
$R_1=R_{\rm e}$ for $(R_1,R_{\rm e})$ 
$\in {\cal R}_{\rm d1e}(\Ch)$. 
Then, it follows from Corollary \ref{co:aaa001} that 
if $\Gamma$ is reversely degraded, we must have 
\beq
I(X,Z|US)=0 \mbox{ for }(U,X,Y)\in{\cal P}_1.
\label{eqn:abz}
\eeq 
This condition is very hard to hold in general. 
Thus the prefect secrecy on private message 
can seldom be attained by the 
deterministic encoding. Another criterion of comparing 
$\Cd$ and $\Cs$ is the maximum equivocation rate in the 
\CEreg region. For $\Cd$ and $\Cs$, those are formally 
defined by 
\beqno
C_{\rm de}(\Ch)&\defeq&
\max_{\scs (R_0,R_1,R_{\rm e})\atop{\scs\in \Cd}} R_{\rm e }
\mbox{ and }
C_{\rm se}(\Ch)\defeq\max_{\scs (R_0,R_1,R_{\rm e})\atop{\scs \in \Cs}}R_{\rm e },
\eeqno
respectively. We describe our results on  
$\Css$, $C_{\rm de}(\Ch)$, $C_{\rm ss}(\Ch)$, and 
$C_{\rm se}(\Ch)$ which are obtained as corollaries 
of Theorems \ref{th:ddirect} and \ref{th:sreg1}. Set    
\beqno
\tCssi
&\defeq &
\ba[t]{l}
\{(R_0,R_1) : R_0,R_1\geq 0\,,
\vSpa\\
R_0 \leq \min \{I(US;Y), I(U;Z|S)\}\,,
\vSpa\\
R_1 \leq [I(X;Y|US)-I(X;Z|US)]^{+}\,,
\vSpa\\
\mbox{ for some }(U,X,S)\in {\cal P}_1\,. \}\,,
\ea
\\
\tCsso
&\defeq &
\ba[t]{l}
\{(R_0,R_1) : R_0,R_1\geq 0\,,
\vSpa\\
R_0 \leq \min \{I(US;Y), I(U;Z|S)\}\,,
\vSpa\\
R_1 \leq I(X;Y|ZUS)\,,
\vSpa\\
\mbox{ for some }(U,X,S)\in {\cal P}_1\,. \}\,.
\ea
\eeqno
Then we have the following.
\begin{co} For any relay channel $\Ch$, 
\beqno
\tCssi\subseteq\Css\subseteq\tCsso\,.
\eeqno
Furthermore, we have 
\beqno
&     & \max_{(X,S)}
\left[I(X;Y|S)-I(X;Z|S)\right]^{+}
\nonumber\\
&\leq & \max_{(U,X,S)\in {\cal P}_1}
\left[I(X;Y|US)-I(X;Z|US)\right]^{+}
\nonumber\\
&\leq & C_{\rm de}(\Ch)\leq C_{\rm ss}(\Ch)\leq C_{\rm se}(\Ch)
\nonumber\\
&\leq & \max_{(U,X,S)\in {\cal P}_1}I(X;Y|ZUS)
= \max_{(X,S)}I(X;Y|ZS)\,.
\eeqno
In particular, if $\Ch$ is reversely degraded, we have 
$$
\tCssi=\Css = \tCsso 
$$
and 
\beqno
& &C_{\rm de}(\Ch)= C_{\rm ss}(\Ch)= C_{\rm se}(\Ch)
\\
&=&\max_{(X,S)}\left[I(X;Y|S)-I(X;Z|S)\right]\,.
\eeqno
\end{co}

\begin{figure}[t]
\bc
\includegraphics[width=8.0cm]{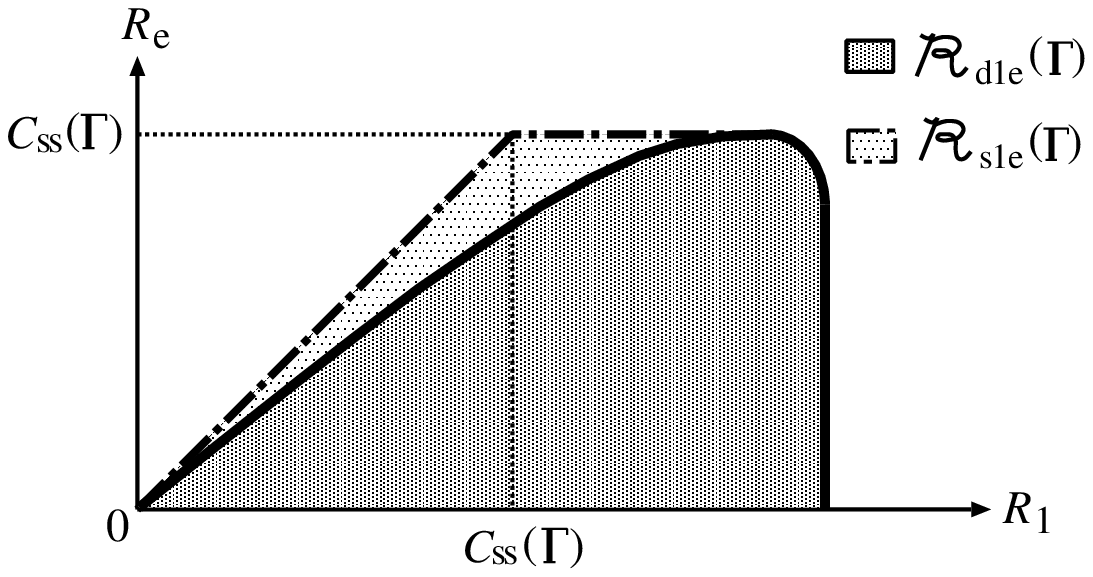}
\caption{The regions $\Cde=\Cdei$ and $\Cse=\Csei$ 
and 
$C_{\rm ss}(\Ch)=$ 
$C_{\rm de}(\Ch)=$ 
$C_{\rm se}(\Ch)$ 
for the reversely degraded relay channels.
} 
\label{fig:Cs_region}
\ec
\vspace*{-3mm}
\end{figure}

Typical shapes of the regions $\Cde$ and $\Cse$ 
in the case of reversely degraded relay channels 
are shown in Fig. \ref{fig:Cs_region}. The secrecy 
capacity $C_{\rm ss}(\Ch)$ is also shown in this figure.
Next, we state a result which is obtained as 
a corollary of Theorems \ref{th:dconv} 
and \ref{th:sreg2}. To state this result, set
\beqno
& &{\cal R}_{\rm d1e}^{\rm (in)}(U,X,S|\Ch)
\\
&\defeq & {\cal R}_{\rm d}^{\rm (in)}(U,X,S|\Ch)
\cap\{(R_0,R_1,R_{\rm e}): R_0=0\}
\\
&=&
\ba[t]{l}
\{(R_1,R_{\rm e}): R_1,R_{\rm e}\geq 0\,,
\vSpa\\
\ba[t]{rcl}
R_1 &\leq & I(X;Y|US)+\min \{I(US;Y),I(U;Z|S)\}\,,
\vSpa\\
R_{\rm e} & \leq & [R_1-I(X;Z|US)]^{+}\,,
\vSpa\\
R_{\rm e} &\leq & [I(X;Y|US)-I(X;Z|US)]^{+}\,.\},
\ea
\ea
\\
& &{\cal R}_{\rm d1e}^{\rm (out)}(U,X,S|\Ch)
\\
&\defeq & {\cal R}_{\rm d}^{\rm (out)}(U,X,S|\Ch)
\cap\{(R_0,R_1,R_{\rm e}): R_0=0\}
\\
&=&
\ba[t]{l}
\{(R_1,R_{\rm e}): R_1,R_{\rm e} \geq 0\,,
\vSpa\\
\ba[t]{rcl}
R_1 &\leq & I(X;Y|US)+\min\{I(U;Z|S),I(US;Y)\}\,,
\vSpa\\
R_{\rm e} &\leq &[R_1-I(X;Z|US)+I(U;Z|XS)]^{+}\,,
\vSpa\\
R_{\rm e} &\leq & [I(X;Y|US)-I(X;Z|US)]^{+}\,.\},
\ea
\ea
\\
& &{\cal R}_{\rm 1e}(U,V,X,S|\Ch)
\\
&\defeq & {\cal R}(U,V,X,S|\Ch)
\cap\{(R_0,R_1,R_{\rm e}): R_0=0\}
\\
&=&
\ba[t]{l}
\{(R_1,R_{\rm e}): 0\leq R_{\rm e}\leq R_1\,,
\vSpa\\
\ba[t]{rcl}
R_{1} &\leq & I(V;Y|US)+\min \{I(US;Y),I(U;Z|S)\}\,,
\vSpa\\
R_{\rm e} &\leq &[I(V;Y|US)-I(V;Z|US)]^{+}\,.\},
\ea
\ea
\\
&&\Cdei\defeq
\bigcup_{(U,X,S)\in {\cal Q}_1}
{\cal R}_{\rm d1e}^{\rm (in)}(U,X,S|\Ch)\,,
\\
&&\Cdeo\defeq
\bigcup_{(U,X,S)\in {\cal Q}_2}
{\cal R}_{\rm d1e}^{\rm (out)}(U,X,S|\Ch)\,,
\\
&&\Csei\defeq
\bigcup_{(U,V,X,S)\in {\cal Q}_1}
{\cal R}_{\rm 1e}(U,V,X,S|\Ch)\,,
\\
&&\Cseo\defeq
\bigcup_{(U,V,X,S)\in {\cal Q}_2}
{\cal R}_{\rm 1e}(U,V,X,S|\Ch)\,.
\eeqno
Furthermore, set
\beqno
& &{\cal C}_{\rm s}(U,V,X,S|\Ch)
\\
&\defeq& {\cal R}(U,V,X,S|\Ch)\cap
\{(R_0,R_1,R_{\rm e}):R_1=R_{\rm e}\}
\\
&=&
\ba[t]{l}
\{(R_0,R_1): R_0,R_1\geq 0\,,
\vSpa\\
R_0 \leq \min \{I(US;Y),I(U;Z|S)\}\,,
\vSpa\\
R_1 \leq [I(V;Y|US)-I(V;Z|US)]^{+}\,\},
\vSpa\\
\ea
\\
&&\Cssi\defeq 
\bigcup_{(U,V,X,S)\in {\cal Q}_1} {\cal C}_{\rm s}(U,V,X,S|\Ch)\,,
\\ 
&&\Csso\defeq 
\bigcup_{(U,V,X,S)\in {\cal Q}_2} {\cal C}_{\rm s}(U,V,X,S|\Ch)\,.
\eeqno
From Theorems \ref{th:dconv} and  \ref{th:sreg2}, 
we have the following corollary.
\begin{co} For any relay channel $\Ch$, 
\beqno
&&\Cdei \subseteq \Cde \subseteq \Cdeo \subseteq \Cseo\,, 
\\
&&\Cdei \subseteq \Csei \subseteq \Cse \subseteq \Cseo\,, 
\\
&&\Cssi \subseteq \Css \subseteq \Csso\,. 
\eeqno
Furthermore, 
\beqno
& &\max_{(U,X,S)\in {\cal P}_1}
\left[I(X;Y|US)-I(X;Z|US)\right]^{+}
\\
&\leq &C_{\rm de}(\Ch)
\\
&\leq & \max_{(U,X,S)\in {\cal P}_2}
\left[
I(V;Y|US)-I(V;Z|US)
\right]^{+}\,,
\\
& &\max_{(U,V,X,S)\in {\cal Q}_1}
\left[I(V;Y|US)-I(V;Z|US)\right]^{+}
\\
&\leq &C_{\rm ss}(\Ch)\leq C_{\rm se}(\Ch)
\\
&\leq & \max_{(U,V,X,S)\in {\cal Q}_2}
\left[
I(V;Y|US)-I(V;Z|US)
\right]^{+}\,.
\eeqno
If $\Ch$ is semi deterministic, then 
\beqno
\Cdei=\Cde=\Cdeo\,, 
\\
\Csei=\Cse=\Cseo\,, 
\\
\Cssi=\Css=\Csso\,. 
\eeqno
Furthermore, 
\beqno
C_{\rm de}(\Ch)
&=&\max_{(U,X,S)\in {\cal P}_1}
\left[I(X;Y|US)-I(X;Z|US)\right]^{+}\,,
\\
C_{\rm ss}(\Ch)&=&C_{\rm se}(\Ch)
\\
&=&\max_{(U,V,X,S)\in {\cal Q}_1}
\left[I(V;Y|US)-I(V;Z|US)\right]^{+}\,.
\eeqno
\end{co}

It can be seen from the above corollary that 
$C_{\rm se}(\Ch)$ may strictly be larger than 
$C_{\rm de}(\Ch)$ unless $\Ch$ is reversely degraded. 
By a simple analytical argument we can show that
$C_{\rm ss}^{\rm (in)}(\Ch)$ can be attained by $S=s^{*}$, 
where $s^*\in {\cal S}$ is the best input alphabet which 
maximizes the secrecy rate
$$
\max_{(V,U,X,S=s^*)\in {\cal Q}_1}
\left\{I(V;Y|US=s^*)- I(V;Z|US=s^*)\right\}\,.
$$
This implies that the coding strategy achieving 
$C_{\rm ss}^{\rm (in)}(\Ch)$ does not help improving the secrecy rate 
compared with the case where the relay is simply a wire-tapper, 
except that the relay may choose 
the best $S=s^*$ to benefit the receiver.
Cover and El Gamal \cite{cg} introduced a transmission scheme of the relay called 
the compress-and-forward scheme, where the relay transmits a quantized 
version of its received signal. This scheme is also applicable to the RCC. 
He and Yener \cite{heye}, \cite{heye3} derived an inner bound of $\Cse$ in the case 
where the relay employs the compress-and-forward scheme to show 
that the relay may improve the secrecy capacity.
%
%
%
%
\section{Gaussian Relay Channels with Confidential Messages}

In this section we study Gaussian relay channels with 
confidential messages, where two channel outputs 
are corrupted by additive white Gaussian noises.
Let $(\xi_1, \xi_2)$ be correlated zero mean Gaussian 
random vector with covariance matrix 
$$
\Sigma=\left(
\ba{cc}
N_1& \rho\sqrt{N_1N_2}
\\
\rho\sqrt{N_1N_2} & N_2
\ea
\right)\,, |\rho| <1\,.
$$
Let $\{(\xi_{1,i},\xi_{2,i})\}_{i=1}^{\infty}$ be a sequence of 
independent identically distributed (i.i.d.) zero mean Gaussian 
random vectors. Each $(\xi_{1,i},\xi_{2,i})$ has the covariance 
matrix $\Sigma$. 
The Gaussian relay channel is specified by the above 
covariance matrix $\Sigma$. Two channel outputs $Y_i$ and $Z_i$ 
of the relay channel at the $i$th transmission are {given} by 
\beqno
Y_i&=&X_i+S_i +\xi_{1,i}\,,
Z_i=X_i+\xi_{2,i}\,.
\eeqno
It is obvious that 
$\Sigma$ belongs to the class NL. In this class of 
Gaussian relay channels we assume that the relay 
encoder $\{ g_i \}_{i=1}$ is allowed to be 
stochastic. Since $(\xi_{1,i},\xi_{2,i}), i=1,2,\cdots,n$ have 
the covariance matrix $\Sigma$, we have   
$$
\xi_{2,i}= \rho \sqrt{\frac{N_2}{N_1}}\xi_{1,i}+ \xi_{2|1,i}\,,
$$
where $\xi_{2|1,i},i=1,2,\cdots,n $ are zero mean Gaussian 
random variable with variance $(1-\rho^2)N_2$ and 
independent of $\xi_{1,i}$. 
In particular if $\Sigma$ satisfies $N_1 \leq N_2$ 
and $\rho=\sqrt{\frac{N_1}{N_2}}$, we have for 
$i=1,2,$ $\cdots, n$,
\beqno
\ba{l}
Y_i=X_i+ S_i +\xi_{1,i},Z_i=X_i+ \xi_{1,i}+ \xi_{2|1,i}
\ea
\eeqno
which implies that for $i=1,2,$ $\cdots, n$,
$Z_i \to (Y_i,S_i) \to X_i$. Hence, the Gaussian relay channel
becomes reversely degraded relay channel.  
Two channel input sequences $\{X_i\}_{i=1}^n$ 
and $\{S_i\}_{i=1}^n$ are subject to 
the following average power constraints:
\beqno
\frac{1}{n}\sum_{i=1}^n{\rm \bf E}\left[X_i^2\right]\leq P_1\,,
\frac{1}{n}\sum_{i=1}^n{\rm \bf E}\left[S_i^2\right]\leq P_2\,.
\eeqno
Let 
${\cal R}_{\rm d}(P_1,P_2|\Sigma)$ 
and  
${\cal R}_{\rm s}(P_1,P_2|\Sigma)$ be 
\CEreg regions for the above Gaussian relay channel when 
we use deterministic and stochastic encoders, respectively. 
To state our results on ${\cal R}_{\rm d}(P_1,P_2|\Sigma)$ 
and ${\cal R}_{\rm s}(P_1,P_2|\Sigma)$, set
\beqno
& & {\cal R}_{\rm d}^{\rm (in)}(P_1,P_2|\Sigma)
\\
&\defeq &
\ba[t]{l}
\{(R_0, R_1, R_{\rm e}): R_0, R_1, R_{\rm e}\geq 0\,,
\vSpa\\
\ba[t]{rcl}
R_0
&\leq &{\ds \max_{0\leq \eta \leq 1}}
            \min
        \ba[t]{l}\left\{
           C\left(\frac{\bar{\theta}P_1+P_2
          +2\sqrt{\bar{\theta}\bar{\eta}P_1P_2}}
          {\theta P_1+N_1}
          \right)\right.\,,
          \vSpa\\
           \quad \! 
        \left.C\left(\frac{\bar{\theta}\eta P_1}{\theta P_1+N_2}\right)
        \right\}\,,
        \ea
        \vSpa\\
R_1 &\leq &
    C \left(\frac{\theta P_1}{N_1}\right)\,,
\vSpa\\
R_{\rm e} &\leq & \left[
                  R_1-C \left(\frac{\theta P_1}{N_2}\right)
                 \right]^{+}\,,
\vspace{2mm}\\
& & \mbox{ for some }0 \leq \theta \leq 1 \,.\}\,,
\ea
\ea
\\
& & {\cal R}_{\rm d}^{\rm (out)}(P_1,P_2|\Sigma)
\\
&\defeq &
\ba[t]{l}
\{(R_0, R_1, R_{\rm e}): R_0, R_1, R_{\rm e}\geq 0\,,
\vSpa\\
\ba[t]{rcl}
R_0
&\leq & \min
        \ba[t]{l}\left\{
           C\left(\frac{\bar{\theta}P_1+P_2
          +2\sqrt{\bar{\theta}\bar{\eta}P_1P_2}}
          {\theta P_1+N_1}
          \right)\right.\,,
          \vSpa\\
           \quad \! \left.C\left(\frac{\bar{\theta}\eta P_1}{\theta
	   P_1+N_2}
        \right)
        \right\}\,,
        \ea
        \vSpa\\
R_1 &\leq &
    C \left(\frac{\theta P_1}
{\frac{(1-\rho^2)N_1N_2 }{ N_1+N_2-2\rho \sqrt{N_1N_2}}}\right)\,,
\vSpa\\
\:\:R_0&+&R_1\leq 
       C\left(\frac{P_1+P_2 
         +2\sqrt{\bar{\theta}\bar{\eta}P_1P_2}}{N_1}
          \right)\,,
\vSpa\\
R_{\rm e} &\leq &\left[
                 R_1-C \left(\frac{\theta P_1}{N_2}\right)
                \right]^{+}\,,
\vSpa\\
& & \mbox{ for some }0 \leq \theta \leq 1, 
0 \leq \eta \leq 1 \,.\}{\,,}
\ea
\ea
\eeqno
where $C(x)\defeq \frac{1}{2}\log (1+x) \,.$
Furthermore, set
\beqno
& & {\cal R}_{\rm s}^{\rm (in)}(P_1,P_2|\Sigma)
\\
&\defeq &
\ba[t]{l}
\{(R_0, R_1, R_{\rm e}): R_0, R_1, R_{\rm e}\geq 0\,,
\vSpa\\
\ba[t]{rcl}
R_0
&\leq &{\ds \max_{0\leq \eta \leq 1}}
            \min
        \ba[t]{l}\left\{
           C\left(\frac{\bar{\theta}P_1+P_2
          +2\sqrt{\bar{\theta}\bar{\eta}P_1P_2}}
          {\theta P_1+N_1}
          \right)\right.\,,
          \vSpa\\
           \quad \! 
        \left.C\left(\frac{\bar{\theta}\eta P_1}{\theta P_1+N_2}\right)
        \right\}\,,
        \ea
\vSpa\\
R_{\rm e} &\leq &R_1\leq C \left(\frac{\theta P_1}{N_1}\right)\,,
\vSpa\\
R_{\rm e}&\leq &
 \left[C \left(\frac{\theta P_1}{N_1}\right)
      -C \left(\frac{\theta P_1}{N_2}\right)\right]^{+}\,,
\vspace{2mm}\\
& & \mbox{ for some }0 \leq \theta \leq 1 \,.\}\,,
\ea
\ea
\\
& & {\cal R}_{\rm s}^{\rm (out)}(P_1,P_2|\Sigma)
\\
&\defeq &
\ba[t]{l}
\{(R_0, R_1, R_{\rm e}): R_0,R_1,R_{\rm e}\geq 0\,,
\vSpa\\
\ba[t]{rcl}
R_0
&\leq & \min
        \ba[t]{l}\left\{
           C\left(\frac{\bar{\theta}P_1+P_2
          +2\sqrt{\bar{\theta}\bar{\eta}P_1P_2}}
          {\theta P_1+N_1}
          \right)\right.\,,
          \vSpa\\
           \quad \! \left.C\left(\frac{\bar{\theta}\eta P_1}{\theta
	   P_1+N_2}
        \right)
        \right\}\,,
        \ea
        \vSpa\\
\:\:R_0&+&R_1\leq  
          C\left(\frac{P_1+P_2 
          +2\sqrt{\bar{\theta}\bar{\eta}P_1P_2}}{N_1}
          \right)\,,
\vSpa\\
R_{\rm e} &\leq &R_1\leq 
    C \left(\frac{\theta P_1}
{\frac{(1-\rho^2)N_1N_2 }{ N_1+N_2-2\rho \sqrt{N_1N_2}}}\right)\,,
\vSpa\\
R_{\rm e}&\leq &
 \left[C \left(\frac{\theta P_1}
 {\frac{(1-\rho^2)N_1N_2}{N_1+N_2-2\rho\sqrt{N_1N_2}}}\right)
-C \left(\frac{\theta P_1}{N_2}\right)\right]^{+}\,,
\vspace{2mm}\\
& & \mbox{ for some }0 
\leq \theta \leq 1, 0 
\leq \eta   \leq 1 \,.\}\,.
\ea
\ea
\eeqno
Our results are the followings.

\begin{Th}{\label{th:ThGauss} 
For any Gaussian relay channel $\Sigma$, 
\beqa
{\cal R}_{\rm d}^{\rm (in)}(P_1,P_2|\Sigma)
\subseteq
{\cal R}_{\rm d}(P_1,P_2|\Sigma)
\subseteq
{\cal R}_{\rm d}^{\rm (out)}(P_1,P_2|\Sigma),\quad
\label{eqn:0aa}
\\
{\cal R}_{\rm s}^{\rm (in)}(P_1,P_2|\Sigma)
\subseteq
{\cal R}_{\rm s}(P_1,P_2|\Sigma)
\subseteq
{\cal R}_{\rm s}^{\rm (out)}(P_1,P_2|\Sigma).\quad
\label{eqn:0aa2}
\eeqa
In particular, if the relay channel is reversely degraded, i.e., 
$N_1 \leq N_2$ and $\rho=\sqrt{\frac{N_1}{N_2}}$, then
\beqno
 {}{\cal R}_{\rm d}^{\rm (in)}(P_1,P_2|\Sigma)
={\cal R}_{\rm d}(P_1,P_2|\Sigma)
={}{\cal R}_{\rm d}^{\rm (out)}(P_1,P_2|\Sigma)\,,
\\
 {}{\cal R}_{\rm s}^{\rm (in)}(P_1,P_2|\Sigma)
={\cal R}_{\rm s}(P_1,P_2|\Sigma)
={}{\cal R}_{\rm s}^{\rm (out)}(P_1,P_2|\Sigma)\,.
\eeqno
}
\end{Th}

Proof of the first inclusions in (\ref{eqn:0aa}) and (\ref{eqn:0aa2})
in the above theorem is standard. The second inclusions in
(\ref{eqn:0aa}) and (\ref{eqn:0aa2}) can be proved by a converse
coding argument similar to the one developed by Liang and Veeravalli
\cite{lv}.  Proof of Theorem \ref{th:ThGauss} will be stated 
in Section VIII.

Next we study the secrecy capacity of 
the Gaussian RCCs. Define two regions by 
\beqno
& &
{\cal R}_{\rm d1e}(P_1,P_2|\Sigma)
\nonumber\\
&\defeq &\left\{(R_1,R_{\rm e}): (0,R_1,R_{\rm e})
    \in {\cal R}_{\rm d}(P_1,P_2|\Sigma) \right\} \,,
\\
& &
{\cal R}_{\rm s1e}(P_1,P_2|\Sigma)
\nonumber\\
&\defeq &\left\{(R_1,R_{\rm e}): (0,R_1,R_{\rm e})
    \in {\cal R}_{\rm s}(P_1,P_2|\Sigma) \right\} \,.
\eeqno
Furthermore, define the secrecy capacity region
${\cal C}_{\rm ss}(P_1,P_2|\Sigma)$ 
and the secrecy capacity ${C}_{\rm ss}(P_1,P_2|\Sigma)$ 
by   
\beqno
& &{\cal C}_{\rm ss}(P_1,P_2|\Sigma)
\nonumber\\
&\defeq &\left\{(R_0,R_1): (R_0,R_1,R_1)
    \in {\cal R}_{\rm s}(P_1,P_2|\Sigma) \right\} \,.
\\
& &
C_{\rm ss}(P_1,P_2|\Sigma)
\\
&\defeq& 
     \max_{(R_1,R_1)\in {\cal R}_{\rm s1e}(P_1,P_2|\Sigma)}R_1
\Irr{=}
\max_{(0,R_1)\in {\cal C}_{\rm ss}(P_1,P_2|\Sigma)}R_1\,.
\eeqno
Maximum equivocation rates for 
${\cal R}_{\rm d}(P_1,$ $P_2|\Sigma)$ and 
${\cal R}_{\rm s}(P_1,$ $P_2|\Sigma)$ are defined by
\beqno
C_{\rm de}(P_1,P_2|\Sigma)
&\defeq& 
\max_{(R_0,R_1,R_{\rm e})\in {\cal R}_{\rm d}(P_1,P_2|\Sigma)}R_{\rm e}\,,
\\
C_{\rm se}(P_1,P_2|\Sigma)
&\defeq& 
\max_{(R_0,R_1,R_{\rm e})\in {\cal R}_{\rm s}(P_1,P_2|\Sigma)}R_{\rm e}\,.
\eeqno
Set
\beqno
{\cal R}_{\rm d1e}^{\rm (in)}(P_1|\Sigma)
&\defeq &
\ba[t]{l}
\{(R_1, R_{\rm e}): R_1, R_{\rm e}\geq 0\,,
\vSpa\\
\ba[t]{rcl}
R_1 &\leq &
    C \left(\frac{\theta P_1}{N_1}\right)\,,
\vSpa\\
R_{\rm e} &\leq & 
\left[R_1-C \left(\frac{\theta P_1}{N_2}\right)\right]^{+}\,,
\vSpa\\
         & &\mbox{ for some }0\leq \theta \leq 1\,.\}\,,
\ea
\ea
\\
{\cal R}_{\rm d1e}^{\rm (out)}(P_1|\Sigma)
&\defeq &
\ba[t]{l}
\{(R_1, R_{\rm e}): R_1, R_{\rm e}\geq 0\,,
\vSpa\\
\ba[t]{rcl}
R_1 &\leq &
    C \left(\frac{\theta P_1}
{\frac{(1-\rho^2)N_1N_2 }{ N_1+N_2-2\rho \sqrt{N_1N_2}}}\right)\,,
\vSpa\\
R_{\rm e} &\leq & \left[R_1-C \left(\frac{\theta P_1}{N_2}\right)\right]^{+}\,,
\vSpa\\
         & &\mbox{ for some }0\leq \theta \leq 1\,.\}\,,
\ea
\ea
\eeqno
\beqno
{\cal R}_{\rm s1e}^{\rm (in)}(P_1|\Sigma)
&\defeq &
\ba[t]{l}
\{(R_1, R_{\rm e}): R_1, R_{\rm e}\geq 0\,,
\vSpa\\
\ba[t]{rcl}
R_{\rm e} &\leq & R_1\leq C \left(\frac{P_1}{N_1}\right)\,,
\vSpa\\
R_{\rm e}&\leq &
 \left[C \left(\frac{P_1}{N_1}\right)
      -C \left(\frac{P_1}{N_2}\right)\right]^{+}\,.\}\,,
\ea
\ea
\eeqno
\beqno
& & {\cal R}_{\rm s1e}^{\rm (out)}(P_1|\Sigma)
\\
&\defeq &
\ba[t]{l}
\{(R_1, R_{\rm e}): R_1, R_{\rm e}\geq 0\,,
\vSpa\\
\ba[t]{rcl}
R_{\rm e} &\leq & R_1 \leq 
    C \left(\frac{P_1}
{\frac{(1-\rho^2)N_1N_2 }{ N_1+N_2-2\rho \sqrt{N_1N_2}}}\right)\,,
\vSpa\\
R_{\rm e}&\leq &
 \left[C \left(\frac{P_1}
 {\frac{(1-\rho^2)N_1N_2}{N_1+N_2-2\rho\sqrt{N_1N_2}}}\right)
-C \left(\frac{P_1}{N_2}\right)\right]^{+}\,.\}\,.
\ea
\ea
\eeqno
Furthermore, set  
\beqno
& & {}{\cal C}_{\rm ss}^{\rm (in)}(P_1,P_2|\Sigma)
\\
&\defeq &
\ba[t]{l}
\{(R_0, R_1): R_0, R_1 \geq 0\,,
\vSpa\\
\ba[t]{rcl}
R_0
&\leq &{\ds \max_{0\leq \eta \leq 1}}
      \min
      \ba[t]{l}\left\{
        C\left(\frac{\bar{\theta}P_1+P_2
        +2\sqrt{\bar{\theta}\bar{\eta}P_1P_2}}
        {\theta P_1+N_1}
        \right)\right.\,,
        \vSpa\\
        \quad \! 
        \left.C\left(\frac{\bar{\theta}\eta P_1}{\theta P_1+N_2}\right)
        \right\}\,,
        \ea
        \vSpa\\
R_{1}&\leq &
 \left[C \left(\frac{\theta P_1}{N_1}\right)
 -C \left(\frac{\theta P_1}{N_2}\right)\right]^{+}\,,
\vspace{2mm}\\
& & \mbox{ for some }0 \leq \theta \leq 1 \,.\}\,,
\ea
\ea
\eeqno
\beqno
& & {\cal C}_{\rm ss}^{\rm (out)}(P_1,P_2|\Sigma)
\\
&\defeq &
\ba[t]{l}
\{(R_0, R_1): R_0, R_1\geq 0\,,
\vSpa\\
\ba[t]{rcl}
R_0
&\leq & {\ds \max_{0\leq \eta\leq 1}}
        \min
        \ba[t]{l}\left\{
          C \left(\frac{\bar{\theta}P_1+P_2
          +2\sqrt{\bar{\theta}\bar{\eta}P_1P_2}}
          {\theta P_1+N_1}
          \right)\right.\,,
          \vSpa\\
          \quad \! \left.
          C \left(\frac{\bar{\theta}\eta P_1}{\theta P_1+N_2}\right)
         \right\}\,,
         \ea
         \vSpa\\
R_{1}&\leq &
 \left[C \left(\frac{\theta P_1}
 {\frac{(1-\rho^2)N_1N_2}{N_1+N_2-2\rho\sqrt{N_1N_2}}}\right)
-C \left(\frac{\theta P_1}{N_2}\right)\right]^{+}\,,
\vspace{2mm}\\
& & \mbox{ for some }0 \leq \theta \leq 1\,.\}\,.
\ea
\ea
\eeqno
We obtain the following two results as a corollary 
of Theorem \ref{th:ThGauss}.
\begin{co} For any Gaussian relay channel $\Sigma$, 
\beqno
{\cal R}_{\rm d1e}^{(\rm in)}(P_1|\Sigma)
\subseteq
{\cal R}_{\rm d1e}(P_1,P_2|\Sigma)
\subseteq
{\cal R}_{\rm d1e}^{(\rm out)}(P_1|\Sigma)\,,
\\
{\cal R}_{\rm s1e}^{(\rm in)}(P_1|\Sigma)
\subseteq
{\cal R}_{\rm s1e}(P_1,P_2|\Sigma)
\subseteq
{\cal R}_{\rm s1e}^{(\rm out)}(P_1|\Sigma)\,.
\eeqno
In particular, if $N_1\leq N_2$ and 
$\rho=\sqrt{\frac{N_1}{N_2}}$, the regions 
${\cal R}_{\rm d1e}(P_1,$ $P_2|\Sigma)$
and 
${\cal R}_{\rm s1e}(P_1,$ $P_2|\Sigma)$
do not depend on $P_2$ and 
\beqno
{\cal R}_{\rm d1e}^{(\rm in)}(P_1|\Sigma)
=
{\cal R}_{\rm d1e}(P_1|\Sigma)
=
{\cal R}_{\rm d1e}^{(\rm out)}(P_1|\Sigma)\,,
\\
{\cal R}_{\rm s1e}^{(\rm in)}(P_1|\Sigma)
=
{\cal R}_{\rm s1e}(P_1|\Sigma)
=
{\cal R}_{\rm s1e}^{(\rm out)}(P_1|\Sigma)\,.
\eeqno
\end{co}
\begin{co}{\rm For any Gaussian relay channel $\Sigma$, 
$$
{\cal C}_{\rm ss}^{(\rm in)}(P_1,P_2|\Sigma)
\subseteq
{\cal C}_{\rm ss}(P_1,P_2|\Sigma)
\subseteq
{\cal C}_{\rm ss}^{(\rm out)}(P_1,P_2|\Sigma)\,.
$$
Furthermore, 
\beqno
&   & \ts \left[C \left(\frac{P_1}{N_1}\right)
               -C \left(\frac{P_1}{N_2}\right)
      \right]^{+}
\nonumber\\
&\leq & C_{\rm de}(P_1,P_2|\Sigma)
        \leq C_{\rm ss}(P_1,P_2|\Sigma)
        \leq C_{\rm se}(P_1,P_2|\Sigma)
\nonumber\\
&\leq &  \ts \left[C \left(\frac{P_1}
          {\frac{(1-\rho^2)N_1N_2}{N_1
           +N_2-2\rho\sqrt{N_1N_2}}}\right)
           -C \left(\frac{ P_1}{N_2}\right)
     \right]^{+} \,. 
\eeqno
In particular, if 
$N_1\leq N_2$ and $\rho=\sqrt{\frac{N_1}{N_2}}$,
\beqno
 {\cal C}_{\rm ss}^{(\rm in)}(P_1,P_2|\Sigma)
={\cal C}_{\rm ss}(P_1,P_2|\Sigma)
={\cal C}_{\rm ss}^{(\rm out)}(P_1,P_2|\Sigma)
\eeqno
and 
\beqno
C_{\rm de}(P_1,P_2|\Sigma)&=&
C_{\rm ss}(P_1,P_2|\Sigma)=
C_{\rm se}(P_1,P_2|\Sigma)
\\
&=&\ts C \left(\frac{P_1}{N_1}\right)
               -C \left(\frac{ P_1}{N_2}\right)
      \,. 
\eeqno
}
\end{co}

\begin{figure}[t]
\bc
\includegraphics[width=8.8cm]{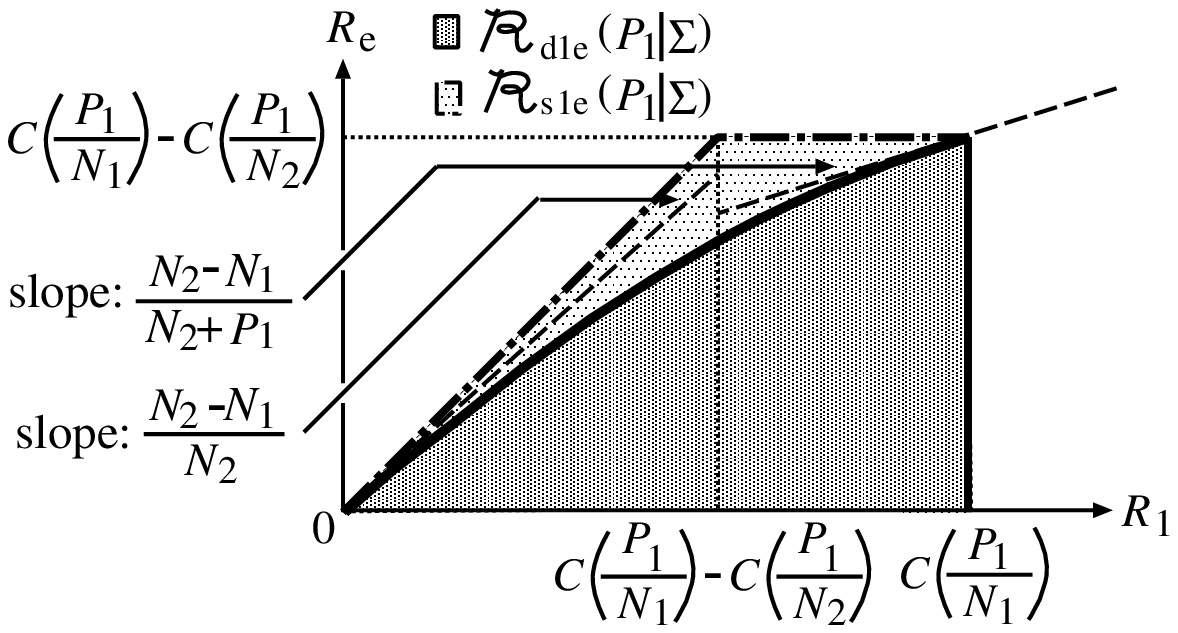}
\caption{
Shapes of 
${\cal R}_{\rm d1e}(P_1|\Sigma)$ and  
${\cal R}_{\rm s1e}(P_1|\Sigma)$ for the reversely 
degraded relay channel $\Sigma$.
}
\label{fig:regG} 
\ec
\vspace*{-4mm}
\end{figure}

Typical shapes of ${\cal R}_{\rm d1e}(P_1|\Sigma)$ 
and ${\cal R}_{\rm s1e}(P_1|\Sigma)$ for the 
reversely degraded relay channel $\Sigma$ 
are shown in Fig. \ref{fig:regG}. 
Note that the secrecy capacity $C_{\rm ss}(P_1,P_2|\Sigma)$ for the
reversely degraded relay channel does not depend on \Irr{the} power constraint
$P_2$ at the relay. This implies that the security of private messages
is not affected by the relay.  Leung-Yan-Cheong and Hellman \cite{lh}
determined the secrecy capacity for the Gaussian wire-tap channel. The
above secrecy capacity is equal to the secrecy capacity of the
Gaussian wire-tap channel derived by them.


\section{
Derivations of the Inner Bounds
}

In this section we prove Theorem \ref{th:ddirect}, 
and the inclusion $\Csi$ $\subseteq \Cs$ 
in Theorem \ref{th:sreg2}. 

\subsection{
Encoding and Decoding Scheme 
} 

We first state an important lemma to derive inner bounds. To describe 
this lemma, we need some preparations. 
Let ${\cal T}_n$, ${\cal J}_n$, and ${\cal L}_n$ 
be three message sets to be transmitted by the sender. 
Let $T_n,$$J_n$, and $L_n$ be uniformly distributed random variable 
on ${\cal T}_n$, ${\cal J}_n$, and ${\cal L}_n$, respectively. Elements of 
${\cal T}_n$ are directed to the receiver and relay. 
Encoder function $f_n$ is a one to one mapping from ${\cal T}_n\times$ ${\cal 
J}_n\times$ ${\cal L}_n$ to ${\cal X}^n$. Using the decoder function 
${\psi}_n$, the receiver outputs an element of ${\cal T}_n\times$ ${\cal 
J}_n\times$ ${\cal L}_n$ from a received message of ${\cal Y}^n$. Using 
the decoder function ${\varphi}_n$, the relay outputs an element of 
${\cal T}_n$ from a received message of ${\cal Z}^n$. 
Formal definitions of ${\psi}_n$ and ${\varphi}_n$ are 
$
 {\psi}_n: {\cal Y}^{n} 
    \to {\cal T}_n \times{\cal J}_n\times{\cal L}_n\,,
{\varphi}_n: {\cal Z}^{n}  \to  {\cal T}_n\,.
$
Error probabilities of decoding at the receiver and the relay 
are defined by 
\beqno
{\mu}_{1}^{(n)}& \defeq & \Pr\{\psi_{n}(Y^n)\neq (T_n,J_n,L_n)\}
\mbox{ and } 
\\
{\mu}_{2}^{(n)}&\defeq &\Pr\{\varphi_{n}(Z^n)\neq T_n\},
\eeqno
respectively.
The following is a key result to derive inner bounds of 
$\Cd$ and $\Cs$.
\begin{lm}\label{lm:direct} \ Choose $(U,X,S)$ $\in{\cal P}_1$ 
such that 
$I(X;Y$ $|{U}S)$ $\geq$ 
$I(X;Z|{U}S)$. Then, there exists 
a sequence of quadruples 
$\{(f_n,\{g_i\}_{i=1}^n,$ 
${\psi }_n ,{\varphi }_n)\}_{n=1}^{\infty}$ 
such that 
\beqno 
   \lim_{n\to\infty}{\mu}_{1}^{(n)}
&=&\lim_{n\to\infty}{\mu}_{2}^{(n)}=0\,, 
\\
\lim_{n\to\infty} \nbn \log \pa {\cal T}_n \pa & = & 
\min\{I(US;Y),I(U;Z|S)\}\,,
\\
\lim_{n\to\infty} 
\nbn \log \pa {\cal J}_n \pa &= & I(X;Y|US)\,,
\\
\lim_{n\to\infty} 
\nbn \log \pa {\cal L}_n \pa &= & I(X;Y|US)-I(X;Z|US)\,,
\\
\lim_{n\to\infty} \nbn H(L_n|Z^{n}) &\geq & 
I(X;Y|US)-I(X;Z|US)\,.
\eeqno
\vspace*{1mm}
\end{lm}

In this subsection we give an encoding and decoding scheme
which attains the transmission and equivocation rates 
described in Lemma \ref{lm:direct}. 
Let
\beqno
& &{\cal T}_n=\{1,2,\cdots, 2^{\lfloor nR_0\rfloor}  \}\,,
\quad {\cal L}_n=\{1,2,\cdots, 2^{\lfloor nr_1 \rfloor}  \}\,,
\\
& &{\cal J}_n=\{1,2,\cdots, 2^{\lfloor nr_2\rfloor} \}\,,
\eeqno
where $\lfloor x \rfloor$ stands for the integer part of $x$ 
for $x>0$. 
We consider a transmission over $B$ 
blocks, each with length $n$. For each $i=1,2, \cdots, B$, let 
$
(t_i,j_i, l_i)\in 
$
$
{\cal T}_n\times 
$
$
{\cal J}_n\times 
$
$
{\cal L}_n
$ 
be a triple of messages to be transmitted at the $i$th block.
A sequence of $B-1$ message triples $(t_i,j_i,l_i),$ 
$i=1,2,\cdots, B-1$ are sent over the channel 
in $nB$ transmission.   
For $i=0$, the constant message pair $(t_0,j_0,l_0)$ $=(1,1,1)$ 
is transmitted. For fixed $n$, the rate triple
$(R_0\frac{B-1}{B}, 
r_1\frac{B-1}{B},r_2\frac{B-1}{B})$        
approaches $(R_0, r_1, r_2)$ as $B$ 
$\to \infty$.

We use random codes for the proof. Fix a joint probability 
distribution of $(U,S,X,Y,Z)$:
\beqno
& &p_{USXYZ}(u,s,x,y,z)\\
&=&p_{S}(s)p_{U|S}(u|s)p_{X|US}(x|u,s){\Ch}(y,z|x,s)\,,
\eeqno
where $U$ is an auxiliary random variable that stands for the
information being carried by the message to be sent 
to the receiver and the relay. 
       
\underline{\bf Random Codebook Generation:} 
We generate a random code book by the following steps.
\begin{itemize} 
\item[1.] Set
$
{\cal W}_n\defeq\{1,2,$ $\cdots, $ $2^{\lfloor nr\rfloor} \}\,.
$
Generate $ 2^{\lfloor nr\rfloor}$ 
i.i.d. ${\vc s}\in {\cal S}^n$ 
each with distribution $\prod_{i=1}^n p_S(s_i).$ 
Index ${\vc s}(w), w\in {\cal W}_n$. 

\item[2.] For each ${\vc s}(w)$, generate 
${2^{\lfloor nR_0\rfloor}}$ i.i.d. 
${\vc u}\in {\cal U}^n$ each with distribution 
$\prod_{i=1}^n p_U(u_i|s_i)$. Index 
${\vc u}(w,t), $ $t$ $\in$ ${\cal T}_n$.

\item[3.] For each ${\vc s}(w)$ and ${\vc u}(w,t)$, generate 
$2^{ \lfloor n r_1\rfloor }$$\cdot 2^{ \lfloor n r_2\rfloor }$ 
i.i.d. ${\vc x}\in {\cal X}^n$ each with distribution  
$\prod_{i=1}^n $ $p_{X|US}(x_i$ $|u_i,s_i)$. Index 
${\vc x}(w,t,$$j,l),$ 
$(w,t,$$j,l)$ 
$\in $ $ {\cal W}_n $ 
$\times {\cal T}_n$
$\times {\cal J}_n$ 
$\times {\cal L}_n$.
\end{itemize}

\underline{\bf Random Partition of ${\cal T}_n$:} \ 
We define the mapping $\phi_n:$ ${\cal T}_n$ $\to {\cal W}_n$ 
in the following manner. For each $t\in {\cal T}_n$, choose 
$w\in {\cal W}_n$ at random according to the uniform distribution 
on ${\cal W}_n$ and map $t$ to $w$. The random choice 
is independent for each $t\in {\cal T}_n$. For each 
$w\in {\cal W}_n$, define ${\cal T}_n(w)\defeq$ 
$\{t\in {\cal T}_n: $ $\phi_n(t)=w \}\,.$
The family of sets $\{{\cal T}_n(w)\}_{w \in {\cal W}_n}$ 
is a partition of ${\cal T}_n$ .

\underline{\bf Encoding:} \ Let $(t_i,j_i,l_i)$ be 
the new message triple to be sent 
from the sender in block $i$ and $(t_{i-1},j_{i-1},l_{i-1})$ 
be the message triple to be sent from the sender in 
previous block $i-1$. 
At the beginning of block $i$, the sender computes 
$w_i=\phi_n(t_{i-1})$ and sends the codeword 
${\vc x}(w_i,t_i,j_i,l_i)\in {\cal X}^n$.     

At the beginning of block $i$, the relay has decoded 
the message $t_{i-1}$. It then computes $w_i=\phi_n(t_{i-1})$ 
and sends the codeword ${\vc s}(w_i)\in {\cal S}^n$.     

\underline{\bf Decoding:} \ Let ${\vc y}_i\in {\cal Y}^n$ 
and ${\vc z}_i\in {\cal Z}^n$  
be the sequences that the reviver and the relay obtain 
at the end of block $i$, respectively. 
The decoding procedures 
at the end of block $i$ are as follows. 

\underline{1. Decoder 2 at the Relay:} \ Define
\beqno
i_{UZ|S}({\vc u};{\vc z}|{\vc s})
&\defeq&
\log
\frac{p_{UZ|S}({\vc u},{\vc z}|{\vc s})}
{p_{U|S}({\vc u}|{\vc s})p_{Z|S}({\vc z}|{\vc s})}\,, 
\\
{\cal A}_{UZ|S,\epsilon}
&\defeq&
\ba[t]{l}
 \{({\vc s},{\vc u},{\vc z}) 
 \in {\cal S}^n \times {\cal U}^n \times {\cal Z}^n: \\
\frac{1}{n}i_{UZ|S}({\vc u};{\vc z}|{\vc s}) 
> R_0+\epsilon \}\,. 
\ea
\eeqno
The relay declares that the message $\hat{t}_i$ is sent 
if there is a unique $\hat{t}_i$ such that 
$$ 
\left(
{\vc s}(w_i),{\vc u}(w_i,\hat{t}_i), {\vc z}_i
\right)
\in {\cal A}_{UZ|S,\epsilon}\,. 
$$
It will be shown that the decoding error in this step 
is small for sufficiently large $n$ if 
$
R_0 < I(U;Z|S)\,.
$

\underline{2. Decoders 1a and 1b at the Receiver:} \ Define
\beqno
i_{SY}({\vc s};{\vc y}) 
&\defeq &
\log
\frac{p_{SY}({\vc s},{\vc y})}
{p_{S}({\vc s})p_{Y}({\vc y})}\,, 
\\
i_{UY|S}({\vc u};{\vc y}|{\vc s}) 
&\defeq &
\log \frac{p_{UY|S}({\vc u},{\vc y}|{\vc s})}
{p_{U|S}({\vc u}|{\vc s})p_{Y|S}({\vc y}|{\vc s})}\,,
\\
{\cal A}_{SY,\epsilon}
&\defeq&
\ba[t]{l}
 \{({\vc s},{\vc y}) 
 \in {\cal S}^n \times {\cal Y}^n: \\
\frac{1}{n}\log i_{SY}({\vc s};{\vc y}) 
> r+\epsilon\}\,, 
\ea
\\
{\cal A}_{UY|S,\epsilon}
&\defeq&
\ba[t]{l}
 \{({\vc s},{\vc u},{\vc y}) 
 \in {\cal S}^n \times {\cal U}^n \times {\cal Y}^n: \\
\frac{1}{n}i_{UY|S}({\vc u};{\vc y}|{\vc s}) 
+r > R_0+\epsilon \}\,. 
\ea
\eeqno
The receiver first declares that the message $\hat{w}_i$ 
is sent if there is a unique $\hat{w}_i$ such that    
$ 
\left({\vc s}(\hat{w}_i),{\vc y}_i\right)
\in {\cal A}_{SY,\epsilon}\,. 
$
It will be shown that the decoding error 
in this step is small for sufficiently 
large $n$ if 
$
r< I(Y;S)\,.
$
Next, the receiver, having known $w_{i-1}$ and $\hat{w}_i$, 
declares that the message $\hat{\hat{t}}_{i-1}$ is sent 
if there is a unique $\hat{\hat{t}}_{i-1}$ such that 
\beqno
& &\left(
{\vc s}(w_{i-1}),
{\vc u}(w_{i-1}, \hat{\hat{t}}_{i-1}),
{\vc y}_{i-1}
\right)
\in {\cal A}_{UY|S,\epsilon}
\\
& &\mbox{ and } 
\hat{\hat{t}}_{i-1}\in {\cal T}_n(\hat{w}_i). 
\eeqno  
It will be shown that the decoding error in this 
step is small for sufficiently large $n$ if  
\beqno
R_0 
& < & I(U;Y|S) + r 
\nonumber\\
& < &I(U;Y|S)+I(Y;S) =I(US;Y)\,.
\eeqno    

\underline{3. Decoder 1c at the Receiver:} \ Define 
\beqno
i_{XY|US}({\vc x};{\vc y}|{\vc u},{\vc s}) 
&\defeq &
\log
\frac{p_{XY|US}({\vc x},{\vc y}|{\vc u},{\vc s})}
{p_{X|US}({\vc x}|{\vc u},{\vc s})
 p_{Y|US}({\vc y}|{\vc u},{\vc s})}\,, 
\\
{\cal A}_{XY|US,\epsilon}
&\defeq&
\ba[t]{l}
 \{({\vc s},{\vc u},{\vc x},{\vc y}) 
 \in {\cal S}^n \times 
     {\cal U}^n \times 
     {\cal X}^n \times 
     {\cal Y}^n:\\
\frac{1}{n}i_{XY|US}({\vc x};{\vc y}|{\vc u},{\vc s}) 
> r_1+r_2+\epsilon \}\,. 
\ea
\eeqno
The receiver, having known $w_{i-1},$ $\hat{\hat{t}}_{i-1}$, 
declares that the message pair $(\hat{j}_{i-1}, \hat{l}_{i-1})$ 
is sent if there is a unique pair $(\hat{j}_{i-1},\hat{l}_{i-1})$ 
such that 
\beqno
& &\left(
{\vc s}(w_{i-1}),
{\vc u}(w_{i-1},\hat{\hat{t}}_{i-1}), 
{\vc x}(w_{i-1},
        \hat{\hat{t}}_{i-1}, 
        \hat{j}_{i-1},
        \hat{l}_{i-1}), {\vc y}_{i-1} 
\right)\\
& &\in {\cal A}_{XY|US,\epsilon}\,.
\eeqno  
It will be shown that the decoding error 
in this step is small for sufficiently large $n$ if  
$
r_1+r_2< I(X;Y|US)\,.
$

\begin{figure}[t]
\bc
\includegraphics[width=8.8cm]{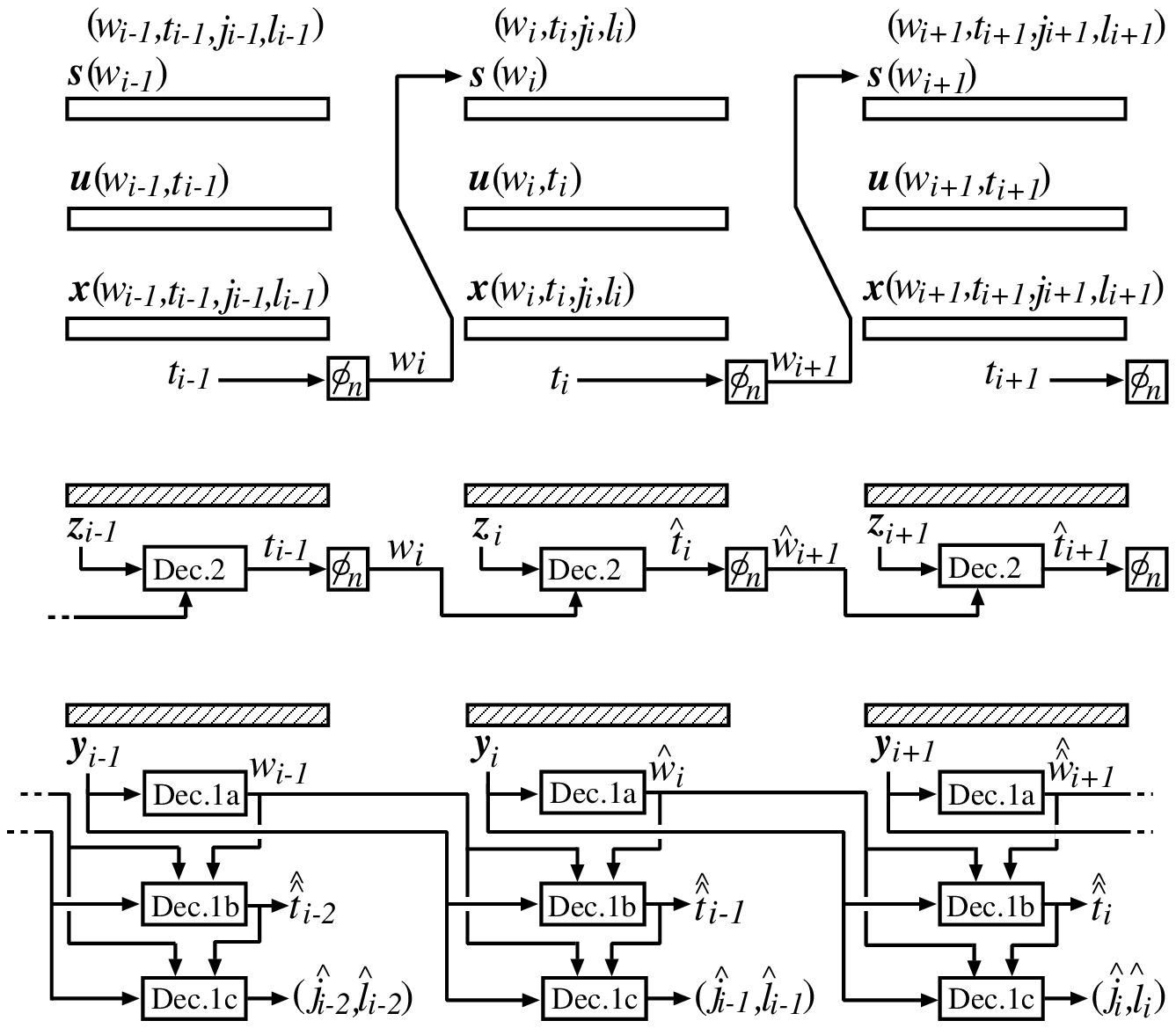}
\caption{
Encoding and decoding processes at the blocks 
$i-1,$ $i,$ and $i+1$.}
\label{fig:cschm} 
\ec
\vspace*{-4mm}
\end{figure}

For convenience we show the encoding and decoding processes 
at the blocks $i-1,$ $i,$, and $i+1$ in Fig. \ref{fig:cschm}.       

\subsection{Computation of Error Probability and Equivocation Rate} 

In this subsection we compute error probabilities of decoding and
equivocation rate for the encoding and decoding scheme stated in the
previous subsection. We will declare an error in block $i$ if one or
more of the following events occurs.

\tabcolsep 0.2mm
\begin{flushleft}
\begin{tabular}{p{0.8cm}p{7.7cm}}
${\cal E}_{{\rm 2},i}$:& Decoder 2 fails. 
Let ${\cal E}_{{\rm 2},i}=
\tilde{\cal E}_{{\rm 2},i}
\cup\hat{\cal E}_{{\rm 2},i}$, where
\\
&\begin{tabular}{p{0.8cm}p{6.8cm}}
$\tilde{\cal E}_{{\rm 2},i}$: & 
$
\left(
{\vc s}(w_i),{\vc u}(w_i,t_i), {\vc z}_i
\right)
\notin {\cal A}_{UZ|S,\epsilon}  
$,\\ 
$\hat{\cal E}_{{\rm 2},i}$: & 
$\exists$ $\hat{t}_i\neq t_i$ such that
$ ({\vc s}(w_i),$
${\vc u}(w_i,\hat{t}_i),$
${\vc z}_i)$
$\in {\cal A}_{UZ|S,\epsilon}\,.$\\
 \end{tabular} 
\\
${\cal E}_{{\rm 1a},i}$: & Decoder 1a fails. 
Let ${\cal E}_{{\rm 1a},i}=
\tilde{\cal E}_{{\rm 1a},i}
\cup \hat{\cal E}_{{\rm 1a},i}$, where
\\
&\begin{tabular}{p{0.8cm}p{6.8cm}}
$\tilde{\cal E}_{{\rm 1a},i}$: & 
$
\left({\vc s}(w_i),{\vc y}_i\right)
\notin {\cal A}_{SY,\epsilon}  
$, \\ 
$\hat{\cal E}_{{\rm 1a},i}$: & 
$\exists$ $\hat{w}_i\neq w_i$ such that
$ ({\vc s}(w_i),$
  ${\vc y}_i)$
$\in {\cal A}_{SY,\epsilon}\,.$ \\
 \end{tabular} 
\\
${\cal E}_{{\rm 1b},i}$: & Decoder 1b fails. 
Let ${\cal E}_{{\rm 1b},i}=
\tilde{\cal E}_{{\rm 1b},i}
\cup \hat{\cal E}_{{\rm 1b},i}$, where
\\
&\begin{tabular}{p{0.8cm}p{6.8cm}}
$\tilde{\cal E}_{{\rm 1b},i}$: & 
$({\vc s}(w_{i-1}),$
 ${\vc u}(w_{i-1},t_{i-1}),$ 
 ${\vc y}_{i-1})$
 $\notin {\cal A}_{UY|S,\epsilon}$,\\ 
$\hat{\cal E}_{{\rm 1b},i}$: & 
$\exists$ $\hat{\hat{t}}_{i-1}\neq t_{i-1}$ 
such that
$({\vc s}(w_{i-1}),$
 ${\vc u}(w_i,$
 $\hat{\hat{t}}_{i-1}),$
 ${\vc y}_{i-1})$
 $\in {\cal A}_{UY|S,\epsilon}$, 
 $\hat{\hat{t}}_{i-1}\in {\cal T}_n({w}_i). $
\end{tabular} 
\\
${\cal E}_{{\rm 1c},i}$: & Decoder 1c fails. 
Let ${\cal E}_{{\rm 1c},i}=
\tilde{\cal E}_{{\rm 1c},i}
\cup\hat{\cal E}_{{\rm 1c},i}$, where
\\
&\begin{tabular}{p{0.8cm}p{6.8cm}}
$\tilde{\cal E}_{{\rm 1c},i}$: & 
$({\vc s}(w_{i-1}),$
${\vc u}(w_{i-1},$ 
$t_{i-1}),$ 
${\vc x}_{i-1}(w_{i-1},$
$t_{i-1},$ 
$j_{i-1},$
$l_{i-1}),$
${\vc y}_{i-1})$
$\notin {\cal A}_{XY|US,\epsilon}$, \\ 
$\hat{\cal E}_{{\rm 1c},i}$: & 
$\exists$ 
$(\hat{j}_{i-1},\hat{l}_{i-1})$ $\neq (j_{i-1},l_{i-1})$ 
such that $({\vc s}($ $w_{i-1}),$
${\vc u}(w_{i-1},$ $t_{i-1}),$
${\vc x}_{i-1}(w_{i-1},$
$t_{i-1},$ 
$\hat{j}_{i-1},$
$\hat{l}_{i-1}),$
${\vc y}_{i-1})$
$\in {\cal A}_{XY|US,\epsilon}\,.$ \\
 \end{tabular} 
\end{tabular}
\end{flushleft}
For each $i=1,2,$ $\cdots, B$, let 
$(T_{n,i},J_{n,i}, $ $L_{n,i})$
$\in {\cal T}_n\times$ ${\cal J}_n\times$ ${\cal L}_n$ 
be a message triple to be transmitted at the block $i$. 
We assume that 
$(T_{n,i},$ $J_{n,i}$, $L_{n,i})$, $i=1,2,$ $\cdots, B$ are  i.i.d. 
random triples uniformly distributed 
on ${\cal T}_n$ $\times {\cal J}_n$ $\times {\cal L}_n$. 
For $i=0$, $T_{n,0}$, $J_{n,0}$ and $L_{n,0}$ are constant. 
For $i=1,2,\cdots, B-1$, define the random variable
$W_{n,i}$ on ${\cal W}_n$ by $W_{n,i}=\phi_n({T}_{n,i-1})$. 
Define the error events ${\cal F}_i$ for decoding errors 
in block $i$ by
\begin{flushleft}
\begin{tabular}{p{0.6cm}p{8.1cm}}
${\cal F}_i$:& 
$\hat{W}_{n,i}\neq {W}_{n,i}$ or  
$\hat{T}_{n,i}\neq {T}_{n,i}$ or 
$\hat{\hat{T}}_{n,i-1}\neq {T}_{n,i-1}$ or 
$(\hat{J}_{n,i-1},\hat{L}_{n,i-1})$
$\neq ({J}_{n,i-1},{L}_{n,i-1})$.  
\end{tabular}
\end{flushleft}
It is obvious that
$
{\cal F}_i\subseteq 
     {\cal E}_{{\rm 2},i}
\cup {\cal E}_{{\rm 1a},i}
\cup {\cal E}_{{\rm 1b},i}
\cup {\cal E}_{{\rm 1c},i}\,.
$
Define
$
e_{{\rm 2},i}^{(n)} \defeq 
\Pr\left\{{\cal E}_{{\rm 2},i}| {\cal F}_{i-1}^{c}\right\}\,.    
$
Definitions of 
$e_{{\rm 1a},i}^{(n)}$,
$e_{{\rm 1b},i}^{(n)}$, and   
$e_{{\rm 1c},i}^{(n)}$ are the same as that of 
$e_{{\rm 2},i}^{(n)}$.  
We further define sets and quantities necessary for 
computation of the equivocation rate. 
%
Define 
\beqno
i_{XZ|US}({\vc x};{\vc z}|{\vc u},{\vc s}) 
&\defeq &
\log
\frac{p_{XZ|US}({\vc x},{\vc z}|{\vc u},{\vc s})}
{p_{X|US}({\vc x}|{\vc u},{\vc s})
 p_{Z|US}({\vc z}|{\vc u},{\vc s})}\,, 
\\
{\cal A}_{XZ|US,\epsilon}
&\defeq&
\ba[t]{l}
 \{({\vc s},{\vc u},{\vc x},{\vc y}) 
 \in {\cal S}^n \times 
     {\cal U}^n \times 
     {\cal X}^n \times 
     {\cal Z}^n:\\
\frac{1}{n}i_{XZ|US}({\vc x};{\vc z}|{\vc u},{\vc s}) 
> r_2+\epsilon \}\,. 
\ea
\eeqno
For given $w_i=\psi_n(t_{i-1})$ $\in {\cal W}_n$,
$(t_i,l_i) \in {\cal T}_n\times {\cal L}_n$ 
and channel output ${\vc z}_i$ 
of ${\vc s}(w_{i})$ and ${\vc x}(w_{i},t_{i},$ 
$j_{i},l_{i})$, define the estimation function
$\tau_n:{\cal W}_n$
$\times {\cal T}_n$ 
$\times {\cal L}_n$
$\times {\cal Z}^n \to {\cal J}_n$      
by
$\tau_n(w_i,t_i,l_i,{\vc z}_i)$ $=\hat{j}_i$
if there is a unique pair $(\hat{j}_{i},{l}_{i})$ 
such that 
\beqno
& &\left(
{\vc s}(w_{i}),
{\vc u}(w_{i},t_{i}), 
{\vc x}(w_{i},t_{i}, 
        \hat{j}_{i},
            {l}_{i}), {\vc z}_{i} 
\right)\in {\cal A}_{XZ|US,\epsilon}\,.
\eeqno  
Define 
$
e_{i}^{(n)}\defeq 
\Pr\left\{
\tau_n(W_{n,i},T_{n,i},L_{n,i},Z^n)\neq J_{n,i}
\right\}\,.    
$
Let \\
\hspace*{2mm}
\begin{tabular}{p{0.6cm}p{7.0cm}}
$\tilde{\cal E}_{i}$: & 
$({\vc s}(w_{i-1}),$
${\vc u}(w_{i-1},$ 
$t_{i-1}),$ 
${\vc x}_{i-1}(w_{i-1},$
$t_{i-1},$ 
$j_{i-1},$
$l_{i-1}),$
${\vc z}_{i-1})$
$\notin {\cal A}_{XZ|US,\epsilon}$, \\ 
$\hat{\cal E}_{i}$: & 
$\exists$ 
$\hat{j}_{i-1}$ $\neq j_{i-1}$ 
such that $({\vc s}($ $w_{i-1}),$
${\vc u}(w_{i-1},$ $t_{i-1}),$
${\vc x}_{i-1}(w_{i-1},$
$t_{i-1},$ 
$\hat{j}_{i-1},$
${l}_{i-1}),$
${\vc z}_{i-1})$
$\in {\cal A}_{XZ|US,\epsilon}\,.$ \\
\end{tabular}
\\ 
Set ${\cal E}_{i}=\tilde{\cal E}_{i}
\cup\hat{\cal E}_{i}$.
Then we have 
$$
e_{i}^{(n)}=\Pr\{{\cal E}_{i}\}\leq 
\Pr\{\tilde{\cal E}_{i}\}+\Pr\{\hat{\cal E}_{i}\}\,.   
$$
It will be shown that the error probability
$e_{i}^{(n)}$ of estimation is small 
for sufficiently large $n$ if 
$
r_2< I(X;Z|US)\,.
$
Set  
\beqno
&&i_{Z|XS}({\vc z}|{\vc x},{\vc s})
\defeq-\log p_{Z|XS}({\vc z}|{\vc x},{\vc s}), 
\\
& &i_{Z|US}({\vc z}|{\vc u},{\vc s})
\defeq-\log p_{Z|US}({\vc z}|{\vc u},{\vc s}), 
\\
&&{\cal B}_{Z|XS,\epsilon}
\defeq 
\ba[t]{l}
\biggl\{({\vc s},{\vc x},{\vc z})\in 
{\cal S}^n\times 
{\cal X}^n\times
{\cal Z}^n: \biggr.
\vSpa\\
\quad 
\ds \left.
\frac{1}{n}i_{Z|XS}({\vc z}|{\vc x},{\vc s})
\geq H(Z|XS)-\epsilon 
      \right\}\,, 
\ea
\\
&&{\cal B}_{Z|US,\epsilon}
\defeq 
\ba[t]{l}
\biggl\{({\vc s},{\vc u},{\vc z})\in 
{\cal S}^n\times 
{\cal U}^n\times 
{\cal Z}^n: 
\biggr.
\vSpa\\
\quad 
\ds \left. 
    \frac{1}{n}i_{Z|US}({\vc z}|{\vc u},{\vc s})
    \leq H(Z|US)+\epsilon \right\}\,,
\ea
\\
&& e_{Z|XS,i}^{(n)}
\defeq \Pr\{{\vc s}(W_{n,i}),{\vc x}(W_{n,i},T_{n,i},L_{n,i}),Z_i^n)
\\
& &\qquad\qquad\qquad\notin {\cal B}_{Z|XS,\epsilon}\}\,, 
\\
& &e_{Z|US,i}^{(n)}
\defeq \Pr\{({\vc s}(W_{n,i}),{\vc u}(T_{n,i}),Z_i^n)
\notin {\cal B}_{Z|US,\epsilon}\}\,. 
\eeqno
The operation ${\sf E}\left[e_{{\rm 2},i}^{(n)}\right]$ 
stands for the expectation of ${e_{{\rm 2},i}^{(n)}}$ 
based on the randomness of code construction.   
Then, we have the following lemma.
\begin{lm}{\label{lm:error}}
For each $i=1,2,\cdots, B-1$, we have 
\beqno
& &{\sf E}\left[e_{{\rm 2},i}^{(n)}\right] 
   \leq \Pr\{(S^n,U^n,Z^n)\notin {\cal A}_{UZ|S,\epsilon}\}
    +2^{-n\epsilon}
\nonumber\\
& &{\sf E}\left[e_{{\rm 1a},i}^{(n)}\right] 
   \leq \Pr\{(S^n,Y^n)\notin {\cal A}_{SY,\epsilon}\}
    +2^{-n\epsilon}
\nonumber\\
& &{\sf E}\left[e_{{\rm 1b},i}^{(n)}\right] 
   \leq \Pr\{(S^n,U^n,Y^n)\notin {\cal A}_{UY|S,\epsilon}\}
    +2 \cdot 2^{-n\epsilon}
\nonumber\\
& &{\sf E}\left[e_{{\rm 1c},i}^{(n)}\right] 
   \leq \Pr\{(S^n,U^n,X^n,Y^n)\notin {\cal A}_{XY|SU,\epsilon}\}
    +2^{-n\epsilon}
\nonumber\\
& &{\sf E}\left[{e}_{i}^{(n)}\right] 
   \leq \Pr\{(S^n,U^n,X^n,Z^n)\notin {\cal A}_{XZ|SU,\epsilon}\}
    +2^{-n\epsilon}
\nonumber\\
& &{\sf E}\left[e_{Z|XS,i}^{(n)}\right] 
   =\Pr\{(S^n,X^n,Z^n)\notin {\cal B}_{Z|XS,\epsilon}\}
\\
& &{\sf E}\left[e_{Z|US,i}^{(n)}\right] 
   =\Pr\{(S^n,U^n,Z^n)\notin {\cal B}_{Z|US,\epsilon}\}\,.
\eeqno
\end{lm}

Proof of this lemma is given in Appendix A.

Next, we state a key lemma useful for the computation 
of the equivocation rate. Set 
$
L_n^{(i)} $ $\defeq (L_{n,1}$ 
$,L_{n,2},\cdots,L_{n,i})\,.$
Then, the equivocation rate over $B$ blocks is 
\beqno
\frac{1}{nB}H(L_n^{(B)}|Z^{nB})
\geq 
\frac{1}{B}\sum_{i=1}^{B-1}\frac{1}{n}H(L_{n,i}|L_n^{(i-1)}Z^{nB}) \,.
\eeqno
For each $i=1,2,\cdots, B-1$, we estimate a lower bound of 
$H(L_{n,i}|$ $L_n^{(i-1)}Z^{nB})$. 
Set 
$
Z_{n(i-1)+1}^{ni}\defeq $ $(Z_{n(i-1)+1},$ $\cdots,$ $Z_{ni})\,. 
$
On a lower bound of $H(L_{n,i}|L_n^{(i-1)}Z^{nB})$, 
we have the following lemma. 

\begin{lm}\label{lm:sec1}
For $i=1,2,$ $\cdots,B-1$, we have 
\beqa
&      & 
\frac{1}{n}H(L_{n,i}|L_{n}^{(i-1)}Z^{nB})
\nonumber\\
& \geq & r_1+r_2-I(X;Z|US)-2\epsilon-\frac{3+\log {\rm e}}{n}
\nonumber\\
&      & -r_2{e}_{i}^{(n)}-(\log |{\cal Z}|)
\left[e_{Z|US,i}^{(n)}+e_{Z|XS,i}^{(n)}\right]\,.
\label{eqn:zz00z}
\eeqa
\end{lm}

Proof of this lemma is given in Appendix B.

{\it Proof of Lemma \ref{lm:direct}: }
Set
\beqno
& &\gamma_{\max}(\epsilon)
\\
&\defeq &
\ba[t]{ll}
  \max\{& \Pr\{(S^n,U^n,Z^n)\notin {\cal A}_{UZ|S,\epsilon}\}
          +2^{-n\epsilon}\,,
  \vspace*{1mm}\\
        & \Pr\{(S^n,Y^n)\notin {\cal A}_{SY,\epsilon}\}
          +2^{-n\epsilon}\,,
\vspace*{1mm}\\
        & \Pr\{(S^n,U^n,Y^n)\notin {\cal A}_{UY|S,\epsilon}\}
          +2^{-n\epsilon}\,,
\vspace*{1mm}\\
        & \Pr\{(S^n,U^n,X^n,Y^n)\notin {\cal A}_{XY|SU,\epsilon}\}
         +2^{-n\epsilon}\,,
\vspace*{1mm}\\
        & \Pr\{(S^n,U^n,X^n,Z^n)\notin {\cal A}_{XZ|SU,\epsilon}\}
         +2^{-n\epsilon}\,,
\vspace*{1mm}\\
        &\Pr\{(S^n,X^n,Z^n)\notin {\cal B}_{Z|XS,\epsilon}\}\,,
\vspace*{1mm}\\
        &\Pr\{(S^n,U^n,Z^n)\notin {\cal B}_{Z|US,\epsilon}\}\:\} \,.
\ea
\eeqno
Then, by Lemma \ref{lm:error}, we obtain
\beqno
& &{\sf E}\left[
\sum_{i=1}^{B-1}
\left\{
 e_{{\rm 2},i}^{(n)}  
+e_{{\rm 1a},i}^{(n)}
+e_{{\rm 1b},i}^{(n)}
+e_{{\rm 1c},i}^{(n)}
+{e}_{i}^{(n)}
\right.\right.
\\
& &\qquad\biggl.\left. +e_{Z|XS,i}^{(n)}+e_{Z|US,i}^{(n)}
   \right\}\biggr]
\\ 
&=& 
\sum_{i=1}^{B-1}
\left\{
{\sf E}\left[e_{{\rm 2},i}^{(n)}\right]  
+{\sf E}\left[e_{{\rm 1a},i}^{(n)}\right]  
+{\sf E}\left[e_{{\rm 1b},i}^{(n)}\right]  
+{\sf E}\left[e_{{\rm 1c},i}^{(n)}\right]
\right.
\\
& &\qquad 
   \left. 
+{\sf E}\left[{e}_{i}^{(n)}\right]
+{\sf E}\left[e_{Z|XS,i}^{(n)}\right]
+{\sf E}\left[e_{Z|US,i}^{(n)}\right]
   \right\}
\\
&\leq &7(B-1)\gamma_{\max}(\epsilon)\,,
\eeqno
from which it follows that 
there exist at least one deterministic code
such that 
\beqa
& &\sum_{i=1}^{B-1}
\left\{
 e_{{\rm 2},i}^{(n)}
+e_{{\rm 1a},i}^{(n)}
+e_{{\rm 1b},i}^{(n)}
+e_{{\rm 1c},i}^{(n)}
+{e}_{i}^{(n)}
\right.
\nonumber\\
& &\qquad 
   \left. 
+e_{Z|XS,i}^{(n)}
+e_{Z|US,i}^{(n)}\right\}
\leq 7(B-1)\gamma_{\max}^{(n)}(\epsilon)\,.
\label{eqn:aszff}
\eeqa
From (\ref{eqn:aszff}), we have
\beqa
&&\mu_1^{(nB)}
       =\sum_{i=1}^{B-1}\left\{
         e_{{\rm 1a},i}^{(n)}
         +e_{{\rm 1b},i}^{(n)}
         +e_{{\rm 1c},i}^{(n)}\right\}
\nonumber\\
& &\qquad\:\:\leq 7(B-1)\gamma_{\max}^{(n)}(\epsilon)\,,
\label{eqn:szff0}\\
& &\mu_2^{(nB)}=\sum_{i=1}^{B-1}e_{2,i}^{(n)}
\leq 7(B-1)\gamma_{\max}^{(n)}(\epsilon)\,,
\label{eqn:szff1}\\
& &
\sum_{i=1}^{B-1}{e}_{i}^{(n)}
\leq 7(B-1)\gamma_{\max}^{(n)}(\epsilon)\,,
\label{eqn:szff1z}\\
& &\sum_{i=1}^{B-1}\left\{
 e_{Z|XS,i}^{(n)}
+e_{Z|US,i}^{(n)}\right\}
\leq 7(B-1)\gamma_{\max}^{(n)}(\epsilon)\,.
\label{eqn:szff2z}
\eeqa
From Lemma \ref{lm:sec1}, (\ref{eqn:szff1z}), 
and (\ref{eqn:szff2z}), we have
\beqa
&     &{\frac{1}{nB}}H(L_{n}^{(B)}|Z^{nB})
\nonumber\\
&\geq & 
{\frac{1}{B}{\ds \sum_{i=1}^{B-1}}
\frac{1}{n}}H(L_{n,i}|L_{n}^{(i-1)}Z^{nB})
\nonumber\\
& \geq & \left(1-\frac{1}{B}\right)
         \left[r_1+r_2-I(X;Z|US)
-2\epsilon-\frac{3+\log{\rm e}}{n}\right]
\nonumber\\
&      & -7\left(1-\frac{1}{B}\right)
[r_2+(\log|{\cal Z}|)]\gamma_{\max}^{(n)}(\epsilon)\,.
\label{eqn:zzz0z}
\eeqa
By the weak law of large numbers, when $n\to\infty$, we have
\beq
\left.
\ba{l}
\frac{1}{n}i_{UZ|S}(U^n;Z^n|S^n) \to I(U;Z|S)
\vspace*{1mm}\\
\frac{1}{n}i_{SY}(S^n;Y^n) \to I(S;Y)
\vspace*{1mm}\\
\frac{1}{n}i_{UY|S}(U^n;Y^n|S^n) \to I(U;Y|S)
\vspace*{1mm}\\
\frac{1}{n}i_{XY|US}(X^n;Y^n|U^nS^n) \to I(X;Y|US)
\vspace*{1mm}\\
\frac{1}{n}i_{XZ|US}(X^n;Z^n|U^nS^n) \to I(X;Z|US)
\vspace*{1mm}\\
\frac{1}{n}i_{Z|XS}(Z^n|X^nS^n) \to H(Z|XS)
\vspace*{1mm}\\
\frac{1}{n}i_{Z|US}(Z^n|U^nS^n) \to H(Z|US)
\ea
\right\}
\label{eqn:szff2}
\eeq 
in probability. Fix $\epsilon>0$ arbitrary and choose
\beq
\left.
\ba{l}
R_0=\min\{I(U;Z|S),I(U;Y|S)+r\}-2\epsilon
\\
r=I(S;Y)-2\epsilon
\\
r_1=I(X;Y|US)-I(X;Z|US)-2\epsilon
\\
r_2=I(X;Z|US)-\epsilon\,.
\ea
\right\}
\label{eqn:szff4a}
\eeq
Then, it follows from (\ref{eqn:szff2}) and the definition of 
$\gamma_{\max}^{(n)}(\epsilon)$ that  for the choice of 
$(R_0,r,r_1,r_2)$ in (\ref{eqn:szff4a}), we have 
\beq
\lim_{n\to\infty}\gamma_{\max}^{(n)}(\epsilon)=0\,.
\label{eqn:dirxx}
\eeq
For $n=1,2,\cdots,$ we choose block $B=B_n$ so that 
$
B_n=\left\lfloor 
\left(\gamma_{\max}^{(n)}(\epsilon)\}\right)^{-1/2} \right\rfloor\,.
$
Define $\{g_i\}_{i=1}^{nB_n}$ by 
$$
g_i \defeq 
\left\{
\ba{ll} 
\phi_n, &\mbox{ if }i\mbox{ mod }n=0\,,\\ 
\mbox{constant}, &\mbox{ otherwise}\,. 
\ea
\right.
$$
Define the sequence of block codes 
$\left\{(f_{\nu},\right.$ 
$\{g_i\}_{i=1}^{\nu},$ 
$\psi_{\nu},$ 
$\varphi_{\nu})$
$\left. \right\}_{\nu=1}^{\infty}$
by
\beqno
(f_{\nu},
  \{g_i\}_{i=1}^{\nu}, \psi_{\nu}, 
   \varphi_{\nu})
&\defeq&
\left\{
\ba{l} 
\mbox{ constant}, \mbox{ if }
1\leq \nu < B_{1}\,,
\vSpa\\ 
(f_{nB_n },\{g_i\}_{i=1}^{nB_n},\psi_{nB_n},\varphi_{nB_n})\,,
\vSpa\\ 
\mbox{ if }nB_n \leq \nu < (n+1)B_{n+1}\,.
\ea
\right.
\eeqno
Combining 
(\ref{eqn:szff0}),
(\ref{eqn:szff1}),
(\ref{eqn:zzz0z}), and 
(\ref{eqn:dirxx}),
we have that there exists a sequence of block codes 
$\left\{(f_{\nu},\right.$ 
$\{g_i\}_{i=1}^{\nu},$ 
$\psi_{\nu},$ $\left.\varphi_{\nu})\right\}_{\nu=1}^{\infty}$
such that
\beqno
& &\lim_{\nu\to\infty}\mu_1^{(\nu)}
 =\lim_{n\to\infty}\mu_1^{(nB_n)}
 \leq \lim_{n\to\infty}7\sqrt{\gamma_{\max}^{(n)}(\epsilon)}=0\,,
\\
& &\lim_{\nu\to\infty}\mu_2^{(\nu)}
=\lim_{n\to\infty}\mu_2^{(nB_n)}
\leq \lim_{n\to\infty}7\sqrt{\gamma_{\max}^{(n)}(\epsilon)}=0\,,
\\
& &\lim_{\nu\to\infty}\frac{1}{\nu} \log |{\cal T}_{\nu}|
=\lim_{n\to\infty}{\frac{1}{nB_n}} \log |({\cal T}_n)^{B_n-1}|
\\
&&=R_0=\min\{I(U;Z|S),I(US;Y)-2\epsilon\}-2\epsilon\,,
\\
& &\lim_{\nu\to\infty}\frac{1}{\nu} \log |{\cal J}_{\nu}|
   =\lim_{n\to\infty}\frac{1}{nB_n}\log |({\cal J}_n)^{B_n-1}|
\\
&&=r_2=I(X;Z|US)-\epsilon\,,
\\
& &\lim_{\nu\to\infty}\frac{1}{\nu} \log |{\cal L}_{\nu}|
  =\lim_{n\to\infty}\frac{1}{nB_n}\log |({\cal L}_n)^{B_n-1}|
\\
&&=r_1=I(X;Y|US)-I(X;Z|US)-2\epsilon\,,
\\
& &
\lim_{\nu\to\infty}\frac{1}{\nu}H(L_{\nu}|Z^{\nu})
=\lim_{n\to\infty}{\frac{1}{nB_n}}H(L_{n}^{(B_n)}|Z^{nB_n})
\\
& &\geq I(X;Y|US)-I(X;Z|US)-5\epsilon\,.
\eeqno
Since $\epsilon$ can be arbitrary small, 
we obtain the desired result for the above sequence of block codes. 
Thus, the proof of Lemma \ref{lm:direct} is completed. 
\hfill\QED

\subsection{Proofs of the Direct Coding Theorems}

\begin{figure}[t]
\bc
\includegraphics[width=7.2cm]{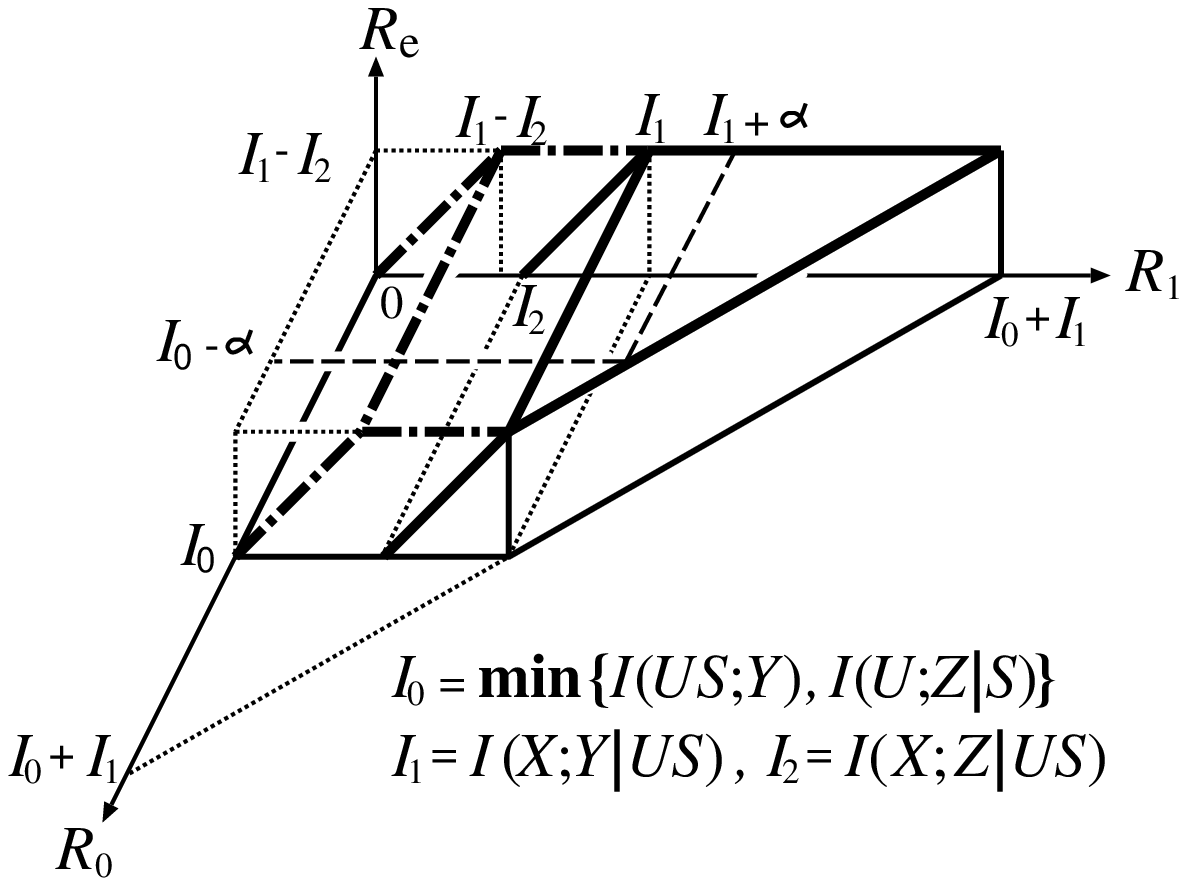}
\caption{
Shapes of 
$\check{\cal R}_{\rm s}^{\rm (in)}(U,X,S|\Ch)$
and ${\cal R}_{\rm d}^{\rm (in)}(U,X,S|\Ch)$.
} 
\label{fig:FigReg}
\ec
\end{figure}

In this subsection we prove 
${\cal R}_{\rm d}^{\rm(in)}(\Ch),
$$\tilde{\cal R}_{\rm d}^{\rm(in)}(\Ch)
\subseteq 
{\cal R}_{\rm d}(\Ch)
$ and $
{\cal R}_{\rm s}^{\rm(in)}(\Ch)
\subseteq 
{\cal R}_{\rm s}(\Ch)\,.$
Set
\beqno
& &\check{\cal R}_{\rm s}^{\rm (in)}(U,X,S|\Ch)
\\
&\defeq &
\ba[t]{l}
\{(R_0,R_1,R_{\rm e}): R_0,R_1,R_{\rm e}\geq 0\,,
\vSpa\\
\ba[t]{rcl}
R_0&\leq &\min \{I(US;Y), I(U;Z|S)\}\,,
\vSpa\\
R_0&+& R_1 \leq I(X;Y|US)
\vSpa\\
& &+\min \{I(U;Z|S),I(US;Y) \}\,,
\vSpa\\
R_{\rm e} &\leq & R_1\,, 
\vSpa\\
R_{\rm e} &\leq & [I(X;Y|US)-I(X;Z|US)]^{+}\,.\}\,,
\ea
\ea
\eeqno
and 
\beqno
\cCsi& \defeq &
\bigcup_{(U,X,S)\in {\cal P}_1}
\check{\cal R}_{\rm s}^{({\rm in})}(U,X,S|\Ch)\,.
\eeqno

\begin{figure}[t]
\bc
\includegraphics[width=8.7cm]{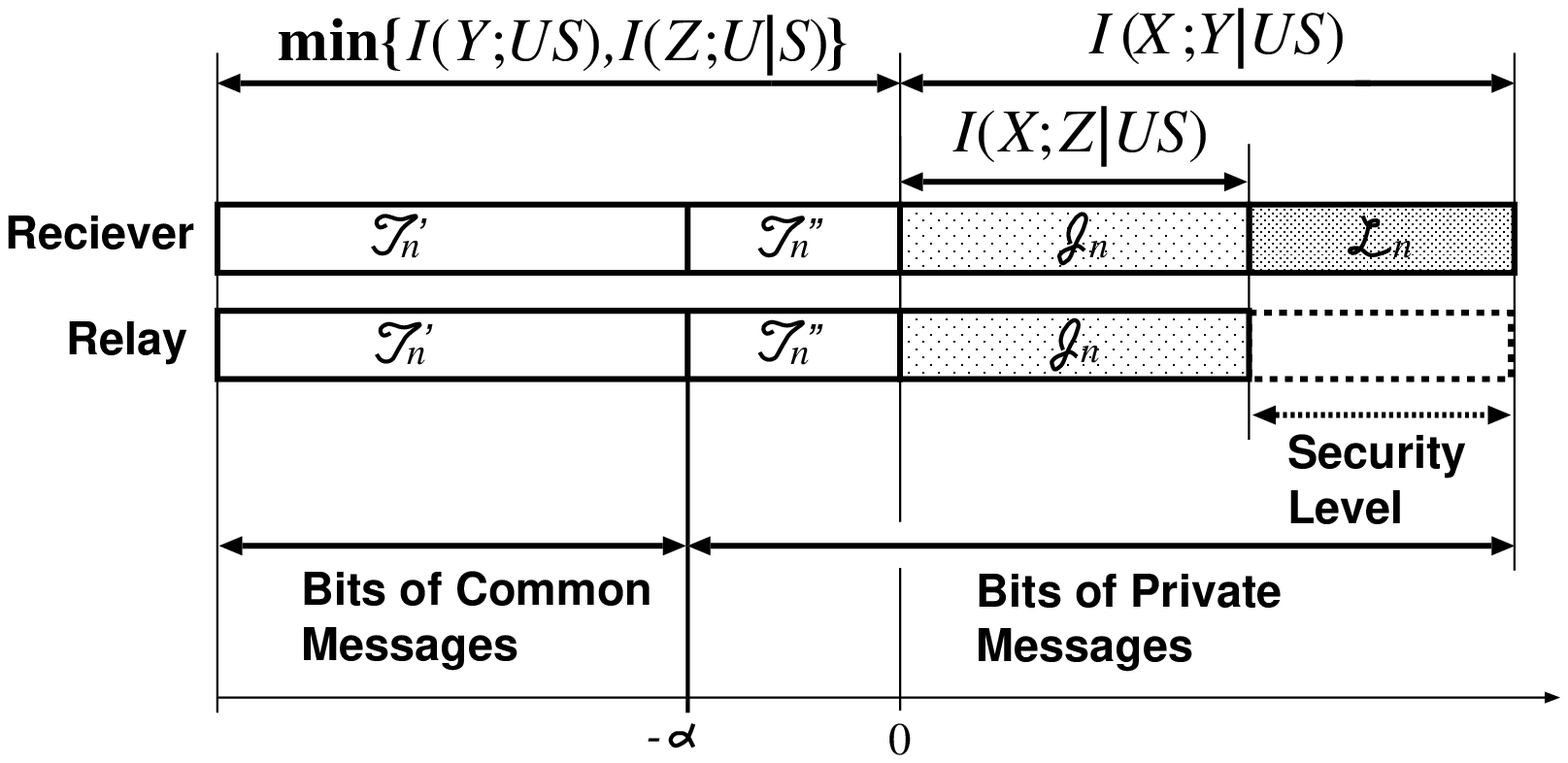}
\caption{
Information contained in the transmitted messages.
} 
\label{fig:info1}
\ec
\vspace*{-5mm}
\end{figure}

{\it Proof of 
  ${\cal R}_{\rm d}^{\rm (in)}(\Ch)\subseteq$ 
  $\Cd$ and   
  $\check{\cal R}_{\rm s}^{\rm (in)}(\Ch)\subseteq$ 
  $\Cs$ :} \ Set
\beqno
I_0&\defeq &\min\{I(US;Y),I(U;Z|S)\}\,,
\\
I_1&\defeq&I(X;Y|US), I_2 \defeq I(X;Z|US)\,.
\eeqno
We consider the case that $I_1\geq I_2$. 
The region $\check{\cal R}(U,X,S|\Ch)$ in this case 
is depicted in Fig. \ref{fig:FigReg}. 
We first prove $\Cdi\subseteq \Cd$. 
From the shape of the region 
${\cal R}_{\rm d}^{\rm (in)}(U,X,S|\Ch)$, 
it suffices to show that for every 
$$
\alpha \in [0, \min\{I(US;Y),I(U;Z|S)\}], 
$$
the following $(R_0,R_1,R_{\rm e})$ is achievable: 
\beqno
R_0&=& \min\{I(US;Y),I(U;Z|S)\}-\alpha\,, 
\\
R_1&=& I(X;Y|US)+\alpha\,,
\\
R_{\rm e}&=& I(X;Y|US)-I(X;Z|US)\,. 
\eeqno
Choose ${\cal T}_n^{\prime}$ 
and ${\cal T}_n^{\prime\prime}$ such that 
\beqno
{\cal T}_n&=&{\cal T}_n^{\prime} 
\times {\cal T}_n^{\prime\prime}\,,
\\
\lim_{n\to\infty}\nbn \log \pa {\cal T}_n^{\prime} \pa 
&=&\min\{I(US;Y),I(U;Z|S)\}-\alpha \,.
\eeqno
We take 
\beqno
{\cal M}_n={\cal T}_n^{\prime}\,,\quad
{\cal K}_n={\cal T}_n^{\prime\prime}
\times {\cal J}_n\times{\cal L}_n\,.
\eeqno
Then, by Lemma \ref{lm:direct}, we have  
\beqno 
\lim_{n\to\infty}{\mu}_{1}^{(n)}
&=&\lim_{n\to\infty}{\mu}_{2}^{(n)}=0, 
\\
\lim_{n\to\infty} \nbn \log \pa {\cal K}_n \pa 
&=& I(X;Y|US)+\alpha\,,
\\
\lim_{n\to\infty} \nbn \log \pa {\cal M}_n \pa 
&=& \min\{I(US;Y),I(U;Z|S)\}-\alpha\,,
\\
\lim_{n\to\infty} \nbn H(K_n|Z^{n}) 
&\geq & 
\lim_{n\to\infty} \nbn H(L_n|Z^{n}) 
\\
&\geq&
I(X;Y|US)-I(X;Z|US)\,.
\eeqno
To help understating the above proof, information 
quantities contained in the transmitted messages are 
shown in Fig. \ref{fig:info1}.
Next we prove $\check{\cal R}_{\rm s}^{\rm (in)}(\Ch)$ 
$\subseteq \Cs$. 
From the shape of the region 
$\check{\cal R}_{\rm s}^{\rm (in)}(U,X,S|\Ch)$, 
it suffices to show that 
the following $(R_0,R_1,R_{\rm e})$ is achievable: 
\beqno
R_0&=& \min\{I(US;Y),I(U;Z|S)\}\,, 
\\
R_1=R_{\rm e}&=& I(X;Y|US)-I(X;Z|US)\,. 
\eeqno
Choose $f_n:$ ${\cal T}_n \times {\cal J}_n $ $\times {\cal L}_n$ 
$\to {\cal X}^n$ specified \Irr{in} Lemma \ref{lm:direct}. 
Set 
${\cal M}_n$$={\cal T}_n$ and 
${\cal K}_n$$={\cal L}_n$. Using this 
$f_n$, for $(m,k)\in {\cal M}_n\times {\cal K}_n$ 
define 
$$
f_n(m,J_n,k)={\vc x}(m,J_n,k)\in{\cal X}^n\,. 
$$
The above $f_n$ is no longer a deterministic function.
It becomes a random function randomized by 
$J_n$ uniformly distributed on ${\cal J}_n$, 
which works as a ``dummy'' random variable. 
It is obvious that this random function attains
\beqno
\lim_{n\to\infty}\frac{1}{n}\log |{\cal M}_n|&=& \min\{I(US;Y),I(U;Z|S)\}\,, 
\\
\lim_{n\to\infty}\frac{1}{n}\log |{\cal K}_n|&=&I(X;Y|US)-I(X;Z|US)\,,
\\
\lim_{n\to\infty}\frac{1}{n}H(K_n|Z^n)& \geq &I(X;Y|US)-I(X;Z|US)\,,
\eeqno
completing the proof. 
\hfill\QED

{\it Proof of $\tCdi\subseteq \Cd$:} \ 
Since $\tilde{\cal R}_{\rm d}^{\rm(in)}(\Ch)\subseteq $ 
${\cal R}^{ \rm (in) }_{\rm d}(\Ch)$, we have 
$\tCdi\subseteq \Cd$. 
\hfill\QED

{\it Proof of $\Csi$ $\subseteq \Cs$:} \ 
%
Choose $(U,V,X,S)\in {\cal Q}_1$. 
The joint distribution of $(U,V,X,S)$ is given by
\beqno  
& & p_{UVXS}(u,v,x,s)
\\
&=&p_{USV}(u,s,v)p_{X|V}(x|v)\,,\:
(u,v,x,s)\in 
{\cal U}\times
{\cal V}\times
{\cal X}\times
{\cal S}\,.
\eeqno
Consider the discrete memoryless channels with input alphabet 
${\cal V}\times{\cal S}$ and output alphabet 
${\cal Y}\times{\cal Z}$, and stochastic 
matrices defined by the conditional 
distribution of $(Y,Z)$ given $V,S$ 
having the form
$$
\Ch^{\prime}(y,z|v,s)=\sum_{x\in {\cal X}}{\Ch}(y,z|x,s)
p_{X|V}(x|v)\,.
$$
Any encoder 
$f_n^{\prime}:$ ${\cal K}_n\times {\cal M}_n$
$\to {\cal V}^n$ for this new RCC determines 
a {\it stochastic} encoder $f_n$ for the 
original RCC by the matrix product of 
$f_n^{\prime}$ with the stochastic matrix
given by 
$p_{X|V}=$
$\{p_{X|V}(x|v)\}_{(v,x)
\in {\cal V}\times{\cal X}}$. Both encoders 
yield the same stochastic connection of 
messages and received sequences, 
so the assertion follows by choosing the 
encoder $f_n^{\prime}$ used for the proof 
of the inclusion $\check{\cal R}_{\rm s}^{\rm (in)}(\Ch^{\prime})
\subseteq$ 
${\cal R}_{\rm s}(\Ch^{\prime})$.   
\hfill\QED

Cardinality bounds of auxiliary random variables 
in ${\cal P}_1$ and ${\cal Q}_1$ can be proved by 
the argument that Csisz\'ar and K\"orner \cite{CsiKor1} 
developed in Appendix in their paper.

\section{Derivations of the Outer Bounds}

In this section we derive the outer bounds stated 
in 
Theorems \ref{th:dconv}-\ref{th:sreg2}. 
We further prove Theorem \ref{th:sdRegssReg}. 
We first remark here that cardinality bounds of auxiliary 
random variables in ${\cal P}_2$ and ${\cal Q}_2$ 
in the outer bounds can be proved by the argument 
that Csisz\'ar and K\"orner \cite{CsiKor1} developed 
in Appendix in their paper. 

The following lemma is a basis on derivations of the outer bounds. 
\begin{lm} \label{lm:conv1} 
We assume $(R_0,R_1,R_{\rm e})\in {\cal R}_{\rm s}^{\ast}(\Ch)$. 
Then, we have
\beq
\left.
\ba{rcl}
R_0&\leq &
\frac{1}{n}\min\{I(M_n;Y^n),I(M_n;Z^n)\}+\delta_{1,n} 
\vSpa\\
R_1&\leq &
\frac{1}{n}I(K_n;Y^n|M_n)+\delta_{2,n}
\vSpa\\
R_{\rm e}&\leq &[R_1-I(K_n;Z^n|M_n)]^{+}+\delta_{3,n}
\vSpa\\
R_{\rm e}&\leq &\left[\frac{1}{n}I(K_n;Y^n|M_n)\right.
\\
& &\qquad -\left.\frac{1}{n}I(K_n;Z^n|M_n)\right]^{+}+\delta_{4,n}\,, 
\ea
\right\}
\label{eqn:convv1}
\eeq
where $\{\delta_{i,n}\}_{n=1}^{\infty}$, $i=1,2,3,4$ 
are sequences that tend to zero as $n\to\infty$.
\end{lm}

The above lemma can be proved by a standard converse 
coding argument using Fano's inequality. 
The detail of the proof is given in Appendix C. 

We first prove $\Cd$ $\subseteq$ $\tCdo$. 
As a corollary of Lemma \ref{lm:conv1}, we have the following lemma.
\begin{lm}\label{lm:coCv1} \ \ We assume that 
$(R_0,R_1,R_{\rm e})\in {\cal R}_{\rm s}^{\ast}(\Ch)$. Then,
\beq
\left.
\ba{rcl}
R_0&\leq &\frac{1}{n}\min\{I(M_n;Y^n),I(M_n;Z^n)\}+\delta_{1,n} 
\vSpa\\
R_1&\leq &\frac{1}{n}I(K_n;Y^n|M_n)+\delta_{2,n}  
\vSpa\\
R_0+R_1&\leq &\frac{1}{n}I(K_nM_n;Y^n) + \tilde{\delta}_{3,n}
\vSpa\\
R_{\rm e}&\leq &[R_1-\frac{1}{n}I(K_n;Z^n|M_n)]^{+}+\delta_{3,n}  
\vSpa\\
R_{\rm e}&\leq &\left[\frac{1}{n}I(K_n;Y^n|M_n)\right.
\\
& &\qquad -\left. 
\frac{1}{n}I(K_n;Z^n|M_n)\right]^{+}+\delta_{4,n}\,, 
\ea
\right\}
\label{eqn:coov1}
\eeq
where $\tilde{\delta}_{3}\defeq$ $\delta_{1,n}+\delta_{2,n}$.
\end{lm} 

By this lemma, it suffices to derive upper 
bounds of 
\beqno
& & I(M_n;Z^n), I(M_n;Y^n), I(K_n;Y^n|M_n), 
\\ 
& & I(K_nM_n;Y^n), I(K_n;Y^n|M_n)-I(K_n;Z^n|M_n) 
\eeqno
to prove $\Cd$ $\subseteq$ $\tCdo$. 
For upper bound\Irr{s} of the above five quantities, we 
have the following Lemma.

\begin{lm} \label{lm:conv2} \ Set 
$$U_i\defeq M_nY^{i-1}Z^{i-1}\,,\quad i=1,2, \cdots, n\,.$$
For $i=1,2, \cdots, n$, $U_i$, $X_iS_i$, and $Y_iZ_i$ 
form a Markov chain $U_i$ $\to X_i S_i$ $\to Y_iZ_i$ 
in this order. Furthermore, we have
\beqa
I(M_n;Y^n)&\leq &\sum_{i=1}^{n}I(U_iS_i;Y_i)\,,
\label{eqn:cv1}
\\
I(M_n;Z^n)&\leq &\sum_{i=1}^{n}I(U_i;Z_i|S_i)\,,
\label{eqn:cv2}
\\
I(K_nM_n;Y^n) &\leq &\sum_{i=1}^{n}I(X_iS_i;Y_i)\,,
\label{eqn:cv2xyz}
\\
I(K_n;Y^n|M_n) &\leq &\sum_{i=1}^{n}
I(X_i;Y_iZ_i|U_iS_i)\,,
\label{eqn:cv3}
\eeqa
\beqa
&  &I(K_n;Y^n|M_n)-I(K_n;Z^n|M_n)
\nonumber\\
&\leq &
\sum_{i=1}^{n}I(X_i;Y_i|Z_iU_iS_i)\,. 
\label{eqn:cv4}
\eeqa
The bounds (\ref{eqn:cv1})-(\ref{eqn:cv3}) 
also hold for any stochastic relay encoder. 
If $\Ch$ belongs to the \Cls, the bound  (\ref{eqn:cv4}) 
also holds for any stochastic relay encoder.  
If $f_n$ is a deterministic encoder, we have 
\beq
I(K_n;Z^n|M_n) \geq \sum_{i=1}^{n}I(X_i;Z_i|U_iS_i)
\label{eqn:cv5}
\eeq
in addition to (\ref{eqn:cv1})-(\ref{eqn:cv4}).
If $\Ch$ belongs to the \Cls, the bound (\ref{eqn:cv5}) 
also hold\Irr{s} for any stochastic relay encoder.
\end{lm}

Proof of Lemma \ref{lm:conv2} is given in Appendix D.

{\it Proof of $\Cd\subseteq \tCdo$ 
and $\Cs\subseteq \tCso$:} \ 
We first assume that 
$(R_0, R_1,R_{\rm e})$$\in\Cs$. Let $Q$ be 
a random variable independent of $K_nM_nX^nY^n$ and 
uniformly distributed on 
$\{1,2,\cdots,n\}$.  Set 
\beqno
X \defeq X_Q, S \defeq S_Q, Y \defeq Y_Q, Z\defeq Z_{Q}. 
\label{convRv}
\eeqno
Furthermore, set
\beqno
U \defeq U_QQ =Z^{Q-1}Y^{Q-1}M_nQ\,. 
\eeqno
Note that $UXSYZ$ satisfies a Markov chain 
$U\to XS\to YZ$. By Lemmas \ref{lm:coCv1} and \ref{lm:conv2}, 
we have 
\beq
\left.
\ba{rcl}
R_0 &\leq &\min\{I(US;Y|Q),I(U;Z|SQ)\} +\delta_{1,n} 
\\
    &\leq &\min\{I(US;Y),I(U;Z|S)\} +\delta_{1,n} 
\\
R_1 &\leq &I(X;YZ|US)     +\delta_{2,n} 
\\
R_0+R_1 &\leq & I(XS;Y|Q)   +\tilde{\delta}_{3,n} 
\\
        &= & I(XS;Y)   +\tilde{\delta}_{3,n} 
\\
R_{\rm e}&\leq &I(X;Y|ZUS)+\delta_{4,n}\,. 
\ea
\right\}
\label{eqn:conv001}
\eeq
Using memoryless character of the 
channel it is straightforward to verify that $U\to XS\to YZ$ and that 
the conditional distribution of $(Y,Z)$ given $XS$ coincides 
with the corresponding channel matrix $\Gamma$.
Hence, by letting $n\to \infty$ in (\ref{eqn:conv001}), we 
obtain $(R_0,R_1,R_{\rm e})\in \Cso$.
Next we assume that $(R_0,R_1,R_{\rm e})\in \Cd$. Then by 
Lemmas \ref{lm:coCv1} and \ref{lm:conv2}, we have
\beq
R_{\rm e}\leq [R_1-I(X;Z|US)]^{+} +\delta_{3,n}
\label{eqn:conv001z}
\eeq
in addition to (\ref{eqn:conv001}). 
Hence by letting $n\to\infty$ in (\ref{eqn:conv001}) 
and (\ref{eqn:conv001z}), we conclude that
$(R_0,R_1,R_{\rm e})$ $\in \tCdo$. 
\hfill\QED

Next we prove the inclusions  
$\Cd$ $\subseteq$ $\Cdo$, 
$\Cd$ $\subseteq$ $\hCdo$, 
and $\Cs $$\subseteq$ $\Cso$. 
As a corollary of Lemma \ref{lm:conv1}, we have 
the following lemma.
\begin{lm}\label{lm:coCv2} \ We assume that 
$(R_0,R_1,R_{\rm e})\in {\cal R}_{\rm s}^{\ast}(\Ch)$. Then,
\beq
\left.
\ba{rcl}
R_0&\leq &\frac{1}{n}\min\{I(M_n;Y^n),I(M_n;Z^n)\}+\delta_{1,n} 
\vSpa\\
R_0+R_1&\leq & \frac{1}{n}I(K_n;Y^n|M_n) 
\vSpa\\
& &+\frac{1}{n}\min\{I(M_n;Y^n),I(M_n;Z^n)\}
+\tilde{\delta}_{3,n}
\vSpa\\
R_{\rm e}&\leq &[R_1-\frac{1}{n}I(K_n;Z^n|M_n)]^{+}
+\delta_{3,n}  
\vSpa\\
R_{\rm e}&\leq &\left[\frac{1}{n}I(K_n;Y^n|M_n)\right.
\\
& &\qquad \left.-\frac{1}{n}I(K_n;Z^n|M_n)\right]^{+}
+\delta_{4,n}\,. 
\ea
\right\}
\label{eqn:cooov1}
\eeq
\end{lm} 

From Lemma \ref{lm:coCv2}, it suffices to derive 
upper bounds of the following five quantities: 
\beqa
& &I(M_n;Z^n),I(M_n;Y^n)\,,
\nonumber\\
& &I(K_n;Y^n|M_n)+I(M_n;Y^n)=I(K_nM_n;Y^n)\,,
\nonumber\\
& &I(K_n;Y^n|M_n)+I(M_n;Z^n)\,,
\label{eqn:conv222}
\\
& &I(K_n;Y^n|M_n)-I(K_n;Z^n|M_n)\,. 
\label{eqn:conv223}
\eeqa
Since
\beqno 
& &I(K_n;Y^n|M_n)+I(M_n;Z^n)
\\
&=&I(K_n;Y^n|M_n)-I(K_n;Z^n|M_n)
+I(K_n M_n;Z^n)\,,
\eeqno
we derive an upper bound of (\ref{eqn:conv222}) 
by estimating upper bounds of $I(K_nM_n;Z^n)$ 
and (\ref{eqn:conv223}).

The following is a key lemma to prove $\Cd$ $\subseteq$ $\Cdo$.
 
\begin{lm}\label{lm:conv3a}\ Set 
$$
U_i\defeq Y_{i+1}^nZ^{i-1}M_n\,,
\quad i=1,2,\cdots,n {\,,} 
$$
where $Y_{i+1}^n$ stands for 
$Y_{i+1}Y_{i+2}\cdots Y_n$.
For $i=1,2, \cdots, n$, $U_i$, $X_iS_iZ_i$, and $Y_i$ 
form a Markov chain $U_i$ $\to X_i Z_iS_i$ $\to Y_i$ 
in this order. Furthermore, we have
\beqa
I(M_n;Y^n)&\leq  &\sum_{i=1}^{n}I(U_iS_i;Y_i)\,,
\label{eqn:cva2q}
\\
I(M_n;Z^n) &\leq &\sum_{i=1}^{n}I(U_i;Z_i|S_i)\,,
\label{eqn:cva1q}
\\
I(K_nM_n;Y^n) &\leq &\sum_{i=1}^{n}I(X_iU_iS_i;Y_i)\,,
\label{eqn:cva22q}
\\
I(K_nM_n;Z^n) &\leq &\sum_{i=1}^{n}I(X_iU_i;Z_i|S_i)\,.
\label{eqn:cva11q}
\eeqa
The bounds (\ref{eqn:cva2q})-(\ref{eqn:cva11q}) 
also hold for any stochastic relay encoder.
If $f_n$ is a deterministic encoder, we have 
\beqa
& &I(K_n;Z^n|M_n) 
\nonumber\\
&\geq &
\sum_{i=1}^{n}
\left\{I(X_i;Z_i|U_iS_i)-I(U_i;Z_i|X_iS_i)\right\}\,,
\label{eqn:cva4ssx}
\\
& &I(K_n;Y^n|M_n{})-I(K_n;Z^n|M_n{})
\nonumber\\
&\leq &
\sum_{i=1}^{n}
\left\{
I(X_i;Y_i|U_iS_i)-I(X_i;Z_i|U_iS_i)
\right\}\,, 
\label{eqn:cva4q}
\eeqa
in addition to (\ref{eqn:cva2q})-(\ref{eqn:cva11q}).
If $\Ch$ belongs to the \Cls,  
the bounds (\ref{eqn:cva4ssx}) and (\ref{eqn:cva4q}) 
also \Irr{hold} for any stochastic relay encoder.  
\end{lm}

Proof of Lemma \ref{lm:conv3a} is in Appendix E. 

{\it Proof of $\Cd$ $\subseteq$ $\Cdo$:} \ We assume that 
$(R_0,$ $R_1,R_{\rm e})$$\in \Cd$.
Let $Q$, $X$, $Y$, $Z$, $S$ be the same random 
variables as those in the proof of $\Cs$ $\subseteq$ $\tCso$. 
Set  
$$
U\defeq U_QQ=Y_{Q+1}^nZ^{Q-1}M_nQ\,. 
$$
Note that $UXSYZ$ satisfies a Markov chain $U\to XSZ\to Y$. 
By Lemmas \ref{lm:coCv2} and \ref{lm:conv3a}, we have 
\beq
\left.
\ba{rcl}
R_0 &\leq &\min\{I(US;Y),I(U;Z|S)\} +\delta_{1,n} 
\\
R_0+R_1&\leq &I(X;Y|US)
\\
       &     &+\min\{I(US;Y),I(U;Z|S)\}+\tilde{\delta}_{3,n} 
\\
R_{\rm e}&\leq &\ba[t]{l}[R_1-I(X;Z|US)\\
                +I(U;Z|XS)]^{+}+\delta_{3,n}\ea
\\
R_{\rm e}&\leq &[I(X;Y|US)-I(X;Z|US)]^{+}+\delta_{4,n}\,. 
\ea
\right\}
\label{eqn:conv002}
\eeq
By letting $n\to \infty$ in (\ref{eqn:conv002}), 
we conclude that $(R_0,R_1,$ $R_{\rm e})$ 
$\in \Cdo$. \hfill\QED

The following is a key lemma to prove $\Cd$ $\subseteq$ $\hCdo$. 
\begin{lm}\label{lm:conv3aa} \ Set 
$$
U_i\defeq Y^{i-1}Z_{i+1}^nM_n\,,\quad i=1,2,\cdots,n\,.
$$
For $i=1,2, \cdots, n$, $U_i$, $X_iS_iZ_i$, and $Y_i$ 
form a Markov 
chain $U_i$ $\to X_i Z_iS_i$ $\to Y_i$ 
in this order. Furthermore, we have
\beqa
I(M_n;Y^n)&\leq &\sum_{i=1}^{n}I(U_i;Y_i)\,,
\label{eqn:cva2r}
\\
I(M_n;Z^n)&\leq &\sum_{i=1}^{n}I(U_i;Z_i|S_i)\,,
\label{eqn:cva1r}
\\
I(K_nM_n;Y^n)&\leq &\sum_{i=1}^{n}I(X_iS_i;Y_i)\,,
\label{eqn:cva22r}
\\
I(K_nM_n;Z^n)&\leq &\sum_{i=1}^{n}I(X_i;Z_i|S_i)\,.
\label{eqn:cva11r}
\eeqa
If $f_n$ is a deterministic encoder,  
we have 
\beqa
& &I(K_n;Z^n|M_n)
\nonumber\\
&\geq&\sum_{i=1}^{n}
   \left\{I(X_i;Z_i|U_iS_i)-I(U_i;Z_i|X_iS_i)\right\}\,,
\label{eqn:cva22zat}
\\
&  &I(K_n;Y^n|M_n{})-I(K_n;Z^n|M_n{})
\nonumber\\
&\leq& \sum_{i=1}^{n}
   \left\{
   I(X_iS_i;Y_i|U_i)-I(X_iS_i;Z_i|U_i)
   \right. 
\nonumber\\
& &\qquad\left. +I(U_i;Z_i|X_iS_i)\right\}
\nonumber\\
&=&\sum_{i=1}^{n}
   \left\{I(X_i;Y_i|U_iS_i)-I(X_i;Z_i|U_iS_i)\right. 
   \nonumber\\
& &\qquad\left.+\ZeTaI+I(U_i;Z_i|X_iS_i)\right\}\,, 
\label{eqn:cva4r}
\eeqa
in addition to (\ref{eqn:cva2r})-(\ref{eqn:cva11r}). 
The bounds (\ref{eqn:cva2r}),(\ref{eqn:cva22r}), 
and (\ref{eqn:cva11r}) also hold for 
any stochastic relay encoder. 
If $\Ch$ belongs to the \Cls, the bound (\ref{eqn:cva1r}) 
also holds for any stochastic relay encoder.  
If $f_n$ is deterministic and $\Ch$ belongs to the \Cls, 
the bounds (\ref{eqn:cva2r})-(\ref{eqn:cva4r}) 
hold for any stochastic relay encoder.  
\end{lm}

Proof of Lemma \ref{lm:conv3aa} is in Appendix F.  

{\it Proof of $\Cd$ $\subseteq$ $\hCdo$:} \ We assume that 
$(R_0,$ $R_1,R_{\rm e})$$\in \Cd$.
Let $Q$, $X$, $Y$, $Z$, $S$ be the same random 
variables as those in the proof
$\Cs$ $\subseteq$ $\tCso$. 
We set  
$$
U\defeq U_QQ=Y^{Q-1}Z_{Q+1}^{n}M_nQ\,. 
$$
Note that $UXSYZ$ satisfies a Markov chain 
$U\to XSZ\to Y$. Furthermore, if $\Ch$ belongs 
to the \Cls, we have 
\beq
U\to XS \to Z,
\label{eqn:Mark}
\eeq
which together with $U\to XSZ\to Y$ yields 
$$
U\to XS \to YZ \,.
$$
By Lemmas \ref{lm:coCv2} and \ref{lm:conv3aa}, we have 
\beq
\left.
\ba{rcl}
R_0 &\leq &\min\{I(U;Y),I(U;Z|S)\} +\delta_{1,n} 
\\
R_0+R_1&\leq &I(X;Y|US)+\min\{I(US;Y),
\\
       &     & I(U;Z|S)+\ZeTa\}+\tilde{\delta}_{3,n} 
\\
R_{\rm e}&\leq & [R_1-I(X;Z|US)
\\ 
         &     & +I(U;Z|XSQ)]^{+}+\delta_{3,n}
\\
         & =   &[R_1-I(X;Z|US)]^{+}+\delta_{3,n}
\\
R_{\rm e}&\leq &[I(XS;Y|U)-I(XS;Z|U)
\\
         &     & +I(U;Z|XSQ)]^{+}+\delta_{4,n}
\\ 
         & =   &[I(XS;Y|U)-I(XS;Z|U)]^{+}+\delta_{4,n}\,. 
\ea
\right\}
\label{eqn:conv002z}
\eeq
Note here that since $I(U;Z|XSQ)$ $\leq$ $I(U;Z|XS)$
and the Markov chain of (\ref{eqn:Mark}), 
the quantity $I(U;Z|XSQ)$ vanishes. 
By letting $n\to \infty$ in (\ref{eqn:conv002z}), 
we conclude that $(R_0,R_1,$ $R_{\rm e})$ 
$\in \hCdo$. \hfill\QED

The following is a key result 
to prove $\Cs \subseteq \Cso$.
\begin{lm}\label{lm:conv3b} 
Let $U_i$, $i=1,2,\cdots,n$ 
be the same random variables as those defined 
in Lemma \ref{lm:conv3a}. We further set $V_i \defeq U_iS_iK_n$. 
For $i=1,2,\cdots,n$, $U_iV_iX_iS_iZ_i$ satisfies 
the following Markov chains 
\beqno
& & U_i   \to V_i\to X_iS_iZ_i\to Y_i\,, 
    U_iS_i\to V_iX_i \to Z_i\,, 
\\
& &U_iS_i\to V_i\to X_i\,.
\eeqno
Furthermore, we have
\beqa
I(M_n;Y^n)&\leq &\sum_{i=1}^{n}I(U_iS_i;Y_i)\,,
\label{eqn:cvza1}
\\
I(M_n;Z^n)&\leq &\sum_{i=1}^{n}I(U_i;Z_i|S_i)\,,
\label{eqn:cvza2}
\\
I(K_nM_n;Y^n)&\leq &\sum_{i=1}^{n}I(V_iU_iS_i;Y_i)\,,
\label{eqn:cvza3}
\\
I(K_nM_n;Z^n)&\leq &\sum_{i=1}^{n}I(V_iU_i;Z_i|S_i)\,,
\label{eqn:cvza4}
\eeqa
\beqa
&  &I(K_n;Y^n|M_n{})-I(K_n;Z^n|M_n{})
\nonumber\\
&=&
\sum_{i=1}^{n}
\left\{
I(V_i;Y_i|U_iS_i)-I(V_i;Z_i|U_iS_i)
\right\}\,. 
\label{eqn:cvza5}
\eeqa
The bounds (\ref{eqn:cvza1})-(\ref{eqn:cvza4}) 
also hold for any stochastic relay encoder. 
If $\Ch$ belongs to the \Cls,  
the bound (\ref{eqn:cvza5}) also hold\Irr{s} 
for any stochastic relay encoder.  
\end{lm}

Proof of Lemma \ref{lm:conv3b} is given in Appendix E.

{\it Proof of $\Cs$ $\subseteq$ $\Cso$: } 
Let $Q$, $X$, $Y$, $Z$, $S$, $U$ be the same 
random variables as those in the proof 
of $\Cd$ $\subseteq$ $\Cdo$. We further set 
$V\defeq USK_n$. Note that $UVXSZ$ satisfies the following 
Markov chains 
\beqno
& & U\to V\to XSZ\to Y\,, 
    US\to VX\to Z\,, 
\\
& &US\to V\to X\,.
\eeqno
By Lemmas \ref{lm:coCv2} and \ref{lm:conv3b},
we have 
\beq
\left.
\ba{rcl}
R_0 &\leq &\min\{I(US;Y),I(U;Z|S)\} +\delta_{1,n} 
\\
R_0+R_1&\leq &I(V;Y|US)
\\
       &     &+\min\{I(US;Y),I(U;Z|S)\}+\tilde{\delta}_{3,n} 
\\
R_{\rm e}&\leq & R_1 +\delta_{3,n} 
\\
R_{\rm e}&\leq &I(V;Y|US)-I(V;Z|US)+\delta_{4,n}\,. 
\ea
\right\}
\label{eqn:conv0022}
\eeq
By letting $n\to \infty$ in (\ref{eqn:conv0022}), 
we conclude that $(R_0,R_1,R_{\rm e})$ $\in \Cso$.
\hfill\QED

{\it Proof of Theorem \ref{th:sdRegssReg}:} \ 
We assume that $\Ch$ belongs to the \Cls. 
By Lemmas \ref{lm:coCv1}-\ref{lm:conv3b} and arguments quite 
parallel with the previous arguments of the derivations 
of outer bounds we can prove that 
$\Cdo$, $\tCdo$, and $\hCdo$ serve 
as outer bounds of ${\cal R}_{\rm d}^{\ast}(\Ch)$ 
and that $\tCso$ and $\Cso$ serve as 
outer bounds of ${\cal R}_{\rm s}^{\ast}(\Ch)$. 
\hfill\QED


\section{
Derivations of the Inner 
and Outer Bounds for the Gaussian Relay Channel
}

In this section we prove Theorem \ref{th:ThGauss}. 
Let $(\xi_1,\xi_2)$ be a zero mean Gaussian random vector with 
covariance $\Sigma$ defined in Section V. By definition, we have 
$$
\xi_2=\rho\sqrt{\frac{N_2}{N_1}}\xi_1
      +\xi_{2|1}\,,
$$
where $\xi_{2|1}$ is a zero mean Gaussian random variable 
with variance $(1-\rho^2)N_2$ and independent of $\xi_{1}$. 
We consider the Gaussian relay channel specified by $\Sigma$. 
For two input random variables $X$ and $S$ 
of this Gaussian relay channel, 
output random variables $Y$ and $Z$ are given by
\beqno
Y&=&X+S+\xi_1\,,\\
Z&=&X+\xi_2=X+ \rho\sqrt{\frac{N_2}{N_1}}\xi_1+\xi_{2|1}\,.
\eeqno
Define two sets of random variables by 
\beqno 
{\cal P}(P_1,P_2)
\defeq \{(U,X,S): 
\ba[t]{l}
{\bf E}[X^2]\leq P_1, {\bf E}[S^2]\leq P_2\,,\\
U\to XS\to YZ\: \}{\,,}
\ea
\eeqno
\beqno 
{\cal P}_G(P_1,P_2)
\defeq \{(U,X,S): 
\ba[t]{l}
U, X, S\mbox{ are zero mean}\\
\mbox{Gaussian random variables.}
\\
{\bf E}[X^2]\leq P_1\,, {\bf E}[S^2]\leq P_2\,,\\
U\to XS\to YZ\:\}\,.  
\ea
\eeqno
Set
\beqno
\tilde{\cal R}_{\rm d}^{(\rm in)}(P_1,P_2|\Sigma)
&\defeq &
\ba[t]{l}
\{(R_0,R_1,R_{\rm e}) : R_0,R_1,R_{\rm e} \geq 0\,,
\vSpa\\
\ba[t]{rcl}
R_0 &\leq & \min \{I(US;Y),I(U;Z|S)\}\,,
\vSpa\\
R_1 &\leq & I(X;Y|US)\,,
\vSpa\\
R_{\rm e} &\leq &[R_1-I(X;Z|US)]^{+}\,,
\ea
\vSpa\\
\mbox{ for some }(U,X,S)
\in {\cal P}_G(P_1,P_2)\,. \}\,,
\ea
\eeqno
\beqno
\tilde{\cal R}_{\rm d}^{(\rm out)}(P_1,P_2|\Sigma)
&\defeq &
\ba[t]{l}
\{(R_0,R_1,R_{\rm e}) : R_0,R_1,R_{\rm e} \geq 0\,,
\vSpa\\
\ba[t]{rcl}
R_0 &\leq & \min \{I(US;Y),I(U;Z|S)\}\,,
\vSpa\\
R_1 &\leq & I(X;YZ|US)\,,
\vSpa\\
R_0&+&R_1 \leq I(XS;Y)\,,
\vSpa\\
R_{\rm e} &\leq &[R_1-I(X;Z|US)]^{+}\,,
\vSpa\\
R_{\rm e} &\leq & I(X;Y|ZUS)\,,
\ea
\vSpa\\
\mbox{ for some }(U,X,S)\in {\cal P}(P_1,P_2)\,.\},
\ea
\eeqno
\beqno
\tilde{\cal R}_{\rm s}^{(\rm in)}(P_1,P_2|\Sigma)
&\defeq &
\ba[t]{l}
\{(R_0,R_1,R_{\rm e}) : R_0,R_1,R_{\rm e}\geq 0,
\vSpa\\
\ba[t]{rcl}
R_0 &\leq & \min \{I(US;Y),I(U;Z|S)\}\,,
\vSpa\\
R_{\rm e}&\leq &R_1\leq I(X;Y|US)\,,
\vSpa\\
R_{\rm e} &\leq & I(X;Y|US)-I(X;Z|US)\,,
\ea
\vSpa\\
\mbox{ for some }(U,X,S)
\in {\cal P}_G(P_1,P_2)\,.\}\,,
\ea
\eeqno
\beqno
\tilde{\cal R}_{\rm s}^{(\rm out)}(P_1,P_2|\Sigma)
&\defeq &
\ba[t]{l}
\{(R_0,R_1,R_{\rm e}): R_0,R_1,R_{\rm e}\geq 0,
\vSpa\\
\ba[t]{rcl}
R_0 &\leq & \min \{I(US;Y),I(U;Z|S)\}\,,
\vSpa\\
R_0&+&R_1 \leq  I(XS;Y)\,,
\vSpa\\
R_{\rm e}&\leq& R_1\leq I(X;YZ|US)\,,
\vSpa\\
R_{\rm e} &\leq & I(X;Y|ZUS)\,,
\ea
\vSpa\\
\mbox{ for some }(U,X,S)\in {\cal P}(P_1,P_2)\,. \}.
\ea
\eeqno

Then we have the following.
\begin{pro}{\label{pro:GaussConv} For any Gaussian 
relay channel we have
\beqno
\tilde{\cal R}_{\rm d}^{(\rm in)}(P_1,P_2| \Sigma)
\subseteq 
{\cal R}_{\rm d}(P_1,P_2|\Sigma)
\subseteq 
\tilde{\cal R}_{\rm d}^{(\rm out)}(P_1,P_2| \Sigma),
\\
\tilde{\cal R}_{\rm s}^{(\rm in)}(P_1,P_2| \Sigma)
\subseteq 
{\cal R}_{\rm s}(P_1,P_2|\Sigma)
\subseteq 
\tilde{\cal R}_{\rm s}^{(\rm out)}(P_1,P_2| \Sigma).
\eeqno
}\end{pro}

{\it Proof: } The first and third inclusions 
in the above proposition can be proved 
by a method quite similar to that in the case 
of discrete memoryless channels. 
In the Gaussian case we replace the entropy $H(Z|XS)$ appearing 
in the definition of $B_{Z|XS, \epsilon}$ by 
the differential entropy $h(Z|XS)$.
Similarly, we replace the entropy $H(Z|US)$ appearing 
in the definition of $B_{Z|US, \epsilon}$ by
the differential entropy $h(Z|US)$. 
On the other hand, 
Lemma \ref{lm:sec1} should be replaced by 
the following lemma.
\begin{lm}\label{lm:Gaulm} For any Gaussian relay channels and for 
$i=1,2,$ $\cdots,B-1$, we have 
\beqa
&      & \frac{1}{n}H(L_{n,i}|L_n^{(i-1)}Z^{nB})
\nonumber\\
& \geq &  r_1-I(X;Z|US)-2\epsilon-\frac{3+\log{\rm e}}{n}
\nonumber\\
&      & -\left[e_{Z|US,i}^{(n)}+\sqrt{(P_1+N_2)e_{Z|US,i}^{(n)}}\right]
\nonumber\\
&      & 
-\left\{{\ts \frac{1}{2}\log(2\pi{\rm e}N_2)}\right\}
         e_{Z|XS,i}^{(n)}\,.
\nonumber
\eeqa
\end{lm}

Proof of this lemma is in Appendix B.
Using Lemma \ref{lm:Gaulm}, we can prove that Lemma \ref{lm:error} 
still holds in the case of Gaussian relay channels.
Using this lemma, we can prove the first and third inclusions 
in Proposition \ref{pro:GaussConv}. 
We omit the detail of the proof. 

We next prove the second and fourth inclusions 
in Proposition \ref{pro:GaussConv}. 
Let $Q$, $X$, $Y$, $Z$, $S$, $U$ be the same random 
variables as those in the proofs of 
$\Cd$$\subseteq$
$\tCdo$ in Theorem \ref{th:ddirect} and 
$\Cs$$\subseteq$$\tCso$ in Theorem \ref{th:sreg1}.
Note that $UXSYZ$ satisfies a Markov chain 
$U\to XS\to YZ$. 
We assume that $(R_0,$ $R_1,R_{\rm e})$ $\in$
${\cal R}_{\rm s}($ $P_1,P_2|\Sigma)$.
On the power constraint on $X$, we have 
\beqno
\hspace*{-4mm}
{\rm \bf E}\left[X^2\right]
&=& 
\frac{1}{n}\sum_{i=1}^n{\rm \bf E}\left[X_i^2\right]\leq P_1\,.
\label{eqn:z00}
\eeqno
Similarly, we obtain
$
{\rm \bf E}\left[S^2\right]\leq P_2\,.
$
Hence, we have $(U,X,$ $S)\in {\cal P}(P_1,P_2)$.
By Lemmas \ref{lm:coCv1} and \ref{lm:conv2}, we have 
(\ref{eqn:conv001}). Hence, by letting $n\to\infty$, 
we obtain $(R_0,R_1,R_{\rm e})\in $ 
$\tilde{\cal R}_{\rm s}^{(\rm out)}$$(P_1,P_2|{\Sigma})\,$.
Next we assume that $(R_0,R_1,R_{\rm e})\in $ 
${\cal R}_{\rm d}$$(P_1,P_2|{\Sigma})\,$.
We also have $(U,X,S)\in {\cal P}(P_1,P_2)$.
By Lemmas \ref{lm:coCv1} and \ref{lm:conv2}, we have 
(\ref{eqn:conv001}) and (\ref{eqn:conv001z}). 
Hence, by letting $n\to\infty$, 
we obtain $(R_0,R_1,R_{\rm e})\in $ 
$\tilde{\cal R}_{\rm d}^{(\rm out)}$$(P_1,P_2|{\Sigma})\,$.
\hfill\QED 

It can be seen from Proposition \ref{pro:GaussConv} that 
to prove Theorem \ref{th:ThGauss}, it suffices to prove 
\beqa
&&\left.
\ba{rcl}
{\cal R}_{\rm d}^{(\rm in)}(P_1,P_2| \Sigma)
&\subseteq& 
   \tilde{\cal R}_{\rm d}^{(\rm in)}(P_1,P_2| \Sigma)
\\
{\cal R}_{\rm s}^{(\rm in)}(P_1,P_2| \Sigma)
&\subseteq& 
   \tilde{\cal R}_{\rm s}^{(\rm in)}(P_1,P_2| \Sigma)
\ea\right\}
\label{eqn:inc1}
\\
&&\left.
\ba{rcl}
\tilde{\cal R}_{\rm d}^{(\rm out)}(P_1,P_2| \Sigma)
&\subseteq& 
   {\cal R}_{\rm d}^{(\rm out)}(P_1,P_2| \Sigma)
\\
\tilde{\cal R}_{\rm s}^{(\rm out)}(P_1,P_2| \Sigma)
&\subseteq& 
   {\cal R}_{\rm s}^{(\rm out)}(P_1,P_2| \Sigma)\,.
\ea\right\}
\label{eqn:inc2}
\eeqa
Proof of (\ref{eqn:inc1})  
is straightforward. 
To prove (\ref{eqn:inc2}), 
we need some preparations. 
Set 
\beqno 
a &\defeq &\ts \frac{N_2-\rho\sqrt{N_1N_2}}
              {N_1+N_2-2\rho\sqrt{N_1N_2}}\,.
\eeqno
Define random variables $\tilde{Y}$, $\tilde{\xi}_1$, 
and $\tilde{\xi}_2$ by  
\beqno
\tilde{Y}&\defeq& a Y+\bar{a}Z\,,
\\
\tilde{\xi}_1
& \defeq &a \xi_1 + \bar{a}\xi_2
=\ts \frac{
(1-\rho^2)N_2\xi_1+(N_1-\rho\sqrt{N_1N_2})\xi_{2|1}
 }{N_1+N_2-2\rho\sqrt{N_1N_2}}\,,\\
\tilde{\xi}_2
& \defeq & \xi_1-\xi_{2}
=\ts \left(1-\rho \sqrt{\frac{N_2}{N_1}}\right)\xi_1-\xi_{2|1}\,.
\eeqno
Let $\tilde{N}_i={\bf E}[\tilde{\xi}_i^2],i=1,2$. 
Then, by simple computation we can show that $\tilde{\xi}_1$ 
and $\tilde{\xi}_2$ are independent Gaussian random variables and 
\beqno
\tilde{N}_1&=&\ts \frac{(1-\rho^2)N_1N_2}{N_1+N_2-2\rho\sqrt{N_1N_2}}\,,
\\
\tilde{N}_2&=&{N_1+N_2-2\rho\sqrt{N_1N_2}}\,.
\eeqno
We have the following relations between $\tilde{Y}$, $Y$, 
and $Z$: 
\beq
\left.
\ba{rcl}
\tilde{Y}&=& X + aS +\tilde{\xi}_1\\
{Y}&=&\tilde{Y}+ {\bar{a}(S + \tilde{\xi}_2) }\\
{Z}&=&\tilde{Y}- {     a (S + \tilde{\xi}_2) }.
\ea
\right\}
\label{eqn:eqqq}
\eeq
The following is a useful lemma to prove (\ref{eqn:inc2}).

\begin{lm}\label{lm:Rndau} Suppose that $(U,X,S)$ 
$\in {\cal P}(P_1, P_2)$. Let $X(s)$ be a random 
variable with a conditional distribution of $X$
for given $S=s$. ${\bf E}_{X(s)}[\cdot]$ stands for 
the expectation with respect to the (conditional) 
distribution of $X(s)$. Then, there exists 
a pair $(\alpha,\beta)\in [0,1]^2$ 
such that
\beqno
& &{\bf E}_S\left({\bf E}_{X(S)}X(S)\right)^2=\bar{\alpha} P_1\,,
\\
h(Y|S)&\leq &\ts \frac{1}{2}
\log\left\{(2\pi{\rm e})({\alpha}P_1+N_1)\right\}\,,
\\
h(Z|S)&\leq &\ts \frac{1}{2}
\log\left\{(2\pi{\rm e})({\alpha}P_1+N_2)\right\}\,,
\\
h(Y)&\leq &\ts \frac{1}{2}\log\left\{(2\pi{\rm e})
(P_1+P_2+2\sqrt{\bar{\alpha} P_1P_2}+N_1)\right\}\,,
\\
h(\tilde{Y}|US)&=&\ts \frac{1}{2}
\log \left\{(2\pi {\rm e})(\beta \alpha P_1+ \tilde{N}_1)\right\}\,,
\\ 
h(Y|US)&\geq &\ts \frac{1}{2}
\log \left\{
(2\pi{\rm e})\left(\beta{\alpha}P_1+N_1\right) 
\right\}\,,
\\
h(Z|US)&\geq &\ts \frac{1}{2}
\log \left\{
(2\pi{\rm e})\left(\beta{\alpha}P_1+N_2\right) 
\right\}\,.
\eeqno
\end{lm}

Proof of Lemma \ref{lm:Rndau} is given in Appendix G. 
Using this lemma, we can prove Theorem \ref{th:ThGauss}. 

{\it Proof of Theorem \ref{th:ThGauss}:} \ We first 
prove (\ref{eqn:inc1}). Choose $(U,$ $X,S)\in$ ${\cal P}_G$ 
such that  
\beqno
& &{\bf E}[X^2]=P_1, \quad{\bf E}[S^2]=P_2,
\\
& &U=\ts \sqrt{\frac{\bar{\theta} \bar{\eta} P_1}{P_2}}S 
+ \tilde{U}, \quad X=U+\tilde{X}, 
\eeqno
where $\tilde{U}$ and $\tilde{X}$ are zero mean Gaussian random 
variables with variance $\bar{\theta}\eta P_1$ and $\theta P_1$, 
respectively. The random variables $S$, $\tilde{U}$, and $\tilde{X}$ 
are independent. For the above choice of 
$(U,X,S)$, we have 
\beqno
I(US;Y)&=& C\left(\ts \frac{\bar{\theta}P_1+P_2
          +2\sqrt{\bar{\theta}\bar{\eta}P_1P_2}}
          {\theta P_1+N_1}
          \right)\,,
\\
I(U;Z|S)&=& C\left(\ts \frac{\bar{\theta}\eta P_1}{\theta P_1+N_2}\right)\,,
\\
I(X;Y|US)&=& C\left(\ts \frac{\theta P_1}{N_1}\right)\,,
\quad
I(X;Z|US)= C\left(\ts \frac{\theta P_1}{N_2}\right)\,.
\eeqno
Thus, (\ref{eqn:inc1}) is proved. 
Next, we prove (\ref{eqn:inc2}). 
By Lemma \ref{lm:Rndau}, we have 
\beqa
I(US;Y)&=&h(Y)-h(Y|US)
\nonumber\\
      &\leq&C\left(\ts \frac{(1-\beta\alpha)P_1+P_2
            +2\sqrt{ \bar{\alpha}P_1P_2}}
            {\beta \alpha P_1+N_1}
            \right)\,,
\label{eqn:bound1}\\
I(U;Z|S)&=&h(Z|S)-h(Z|US)
\nonumber\\
       &\leq& C\left(\ts \frac{\bar{\alpha}P_1}
            {\beta \alpha P_1+N_2}\right)\,,
\label{eqn:bound2}\\
I(XS;Y)&=&h(Y)-h(Y|XS)\nonumber\\
       &\leq& C\left(\ts \frac{(1-\beta \alpha)P_1+P_2
            +2\sqrt{\bar{\alpha}P_1P_2}}{N_1}
        \right)\,,
\label{eqn:bound3}\\
I(X;Z|US)&=& h(Z|US)-h(Z|XS)
\nonumber\\
       &\geq& C\left(\ts\frac{\beta \alpha  P_1}{N_2}\right)\,,
\label{eqn:bound31}\\
I(X;YZ|US)&=&h(YZ|US)-h(YZ|XS)\nonumber\\
          &=&h(\tilde{Y}Z|US)-h(\tilde{Y}Z|XS) \nonumber\\
          &=&h(\tilde{Y}|US)+ h(Z|\tilde{Y}US) \nonumber\\
          & &-h(\tilde{Y}|XS)-h(Z|\tilde{Y}XS) \nonumber\\
          &{\MEq{a}}&h(\tilde{Y}|US)-h(\tilde{Y}|XS)  
\nonumber\\
          &=&C \left(\ts \frac{\beta \alpha  P_1}
 {\frac{(1-\rho^2)N_1N_2}{N_1+N_2-2\rho\sqrt{N_1N_2}}}\right)\,, 
\label{eqn:bound4}
\eeqa
where {\rm (a)} follows from 
\beqno
& &h(Z|\tilde{Y}US)=h(Z|\tilde{Y}XS)
=h(Z|\tilde{Y}S)\\
&=&\ts \frac{1}{2}
\log \left\{(2\pi{\rm e}) a^2 \tilde{N_2} \right\}\,. 
\eeqno
From (\ref{eqn:bound31}) and (\ref{eqn:bound4}), we have 
\beqa
\hspace*{-4mm}I(X;Y|ZUS)
&\leq & C \left(\ts \frac{\beta \alpha  P_1}
 {\frac{(1-\rho^2)N_1N_2}{N_1+N_2-2\rho\sqrt{N_1N_2}}}\right)
  -C\left(\ts\frac{\beta \alpha  P_1}{N_2}\right)\,.
\label{eqn:bound5}
\eeqa
Here we transform the variable pair 
$(\alpha,\beta)\in [0,1]^2$ into 
$(\eta, \theta)\in [0,1]^2$ in the following manner:
\beq
\theta=\beta\alpha, \quad  
\eta=1-\frac{\bar{\alpha}}{\bar{\theta}}
=\frac{\alpha-\theta}{1-\theta}\,. 
\label{eqn:trans0}
\eeq
This map is a bijection because from $(\ref{eqn:trans0})$, 
we have 
\beq
\alpha=1-\bar{\theta}\bar{\eta}\geq \theta, \quad
\beta=\frac{\theta}{\alpha}\,.
\label{eqn:trans1}
\eeq
Combining (\ref{eqn:bound1})-(\ref{eqn:bound3}), (\ref{eqn:bound4}), 
(\ref{eqn:bound5}), and (\ref{eqn:trans1}),
we have (\ref{eqn:inc2}).
\hfill\QED 

{
\section{Conclusion}}
{
We have considered the coding problem of the RCC, where the relay acts 
as both a helper and a wire-tapper. We have derived the inner and outer 
bounds of the deterministic and stochastic \CEreg regions of the RCC 
and have established the deterministic rate region in the case where 
the relay channel is reversely degraded. 
Furthermore, we have computed the inner and outer bounds of 
the deterministic and stochastic secrecy capacities and 
have determined the deterministic secrecy 
capacity for the class of reversely degraded relay channels. 
We have also evaluated the \CEreg region 
and secrecy capacity in the case of Gaussian 
relay channels.}

In this paper, we have focused purely on the derivation 
of information-theoretic bounds on the RCC. Problem 
of practical constructions 
of codes achieving the derived inner bounds of the RCC is left 
to us as a further study. Applications of LDPC codes to the wire-tap   
channel were studied in \cite{tdcmj}. This work may provide 
some key ideas to investigate the code design problem for the RCC.

\section*{\empty}
\appendix

In the following arguments, $X_{[i]}$ stands for $(X^{i-1},$ $X_{i+1}^n)$. 
Similar notations are used for other random variables.

\subsection{
Proof of Lemma \ref{lm:error}
}

In this appendix we prove Lemma \ref{lm:error}.

{\it Proof of Lemma \ref{lm:error}:}
We first derive the upper bound of 
${\sf E}\left[\hat{e}_{2,i}^{(n)}(1|w_i)\right]$ 
in Lemma \ref{lm:error}. Set
\beqno
& &\tilde{e}_{2,i}^{(n)}
\defeq \Pr\{\tilde{\cal E}_{2,i}|{\cal F}^{c}_{i-1} \}
\,,
\hat{e}_{2,i}^{(n)}
\defeq  \Pr\{\hat{\cal E}_{2,i}|{\cal F}^{c}_{i-1} \}
\,,
\\
& &\tilde{e}_{2,i}^{(n)}(t_{i}|\phi_n(t_{i-1}),l_i) 
\\
&\defeq&\Pr\{\tilde{\cal E}_{2,i}|
{\cal F}^{c}_{i-1},\ba[t]{l} T_{n,i}=t_{i}, T_{n,i-1}=t_{i-1},\\
                     J_{n,i}=j_i,L_{n,i}=l_i\}\,,\ea
\\
& &
\hat{e}_{2,i}^{(n)}(t_{i}|\phi_n(t_{i-1})) 
\\
&\defeq& \Pr\{\hat{\cal E}_{2,i}|{\cal F}^{c}_{i-1},
T_{n,i}=t_{i},T_{n,i-1}=t_{i-1}\}.
\eeqno
Similar notations are used for other error probabilities.
By definition of 
$\tilde{e}_{2,i}^{(n)}$ and 
$\hat{e}_{2,i}^{(n)}$, we have  
\beq
\left.
\ba{rcl}
{\sf E}\left[{e}_{2,i}^{(n)}\right]
&\leq& 
 {\sf E}\left[\hat{e}_{2,i}^{(n)}\right]
+{\sf E}\left[\tilde{e}_{2,i}^{(n)}\right]
\vSpa\\
{\sf E}\left[\tilde{e}_{2,i}^{(n)}\right]
&=&\ds \frac{1}{|{\cal T}_n|^2|{\cal J}_n||{\cal L}_n|}
\vSpa\\
& & \times 
\ds \sum_{\scs (t_i,t_{i-1},j_i,l_i)
          \atop{\scs \in{\cal T}_n^2
          \times {\cal J}_n\times {\cal L}_n}}
\hspace*{-2mm}
{\sf E}\left[\tilde{e}_{2,i}^{(n)}
     (t_i|\phi_n(t_{i-1}),j_i,l_i)\right]
\vSpa\\
{\sf E}\left[\hat{e}_{2,i}^{(n)}\right]
&=&\ds \frac{1}{|{\cal T}_n|^2}
\sum_{(t_i,t_{i-1})\in {\cal T}_n^2 }
\hspace*{-2mm}
{\sf E}\left[
\hat{e}_{2,i}^{(n)}(t_i|\phi_n(t_{i-1}))
\right].
\ea\right\}
\label{eqn:bsqz1} 
\eeq
By the symmetrical property of random coding, 
it suffices to evaluate 
${\sf E}[\tilde{e}_{{\rm 2},i}^{(n)}$ 
$(1|\phi_n(t_{i-1}),1,1)]$
and ${\sf E}[\hat{e}_{{\rm 2},i}^{(n)}$ $(1|\phi_n(t_{i-1}))]$. 
\newcommand{\tI}{\hat{t}_{i}}
Note that
\beqa
& &{\sf E}\left[\tilde{e}_{{\rm 2},i}^{(n)}(1|\phi_n(t_{i-1}),1,1)\right]
\nonumber\\
&=&\sum_{w_i\in{\cal W}_n}
{\sf E}\left. \left[\tilde{e}_{{\rm 2},i}^{(n)}
(1|w_{i},1,1) \right| \phi_n(t_{i-1})=w_i\right]
\frac{1}{|{\cal W}_n|}\,,
\label{eqn:assq02} 
\\
& &{\sf E}\left[\hat{e}_{{\rm 2},i}^{(n)}(1|\phi_n(t_{i-1}))\right]
\nonumber\\
&=&\sum_{w_i\in{\cal W}_n}
{\sf E}\left. \left[\hat{e}_{{\rm 2},i}^{(n)}
(1|w_{i}) \right| \phi_n(t_{i-1})=w_i\right]
\frac{1}{|{\cal W}_n|}\,.
\label{eqn:assq02z} 
\eeqa
On ${\sf E}[\tilde{e}_{{\rm 2},i}^{(n)}$ $(1|w_{i},1,1)$ 
$|\phi_n(t_{i-1})=w_i]$,
we have the following. 
\beqa
& &{\sf E}
   \left[ \left. \tilde{e}_{2,i}^{(n)}(1|w_i,1,1) \right|
   \phi_n(t_{i-1})=w_i \right]
\nonumber\\
&=&\sum_{\scs ({\svc s}(w_i),{\svc u}(w_i,1),
   \atop {\scs {\svc z}_i) \notin {\cal A}_{UZ|S,\epsilon}}}
   \sum_{{\svc x}(w_i,1,1,1)\in{\cal X}^n}
\hspace*{-6mm}p_S({\vc s}(w_i))p_{U|S}({\vc u}(w_i,1)|{\vc s}(w_i))
\nonumber\\
& &\qquad \qquad \qquad \times 
p_{X|US}({\vc x}(w_i,1,1,1)|{\vc u}(w_i,1),{\vc s}(w_i))
\nonumber\\
& &\qquad \qquad \qquad \times 
p_{Z|XS}({\vc z}_{i}|{\vc x}(w_i,1,1,1),{\vc s}(w_i) )
\nonumber\\
&=& \sum_{\scs ({\svc s}(w_i),{\svc u}(w_i,1),
    \atop {\scs {\svc z}_i) \notin {\cal A}_{UZ|S,\epsilon}}}
\hspace*{-6mm}p_S({\vc s}(w_i))
p_{UZ|S}({\vc u}(w_i,1),{\vc z}_i|{\vc s}(w_i))
\nonumber\\
&= & \Pr\left\{(S^n,U^n,Z^n)
\notin {\cal A}_{UZ|S,\epsilon} \right\}\,.
\label{eqn:sddf0}
\eeqa
From (\ref{eqn:assq02}) and (\ref{eqn:sddf0}), 
we have
\beq
{\sf E}\left[\tilde{e}_{{\rm 2},i}^{(n)}(1|\phi_n(t_{i-1}),1)\right]
=\Pr\left\{(S^n,U^n,Z^n) \notin {\cal A}_{UZ|S,\epsilon} \right\}\,.
\label{eqn:sddf001}
\eeq
On ${\sf E}[\hat{e}_{{\rm 2},i}^{(n)}$ $(1|w_{i})$ 
$|\phi_n(t_{i-1})=w_i]$, we have the following. 
\beqa
& &{\sf E}\left[\left. 
                \hat{e}_{2,i}^{(n)}(1|w_i)\right|
                \phi_n(t_{i-1})=w_i\right]
\nonumber\\
&\leq &\sum_{\tI\neq 1}
           \sum_{\scs ({\svc s}(w_i),{\svc u}(w_i,\tI),
           \atop {\scs {\svc z}_i) \in {\cal A}_{UZ|S,\epsilon}}}
\hspace*{-2mm}p_S({\vc s}(w_i))p_{U|S}({\vc u}(w_i,\tI)|{\vc s}(w_i))
\nonumber\\
& &\qquad\qquad \qquad\qquad \times p_{Z|S}({\vc z}_i|{\vc s}(w_i))
\nonumber\\
&\MLeq{a} & \sum_{\tI\neq 1}
           \sum_{\scs ({\svc s}(w_i),{\svc u}(w_i,\tI),
           \atop {\scs {\svc z}_i) \in {\cal A}_{UZ|S,\epsilon}}}
\hspace*{-2mm}
p_S({\vc s}(w_i))
p_{UZ|S}({\vc u}(w_i,t_i),{\vc z}_i|{\vc s}(w_i))
\nonumber\\
& &\qquad\qquad \qquad\qquad \times 2^{-n[R_0+\epsilon]}
\nonumber\\
&=&\sum_{\tI\neq 1}2^{-n[R_0+\epsilon]}
\hspace*{-2mm}
           \sum_{\scs ({\svc s}(w_i),{\svc u}(w_i,\tI),
           \atop {\scs {\svc z}_i) \in {\cal A}_{UZ|S,\epsilon}}}
\hspace*{-2mm}p_{SUZ}({\vc s}(w_i), {\vc u}(w_i,\tI),{\vc z}_i)
\nonumber\\
&\leq& 2^{-n[R_0+\epsilon]}(2^{nR_0}-1)\leq 2^{-n\epsilon}\,.
\label{eqn:sddf2}
\eeqa
Step (a) follows from the definition of 
${\cal A}_{UZ|S,\epsilon}$.
From (\ref{eqn:assq02z}) and (\ref{eqn:sddf2}), we have 
\beq
{\sf E}\left[\hat{e}_{{\rm 2},i}^{(n)}(1|\phi_n(t_{i-1}))\right]
\leq 2^{-n\epsilon}\,.
\label{eqn:sddf002}
\eeq
Hence, form 
(\ref{eqn:bsqz1}), 
(\ref{eqn:sddf001}), and (\ref{eqn:sddf002}) 
we have 
$$ 
{\sf E}\left[\hat{e}_{{\rm 2},i}^{(n)}\right]
\leq \Pr\left\{(S^n,U^n,Z^n)
\notin {\cal A}_{UZ|S,\epsilon} \right\}+2^{-n\epsilon}\,.
$$
In a manner quite similar to the above argument,
we can derive the upper bounds of 
${\sf E}\left[{e}_{{\rm 1a},i}^{(n)}\right]$ and 
${\sf E}\left[{e}_{{\rm 1c},i}^{(n)}\right]$ stated 
in Lemma \ref{lm:error}. 

Next, we derive the upper bound of 
${\sf E}\left[{e}_{{\rm 1b},i}^{(n)}\right]$.
By definition of $\tilde{e}_{{\rm 1b},i}^{(n)}$ and 
$\hat{e}_{{\rm 1b},i}^{(n)}$, we have  
\beq
\left.
\ba{rcl}
 {\sf E}\left[{e}_{{\rm 1b},i}^{(n)}\right]
&\leq& 
 {\sf E}\left[  \hat{e}_{{\rm 1b},i}^{(n)}\right]
+{\sf E}\left[\tilde{e}_{{\rm 1b},i}^{(n)}\right]
\vSpa\\
{\sf E}\left[\tilde{e}_{{\rm 1b},i}^{(n)}\right]
&=&\ds \frac{1}{|{\cal T}_n|^2|{\cal L}_n|}
\vSpa\\
&&\ds \times \hspace*{-2mm}\sum_{\scs (t_{i-1},t_{i-2},
\atop{\scs l_{i-1})\in {\cal T}_n^2 \times {\cal L}_n}}
\hspace*{-3mm}
{\sf E}\left[
\tilde{e}_{{\rm 1b},i}^{(n)}(t_{i-1}|\phi_n(t_{i-2}),l_{i-1})
\right]
\vSpa\\
{\sf E}\left[\hat{e}_{{\rm 1b},i}^{(n)}\right]
&=&\ds \frac{1}{|{\cal T}_n|^2}
\sum_{(t_{i-1},t_{i-2})\in{\cal T}_n^2}
\hspace*{-3mm}
{\sf E}\left[\hat{e}_{{\rm 1b},i}^{(n)}
(t_{i-1}|\phi_n(t_{i-2}))\right].
\ea
\right\}
\label{eqn:zoz}
\eeq
By the same argument as that of the derivation 
of (\ref{eqn:sddf001}), we have 
\beqa
\lefteqn{{\sf E}\left[\tilde{e}_{{\rm 1b},i}^{(n)}
(t_{i-1}|\phi_n(t_{i-2}),l_{i-1})\right]}
\nonumber\\
&=&
\Pr\left\{(S^n,U^n,Y^n)
\notin {\cal A}_{UY|S,\epsilon} \right\}
\label{eqn:zasa}
\eeqa
for any $(t_{i-1}, t_{i-2},l_{i-1})
\in {\cal T}_n^2\times {\cal L}_n$.
Then, from (\ref{eqn:zoz}) and (\ref{eqn:zasa}), we have 
\beq
{\sf E}\left[\tilde{e}_{{\rm 1b},i}^{(n)}\right]
=\Pr\left\{(S^n,U^n,Y^n)\notin {\cal A}_{UY|S,\epsilon} \right\}\,.
\label{eqn:zasa2}
\eeq
Next, we evaluate 
${\sf E}\left[
\hat{e}_{{\rm 1b},i}^{(n)}(t_{i-1}|\phi_n(t_{i-2}))
\right]\,.$
By the symmetrical property of random coding, it suffices 
to evaluate the above quantity 
for $(t_{i-1}, t_{i-2})=(1,1)$ or $(1,2)$. 
When $(t_{i-1}, t_{i-2})=(1,1)$, set $\phi_{n}(1)=w_i$. 
\newcommand{\hT}{\hat{\hat{t}}_{i-1}}
Then, we have 
\beqa
& &
{\sf E}\left[
\hat{e}_{{\rm 1b},i}^{(n)}(1|\phi_n(1))\right]
=
{\sf E}\left[
\hat{e}_{{\rm 1b},i}^{(n)}(1|w_i)\right]
\nonumber\\
&\leq&
\sum_{\hT\neq 1}
\sum_{w_{i}\in{\cal W}_n}
{\sf E}
\left[\left.\hat{e}_{{\rm 1b},i}^{(n)}(1|w_{i})\right|
\phi_n(1)=\phi_{n}(\hT)=w_{i}\right]
\nonumber\\
& &\quad \times {\sf Pr}\left\{
\phi_n(1)=\phi_{n}(\hT)=w_{i}\right\}
\nonumber\\
&=&
\sum_{\hT\neq 1}
\sum_{w_{i}\in{\cal W}_n}
{\sf E}
\left[\left.\hat{e}_{{\rm 1b},i}^{(n)}(1|w_{i})\right|
\phi_n(1)=\phi_{n}(\hT)=w_{i}\right]
\nonumber\\
& &\quad \times \frac{1}{|{\cal W}_n|^2}\,.
\label{eqn:sqpp1}
\eeqa
On upper bound of 
$$
{\sf E}\left[\left.\hat{e}_{{\rm 1b},i}^{(n)}(1|w_{i})\right|
       \phi_n(1)=\phi_{n}(\hT)=w_{i}\right]\,,
$$
we have the following chain of inequalities:
\beqa
&&
{\sf E}\left[\left.\hat{e}_{{\rm 1b},i}^{(n)}(1|w_{i})\right|
       \phi_n(1)=\phi_{n}(\hT)=w_{i}\right]
\nonumber\\
&\leq&  \sum_{\scs ({\lvc s}(w_{i}),{\lvc u}(w_{i},\hT),
      \atop {\scs {\lvc y}_{i-1}) \in {\cal A}_{UY|S,\epsilon}}
          }
p_S({\vc s}(w_{i}))
\nonumber\\
& &
\times
p_{U|S}({\vc u}(w_{i},\hT)|{\vc s}(w_{i}))
p_{Y|S}({\vc y}_i|{\vc s}(w_{i}))
\nonumber\\
&\MLeq{a}& 
         \sum_{\scs({\lvc s}(w_{i}),{\lvc u}(w_{i},\hT),
         \atop {\scs {\lvc y}_{i-1})\in {\cal A}_{UY|S,\epsilon}}
          }
p_S({\vc s}(w_{i}))
\nonumber\\
& &
\times
 p_{UY|S}({\vc u}(w_{i},\hT),{\vc y}_{i-1}|
{\vc s}(w_{i}))2^{-n[R_0-r+\epsilon]}
\nonumber\\
&=& 2^{-n[R_0-r+\epsilon]}
\nonumber\\
& & \times 
    \hspace*{-2mm}
       \sum_{\scs ({\lvc s}(w_{i}),{\lvc u}(w_{i},\hT),
        \atop {\scs{\lvc y}_{i-1})\in {\cal A}_{UY|S,\epsilon}}}
\hspace*{-2mm}
p_{SUY}({\vc s}(w_{i}),{\vc u}(w_{i},\hT),{\vc y}_{i-1})
\nonumber\\
&\leq &2^{-n[R_0-r+\epsilon]}\,.
\label{eqn:sdd00f5}
\eeqa
Step (a) follows from the definition of 
${\cal A}_{UY|S,\epsilon}$. 
It follows  from (\ref{eqn:sqpp1}) and (\ref{eqn:sdd00f5}) 
that when 
$(t_{i-1},t_{i-2})=(1,1)$, we have 
\beqa
& & 
{\sf E}\left[ \hat{e}_{{\rm 1b},i}^{(n)}(1|\phi_{n}(1)) \right]
\leq 
\sum_{\hT\neq 1}
\sum_{w_{i}\in{\cal W}_n}
\frac{2^{-n[R_0-r+\epsilon]}}{|{\cal W}_n|^2}
\nonumber\\
&\leq &(2^{nR_0}-1) \frac{2^{-n[R_0-r+\epsilon]}}{|{\cal W}_n|}
\leq 2\cdot 2^{-n\epsilon}\,.
\label{eqn:sqpp2}
\eeqa
When $(t_{i-1}, t_{i-2})=(1,2)$, set 
$\phi_{n}(1)=w_i$ and $\phi_{n}(2)=w_{i-1}$. Then, we have 
\beqa
& &
{\sf E}\left[
\hat{e}_{{\rm 1b},i}^{(n)}(1|\phi_{n}(2))\right]
=
{\sf E}\left[
\hat{e}_{{\rm 1b},i}^{(n)}(1|w_{i-1})\right]
\nonumber\\
&\leq&
\sum_{\hT\neq 1}
\sum_{(w_{i},w_{i-1}) \in{\cal W}_n^2}
\hspace*{-4mm}
{\sf E}\left[\left.\hat{e}_{{\rm 1b},i}^{(n)}(1|w_{i-1})\right|\right.
\hspace*{-1mm}\ba[t]{l}
\phi_n(1)=\phi_{n}(\hT)
\vSpa\\
=w_{i},
\vSpa\\
\Bigl. \phi_n(2)=w_{i-1} \Bigr]
\ea
\nonumber\\
& &
\qquad\quad 
\times {\sf Pr}\left\{\phi_n(1)=\phi_{n}(\hT)=w_{i},
 \phi_{n}(2)=w_{i-1}\right\}
\nonumber\\
&=&
\sum_{\hT\neq 1,2}
\sum_{(w_{i},w_{i-1}) \in{\cal W}_n^2}
\hspace*{-4mm}
{\sf E}\left[\left.\hat{e}_{{\rm 1b},i}^{(n)}(1|w_{i-1})\right|\right.
\hspace*{-1mm}
\ba[t]{l}
\phi_n(1)=\phi_{n}(\hT)
\vSpa\\
=w_{i},
\vSpa\\
\Bigl.\phi_n(2)=w_{i-1} \Bigr]
\ea
\nonumber\\
& & \qquad\quad \times \frac{1}{ |{\cal W}_n|^3}
\nonumber\\
& &+\sum_{w_{i}=w_{i-1}\in{\cal W}_n}
\hspace*{-4mm}
{\sf E}\left[\left.\hat{e}_{{\rm 1b},i}^{(n)}(1|w_{i-1})\right|\right.
\hspace*{-1mm}
\ba[t]{l}
\phi_n(1)=\phi_{n}(2)=w_{i},
\vSpa\\
\Bigl.\phi_n(2)=w_{i-1} \Bigr]
\ea
\nonumber\\
& & \qquad\quad \times \frac{1}{ |{\cal W}_n|^2}\,.
\label{eqn:sqppzz1}
\eeqa
On upper bound of 
$$
{\sf E}\left[\left.\hat{e}_{{\rm 1b},i}^{(n)}(1|w_{i-1})\right|
       \phi_n(1)=\phi_{n}(\hT)=w_{i},\phi_n(2)=w_{i-1}
\right]\,,
$$
we have the following chain of inequalities:
\beqa
&&\hspace*{-4mm}
{\sf E}\left[\left.\hat{e}_{{\rm 1b},i}^{(n)}(1|w_{i-1})\right|
        \phi_n(1)=\phi_{n}(\hT)=w_{i},\phi_n(2)=w_{i-1}\right]
\nonumber\\
&\leq&  \sum_{\scs ({\lvc s}(w_{i-1}),{\lvc u}(w_{i-1},\hT),
      \atop {\scs {\lvc y}_{i-1}) \in {\cal A}_{UY|S,\epsilon}}
          }
p_S({\vc s}(w_{i-1}))
\nonumber\\
& &
\times
p_{U|S}({\vc u}(w_{i-1},\hT)|{\vc s}(w_{i-1}))
p_{Y|S}({\vc y}_i|{\vc s}(w_{i-1}))
\nonumber\\
&\MLeq{a}& 
         \sum_{\scs({\lvc s}(w_{i-1}),{\lvc u}(w_{i-1},\hT),
         \atop {\scs {\lvc y}_{i-1})\in {\cal A}_{UY|S,\epsilon}}
          }
p_S({\vc s}(w_{i-1}))
\nonumber\\
& &
\times
 p_{UY|S}({\vc u}(w_{i-1},\hT),{\vc y}_{i-1}|
{\vc s}(w_{i-1}))2^{-n[R_0-r+\epsilon]}
\nonumber\\
&=&
    2^{-n[R_0-r+\epsilon]}
\nonumber\\
& & \times 
       \hspace*{-1mm}
       \sum_{\scs ({\lvc s}(w_{i-1}),{\lvc u}(w_{i-1},\hT),
        \atop {\scs{\lvc y}_{i-1})\in {\cal A}_{UY|S,\epsilon}}}
       \hspace*{-4mm}
p_{SUY}({\vc s}(w_{i-1}),{\vc u}(w_{i-1},\hT),{\vc y}_{i-1})
\nonumber\\
&\leq &2^{-n[R_0-r+\epsilon]}\,.
\label{eqn:d0f6}
\eeqa
Step (a) follows from the definition of 
${\cal A}_{UY|S,\epsilon}$. On upper bound of 
$$
{\sf E}\left[\left.\hat{e}_{{\rm 1b},i}^{(n)}(1|w_{i-1})\right|
       \phi_n(1)=\phi_{n}(2)=w_{i},\phi_n(2)=w_{i-1}
\right]\,,
$$
we have the following chain of inequalities:
\beqa
&&\hspace*{-4mm}
{\sf E}\left[\left.\hat{e}_{{\rm 1b},i}^{(n)}(1|w_{i-1})\right|
        \phi_n(1)=\phi_{n}(2)=w_{i},\phi_n(2)=w_{i-1}\right]
\nonumber\\
&=&  \sum_{\scs ({\lvc s}(w_{i-1}),{\lvc u}(w_{i-1},2),
      \atop {\scs {\lvc y}_{i-1}) \in {\cal A}_{UY|S,\epsilon}}
          }
p_S({\vc s}(w_{i-1}))
\nonumber\\
& &
\times
p_{U|S}({\vc u}(w_{i-1},2)|{\vc s}(w_{i-1}))
p_{Y|S}({\vc y}_i|{\vc s}(w_{i-1}))
\nonumber\\
&\MLeq{a}& 
         \sum_{\scs({\lvc s}(w_{i-1}),{\lvc u}(w_{i-1},2),
         \atop {\scs {\lvc y}_{i-1})\in {\cal A}_{UY|S,\epsilon}}
          }
p_S({\vc s}(w_{i-1}))
\nonumber\\
& &
\times
 p_{UY|S}({\vc u}(w_{i-1},2),{\vc y}_{i-1}|
{\vc s}(w_{i-1}))2^{-n[R_0-r+\epsilon]}
\nonumber\\
&=&2^{-n[R_0-r+\epsilon]}
\nonumber\\
& & \times 
       \hspace*{-1mm}
       \sum_{\scs ({\lvc s}(w_{i-1}),{\lvc u}(w_{i-1},2),
        \atop {\scs{\lvc y}_{i-1})\in {\cal A}_{UY|S,\epsilon}}}
       \hspace*{-4mm}
p_{SUY}({\vc s}(w_{i-1}),{\vc u}(w_{i-1},2),{\vc y}_{i-1})
\nonumber\\
&\leq &2^{-n[R_0-r+\epsilon]}\,.
\label{eqn:d0f6z}
\eeqa
Step (a) follows from the definition of 
${\cal A}_{UY|S,\epsilon}$. 
It follows  from (\ref{eqn:sqppzz1})-(\ref{eqn:d0f6z}) that when 
$(t_{i-1},t_{i-2})=(1,2)$, we have 
\beqa
& & 
{\sf E}\left[
\hat{e}_{{\rm 1b},i}^{(n)}(1|\phi_n(2))\right]
\leq 
\sum_{\hT\neq 1,2}
\sum_{(w_{i}, w_{i-1})\in{\cal W}_n^2 }
\frac{2^{-n[R_0-r+\epsilon]}}{|{\cal W}_n|^3}
\nonumber\\
& & +\sum_{w_{i}=w_{i-1}\in{\cal W}_n}
\frac{2^{-n[R_0-r+\epsilon]}}{|{\cal W}_n|^2}
\nonumber\\
& &=(2^{nR_0}-1) \frac{2^{-n[R_0-r+\epsilon]}}{|{\cal W}_n|}
\leq 2\cdot 2^{-n\epsilon}\,.
\label{eqn:sqpp5}
\eeqa
From 
(\ref{eqn:zoz}), 
(\ref{eqn:zasa2}),
(\ref{eqn:sqpp2}), and (\ref{eqn:sqpp5}), 
we have 
$$
{\sf E}\left[{e}_{{\rm 1b},i}^{(n)}\right]
\leq \Pr\left\{(S^n,U^n,Y^n)\notin {\cal A}_{UY|S,\epsilon} \right\} 
+2\cdot2^{-n\epsilon}\,.
$$
To derive the upper bound of 
${\sf E}\left[{e}_{i}^{(n)}\right]$ 
in Lemma \ref{lm:error}, set
\beqno
& &\tilde{e}_{i}^{(n)}
\defeq \Pr\{\tilde{\cal E}_{i}\}
\,,
\hat{e}_{i}^{(n)}
\defeq \Pr\{\hat{\cal E}_{i}\}\,,
\\
& &\tilde{e}_{i}^{(n)}(j_{i}|\phi_n(t_{i-1}),t_i, l_i)
\\ 
&\defeq &\Pr\{\tilde{\cal E}_{i}|
\ba[t]{l} T_{n,i}=t_{i}, T_{n,i-1}=t_{i-1},
          J_{n,i}=j_i, L_{n,i}=l_i\}\,,\ea
\\
& &\hat{e}_{i}^{(n)}(j_{i}|\phi_n(t_{i-1}),t_i,l_i) 
\\
&\defeq& \Pr\{\hat{\cal E}_{i}|
\ba[t]{l} T_{n,i}=t_{i}, T_{n,i-1}=t_{i-1},
          J_{n,i}=j_i, L_{n,i}=l_i\}\,.\ea
\eeqno
By definition of 
$\tilde{e}_{i}^{(n)}$ and 
$\hat{e}_{i}^{(n)}$, we have  
\beq
\left.
\ba{rcl}
{\sf E}\left[{e}_{i}^{(n)}\right]
&\leq& 
 {\sf E}\left[\hat{e}_{i}^{(n)}\right]
+{\sf E}\left[\tilde{e}_{i}^{(n)}\right]
\vSpa\\
{\sf E}\left[\tilde{e}_{i}^{(n)}\right]
&=&\ds \frac{1}{|{\cal T}_n|^2|{\cal J}_n||{\cal L}_n|}
\vSpa\\
& & \times 
\ds \sum_{\scs (t_i,t_{i-1},j_i,l_i)
          \atop{\scs \in{\cal T}_n^2
          \times {\cal J}_n\times {\cal L}_n}}
\hspace*{-2mm}
{\sf E}\left[\tilde{e}_{i}^{(n)}
     (j_i|\phi_n(t_{i-1}),t_i,l_i)\right]
\vSpa\\
{\sf E}\left[\hat{e}_{i}^{(n)}\right]
&=&\ds \frac{1}{|{\cal T}_n|^2|{\cal J}_n||{\cal L}_n|}
\vSpa\\
& & \times 
\ds \sum_{\scs (t_i,t_{i-1},j_i,l_i)
          \atop{\scs \in{\cal T}_n^2
          \times {\cal J}_n\times {\cal L}_n}}
\hspace*{-2mm}
{\sf E}\left[\hat{e}_{i}^{(n)}
     (j_i|\phi_n(t_{i-1}),t_i,l_i)\right].
\ea\right\}
\label{eqn:bsqz10} 
\eeq
By the symmetrical property of random coding 
it suffices to evaluate 
${\sf E}[\tilde{e}_{i}^{(n)}$ 
$(1|\phi_n($ $t_{i-1}),1,1)]$
and 
${\sf E}[\hat{e}_{i}^{(n)}$ 
$(1|\phi_n($ $t_{i-1}),1,1)]$.
In a manner quite similar 
to that of the derivation of the upper bound of 
${\sf E}[\tilde{e}_{2,i}^{(n)}$ 
$(1|\phi_n(t_{i-1})$ $,1,1)]$
and 
${\sf E}[\tilde{e}_{2,i}^{(n)}$ 
$(1|\phi_n(t_{i-1}))]$,
we obtain 
\beqno
& &{\sf E}[\tilde{e}_{i}^{(n)}(1|\phi_n(t_{i-1}),1,1)]
\\
&=&\Pr\left\{(S^n,U^n,X^n,Z^n)
\notin {\cal A}_{XZ|US,\epsilon} \right\}
\\
& &{\sf E}[\hat{e}_{i}^{(n)}(1|\phi_n(t_{i-1}),1,1)]
\leq 2^{-n\epsilon}\,.
\eeqno
Hence we have 
$$ 
{\sf E}\left[{e}_{i}^{(n)}\right]
\leq \Pr\left\{
(S^n,U^n,X^n,Z^n)
\notin {\cal A}_{XZ|US,\epsilon}\right\}+2^{-n\epsilon}\,.
$$
By an argument quite similar to that of the derivation of 
(\ref{eqn:sddf001}), we can prove the formulas of 
${\sf E}\left[{e}_{ZX|S,i}^{(n)}\right]$ and 
${\sf E}\left[{e}_{ZU|S,i}^{(n)}\right]$ stated in Lemma \ref{lm:error}. 
We omit the proofs. \hfill\QED

\subsection{Proofs of Lemmas \ref{lm:sec1} and \ref{lm:Gaulm} }

{\it Proof of Lemma \ref{lm:sec1}:} 
On a lower bound of $H(L_{n,i}|$$L_n^{(i-1)}$\\$Z^{nB})$, 
we have the following chain of inequalities: 
\beqa
& &H(L_{n,i}|L_n^{(i-1)}Z^{nB})
\nonumber\\
&\geq & H(L_{n,i}|L_n^{(i-1)}Z^{nB}{\Rw}_{n,i}T_{n,i})
\nonumber\\
&=& H(J_{n,i}L_{n,i}|L_n^{(i-1)}Z^{nB}{\Rw}_{n,i}T_{n,i})
\nonumber\\
& &-H(J_{n,i}|L_{n,i}L_n^{(i-1)}Z^{nB}{\Rw}_{n,i}T_{n,i})
\nonumber\\
&\geq& H(J_{n,i}L_{n,i}|L_n^{(i-1)}Z^{nB}{\Rw}_{n,i}T_{n,i})
\nonumber\\
& &-H(J_{n,i}|Z_{ni+1}^{n(i+1)}{\Rw}_{n,i}T_{n,i}L_{n,i})\Irr{.}
\label{eqn:zzapp0}
\eeqa
By Fano's inequality, we have   
\beq
\frac{1}{n}H(J_{n,i}|Z_{ni+1}^{n(i+1)}{\Rw}_{n,i}T_{n,i}L_{n,i}) 
\leq r_2{e}_i^{(n)}+\frac{1}{n}\,. 
\label{eqn:zzapp}
\eeq
From (\ref{eqn:zzapp0}) and (\ref{eqn:zzapp}), we have
\beqa
& &H(L_{n,i}|L_n^{(i-1)}Z^{nB})
\nonumber\\
&\geq& H(J_{n,i}L_{n,i}|L_n^{(i-1)}Z^{nB}{\Rw}_{n,i}T_{n,i})
-nr_2{e}_i^{(n)}-1\,.
\label{eqn:zzapp2}
\eeqa
On the first quantity in the right members of (\ref{eqn:zzapp2}), 
we have the following chain of inequalities:  
\beqa
& &H(J_{n,i}L_{n,i}|L_n^{(i-1)}Z^{nB}{\Rw}_{n,i}T_{n,i})
\nonumber\\
&=& H(J_{n,i}L_{n,i}|L_n^{(i-1)}Z_{[i]}^{nB}{\Rw}_{n,i}T_{n,i})
\nonumber\\
& &
   -I(Z_{n(i-1)+1}^{ni};J_{n,i}L_{n,i}|L_n^{(i-1)}Z_{[i]}^{nB}{\Rw}_{n,i}T_{n,i})
\nonumber\\
&=& H(J_{n,i}L_{n,i}|L_n^{(i-1)}Z_{[i]}^{nB}{\Rw}_{n,i}T_{n,i})
\nonumber\\
& &
+H(Z_{n(i-1)+1}^{ni}|
      Z_{[i]}^{nB}
     {\Rw}_{n,i}
         T_{n,i}
         J_{n,i}
         L_{n}^{(i)})
\nonumber\\
& &-H(Z_{n(i-1)+1}^{ni}|Z_{[i]}^{nB}{\Rw}_{n,i}T_{n,i}J_{n,i}L_n^{(i-1)})
\nonumber\\
&=& \log \left(\left|{\cal J}_{n}\right| \left|{\cal L}_{n}\right|\right)
+H(Z_{n(i-1)+1}^{ni}|
      Z_{[i]}^{nB}
{\Rw}_{n,i}
    T_{n,i}
    J_{n,i}
    L_{n}^{(i)})
\nonumber\\
&  &-H(Z_{n(i-1)+1}^{ni}|L_n^{(i-1)}Z_{[i]}^{nB}{\Rw}_{n,i}T_{n,i}J_{n,i})
\nonumber\\
&\geq& n(r_1+r_2)-2
+H(Z_{n(i-1)+1}^{ni}|Z_{[i]}^{nB}
{\Rw}_{n,i}
    T_{n,i}
    J_{n,i}
    L_{n}^{(i)})
\nonumber\\ 
& &
-H(Z_{n(i-1)+1}^{ni}|{\Rw}_{n,i}T_{n,i})
\nonumber\\
&\MEq{a}& n(r_1+r_2)
+H(Z_{n(i-1)+1}^{ni}|
{\Rw}_{n,i}
    T_{n,i}
    J_{n,i}
    L_{n,i})
\nonumber\\ 
& &
-H(Z_{n(i-1)+1}^{ni}|{\Rw}_{n,i}T_{n,i})-2\,.
\label{eqn:za}
\eeqa
Equality \mbox{\rm (a)} follows from the following 
Markov chain: 
$$
Z_{n(i-1)+1}^{ni}
\to {\Rw}_{n,i}T_{n,i}J_{n,i}L_{n,i}\to 
Z_{[i]}^{nB} L_{n}^{(i-1)}\,.
$$
To derive a lower bound of 
$H(Z_{n(i-1)+1}^{ni}|{\Rw}_{n,i}T_{n,i}J_{n,i},L_{n,i})$, 
set  
\beqno
{\cal B}_1^{*}
&\defeq &
\ba[t]{l}\left\{( w,t,j,l,{\vc z}): \right.\\
\quad \left. ({\vc s}(w),{\vc x}(w,t,j,l),{\vc z}) 
\in {\cal B}_{Z|XS,\epsilon}\right\}\,.
\ea
\eeqno
By definition of ${\cal B}_1^*$, if $(w,t,j,l,{\vc z})$
$\in {\cal B}_1^*$, we have 
\beqno
  -\frac{1}{n}\log p_{Z|XS}
  ({\vc z}|{\vc x}(w,t,j,l),{\vc s}(w))
&\geq &H(Z|XS)-\epsilon\,. 
\eeqno
By definition of $e_{Z|XS,i}^{(n)}$, we have
$$
\Pr\{({\Rw}_{n,i},T_{n,i},J_{n,i},L_{n,i}, 
      Z_{n(i-1)+1}^{ni})\notin {\cal B}_1^{*} \} 
=e_{Z|XS,i}^{(n)}\,. 
$$ 
Then, we have  
\beqa
&     &  H(Z_{n(i-1)+1}^{ni}|{\Rw}_{n,i}T_{n,i}J_{n,i}L_{n,i})
\nonumber\\
&\geq & n[H(Z|XS)-\epsilon]
\nonumber\\
& &\times 
\Pr\{({\Rw}_{n,i},T_{n,i},J_{n,i},L_{n,i}, 
      Z_{n(i-1)+1}^{ni})\in {\cal B}_1^{*}\} 
\nonumber\\
&\geq & n[H(Z|XS)-\epsilon](1-e_{Z|XS,i}^{(n)})
\nonumber\\
&\geq & n[H(Z|XS)-\epsilon]-nH(Z|XS)e_{Z|XS,i}^{(n)}\,.
\label{eqn:aa0}
\eeqa
To derive an upper bound of 
$H(Z_{n(i-1)+1}^{ni}|{\Rw}_{n,i}T_{n,i})$, set  
\beqno
{\cal B}_2^{*}
\defeq \left\{
(w,t,{\vc z}): ({\vc s}(w),{\vc u}(w,t),{\vc z}) 
\in {\cal B}_{Z|US,\epsilon}
\right\}\,.
\eeqno
By definition of ${\cal B}_2^*$, if 
$(w,t,{\vc z})$
$\in {\cal B}_2^*$, we have      
\beqno
-\frac{1}{n}
   \log p_{Z|US}
({\vc z}|{\vc u}(w,t),{\vc s}(w))
&\leq& H(Z|US)+ \epsilon\,. 
\eeqno
By definition of $e_{Z|US,i}^{(n)}$, 
we have
$$
\Pr\{({\Rw}_{n,i},T_{n,i}, Z_{n(i-1)+1}^{ni})
        \notin {\cal B}_2^{*} \}
=e_{Z|US,i}^{(n)}\,.
$$
Set 
\beqno
{\cal D}\defeq
\left\{(w,t):
(w,t,{\vc z})\in ({\cal B}_2^{*})^{c}\mbox{ for some }{\vc z}
\right\}
\eeqno
and for $(w,t)\in {\cal D}$, set
\beqno
{\cal D}(w,t)\defeq
\left\{{\vc z}:
(w,t,{\vc z})\in ({\cal B}_2^{*})^{c}
\right\}\,.
\eeqno
Then, we have  
\beqa
&     &  
H(Z_{n(i-1)+1}^{ni}|{\Rw}_{n,i}T_{n,i})
\nonumber\\
&\leq & n[H(Z|US)+\epsilon]
-\sum_{(w,t)\in{\cal D}}\sum_{{\lvc z}\in {\cal D}(w,t)}
p_{Z^nW_nT_n}({\vc z},w,t) 
\nonumber\\
& &\times \log p_{Z^n|W_nT_n}({\vc z}|w,t)\,. 
\label{eqn:aabbb20}
\eeqa
We derive an upper bound of the second term in the 
right member of (\ref{eqn:aabbb20}). Let $\bar{Z}$ be  
a random variable uniformly distributed on ${\cal Z}$.
Let $\bar{Z}^n=(\bar{Z}_1,\bar{Z}_2,$ $\cdots, \bar{Z}_n)$ 
be $n$ independent copies of $\bar{Z}$. 
We assume that $\bar{Z}^n$ is independent of 
$W_n$ and $T_n$. 
We first observe that
\beqa
&&-\sum_{(w,t)\in {\cal D}}
\sum_{{\lvc z} \in {\cal D}(w,t)}
\hspace*{-3mm}p_{Z^nW_nT_n}({\vc z},w,t)
\log 
\frac{p_{Z^n|W_nT_n}({\vc z}|w,t)}
{p_{\bar{Z}^n}({\vc z})} 
\nonumber\\
&=&\sum_{(w,t)\in {\cal D}}\sum_{{\lvc z} \in {\cal D}(w,t)}
\hspace*{-3mm}p_{Z^nW_nT_n}({\vc z},w,t)
\log 
\frac{p_{\bar{Z}^n}({\vc z})}
{p_{Z^n|W_nT_n}({\vc z}|w,t)}
\nonumber\\
&\MLeq{a}&
\left(\log {\rm e}\right)\cdot\sum_{(w,t)\in {\cal D}}
\sum_{{\lvc z} \in {\cal D}(w,t)}\hspace*{-3mm}
p_{Z^nW_nT_n}({\vc z},w,t)
\nonumber\\
& & \qquad
\times
\left[
\frac{p_{\bar{Z}^n}({\vc z})}
{p_{Z^n|W_nT_n}({\vc z}|w,t)}-1
\right]
\nonumber\\
&=&\left(\log {\rm e}\right)\cdot
\sum_{(w,t)\in {\cal D}}\sum_{{\lvc z} \in {\cal D}(w,t)}
\hspace*{-1mm}
\left[
p_{\bar{Z}^n}({\vc z}){p_{W_nT_n}(w,t)}
\right.
\nonumber\\
& &\qquad \left. 
-{p_{Z^nW_nT_n}({\vc z},w,t)}
\right]
\nonumber\\
&=&
\left(\log {\rm e}\right)\cdot
\left[
 p_{\bar{Z}^nW_nT_n}\left({\cal B}_2^{*}\right)
-p_{{Z}^nW_nT_n}\left({\cal B}_2^{*}\right)
\right]
\leq \log {\rm e}\,.
\label{eqn:za0q0}
\eeqa
Step (a) follows from the inequality 
$\log a \leq (\log {\rm e})(a-1)$.  
From (\ref{eqn:za0q0}), we have
\beqa
& &-\sum_{(w,t)\in {\cal D}}
\sum_{{\lvc z}\in {\cal D}(w,t)}\hspace*{-3mm}
p_{Z^nW_nT_n}({\vc z},w,t)
\log p_{Z^n|W_nT_n}({\vc z}|w,t) 
\nonumber\\
&\leq&-\sum_{(w,t)\in {\cal D}}\sum_{{\lvc z}\in{\cal D}(w,t)}
\hspace*{-3mm}p_{Z^nW_nT_n}({\vc z},w,t)
\log p_{\bar{Z}^n}({\vc z}) 
+\log{\rm e}
\nonumber\\
&=& n\sum_{(w,t)\in {\cal D}}
\sum_{{\lvc z}\in {\cal D}(w,t)}\hspace*{-3mm}
    p_{Z^nW_nT_n}({\vc z},w,t)\log |{\cal Z}|
   +\log{\rm e}
\nonumber\\
&=& ne_{Z|US,i}^{(n)}\log |{\cal Z}|+\log{\rm e}\,.
\label{eqn:xcc220}
\eeqa
Combining (\ref{eqn:zzapp2})-(\ref{eqn:aabbb20}) 
and (\ref{eqn:xcc220}), we have 
\beqa
&      & \frac{1}{n}H(L_{n,i}|L_n^{(i-1)}Z^{nB})
\nonumber\\
& \geq &  r_1+r_2-I(X;Z|US)-2\epsilon-\frac{3+\log {\rm e}}{n}
\nonumber\\
&      &  -\left[(\log |{\cal Z}|)e_{Z|US,i}^{(n)}
          +H(Z|XS)e_{Z|XS,i}^{(n)}\right]\,,
\nonumber\\
& \geq &  r_1+r_2-I(X;Z|US)-2\epsilon-\frac{3+\log {\rm e}}{n}
\nonumber\\
&      &  -r_2{e}_i^{(n)}-(\log |{\cal Z}|)
          \left[e_{Z|US,i}^{(n)}+e_{Z|XS,i}^{(n)}\right]\,,
\nonumber
\eeqa
completing the proof.\hfill \QED

{\it Proof of Lemma \ref{lm:Gaulm}:}
In a manner quite similar to the derivation of 
(\ref{eqn:zzapp2}) and (\ref{eqn:za}) 
in the proof of Lemma \ref{lm:sec1}, we have 
\beqa
& &H(L_{n,i}|L_n^{(i-1)}Z^{nB})
\nonumber\\
&\geq& H(J_{n,i}L_{n,i}|L_n^{(i-1)}Z^{nB}{\Rw}_{n,i}T_{n,i})
       -nr_2{e}_i^{(n)}-1,\quad
\label{eqn:za002}\\
& &H(J_{n,i}L_{n,i}|L_n^{(i-1)}Z^{nB}{\Rw}_{n,i}T_{n,i})
\nonumber\\
& \geq & n(r_1+r_2)
+h(Z_{n(i-1)+1}^{ni}|
{\Rw}_{n,i}
    T_{n,i}
    J_{n,i}
    L_{n,i})
\nonumber\\ 
& &
-h(Z_{n(i-1)+1}^{ni}|{\Rw}_{n,i}T_{n,i})-2.
\label{eqn:za2}
\eeqa
On a lower bound of 
$h(Z_{n(i-1)+1}^{ni}|{\Rw}_{n,i}T_{n,i}J_{n,i}L_{n,i})$, we have  
\beqa
&     &  h(Z_{n(i-1)+1}^{ni}|{\Rw}_{n,i}T_{n,i}J_{n,i}L_{n,i})
\nonumber\\
&\geq & n[h(Z|XS)-\epsilon]
\nonumber\\
& &\times 
\Pr\{({\Rw}_{n,i},T_{n,i},J_{n,i},L_{n,i}, 
      Z_{n(i-1)+1}^{ni})\in {\cal B}_1^{*}\} 
\nonumber\\
&\geq & n[h(Z|XS)-\epsilon](1-e_{Z|XS,i}^{(n)})
\nonumber\\
&\geq & n[h(Z|XS)-\epsilon]-nh(Z|XS)e_{Z|XS,i}^{(n)}
\nonumber\\
&=    & n[h(Z|XS)-\epsilon]-n
\left\{{\ts \frac{1}{2}\log(2\pi{\rm e}N_2)}\right\}
e_{Z|XS,i}^{(n)}\,.
\label{eqn:aa02}
\eeqa
Next, we derive an upper bound of 
$h(Z_{n(i-1)+1}^{ni}|{\Rw}_{n,i}T_{n,i})$.
By definition of ${\cal B}_2^*$, if 
$(w,t,{\vc z})$
$\in {\cal B}_2^*$, we have      
\beqno
-\frac{1}{n}
   \log p_{Z|US}
({\vc z}|{\vc u}(w,t),{\vc s}(w))
&\leq& h(Z|US)+ \epsilon\,. 
\eeqno
Then we have  
\beqa
&     &  
h(Z_{n(i-1)+1}^{ni}|{\Rw}_{n,i}T_{n,i})
\nonumber\\
&\leq & n[h(Z|US)+\epsilon]
-\sum_{(w,t)\in{\cal D}}\int_{{\cal D}(w,t)}\hspace*{-3mm}
p_{Z^nW_nT_n}({\vc z},w,t) 
\nonumber\\
& &\times \log p_{Z^n|W_nT_n}({\vc z}|w,t) {\rm d}{\vc z}\,.
\label{eqn:aabbb2}
\eeqa
We derive an upper bound of the second term in the 
right member of (\ref{eqn:aabbb2}). Let $\bar{Z}$ be  
a random variable whose density function denoted 
by $p_{\bar{Z}}(z)$ is      
$$
p_{\bar{Z}}(z)=\frac{1}{2}{\rm e}^{-|z|}\,. 
$$
Let $\bar{Z}^n=(\bar{Z}_1,\bar{Z}_2,$ $\cdots, \bar{Z}_n)$ 
be $n$ independent copies of $\bar{Z}$. 
We assume that $\bar{Z}^n$ is independent of 
$W_n$ and $T_n$. For 
${\vc z}\defeq (z_1,$ $z_2,\cdots,z_n)$,       
the density function $p_{\bar{Z}^n}({\vc z})$ of $\bar{Z}^n$ is 
$$
p_{\bar{Z}^n}({\vc z})
=\left(\frac{1}{2}\right)^n
\prod_{i=1}^n{\rm e}^{-|z_i|}\,. 
$$
In a manner quite similar to the derivation 
of (\ref{eqn:za0q0}) in the proof of Lemma \ref{lm:sec1}, 
we have 
\beqa
&&-\sum_{(w,t)\in {\cal D}}
\int_{{\cal D}(w,t)}\hspace*{-4mm}
p_{Z^nW_nT_n}({\vc z},w,t)
\log 
\frac{p_{Z^n|W_nT_n}({\vc z}|w,t)}
{p_{\bar{Z}^n}({\vc z})} {\rm d}{\vc z}
\nonumber\\
& &\leq \log {\rm e}\,.
\label{eqn:za0q}
\eeqa
From (\ref{eqn:za0q}), we have
\beqa
& &-\sum_{(w,t)\in {\cal D}}\int_{{\cal D}(w,t)}\hspace*{-4mm}
p_{Z^nW_nT_n}({\vc z},w,t)
\log p_{Z^n|W_nT_n}({\vc z}|w,t) {\rm d}{\vc z}
\nonumber\\
&\leq&-\sum_{(w,t)\in {\cal D}}\int_{{\cal D}(w,t)}\hspace*{-4mm}
p_{Z^nW_nT_n}({\vc z},w,t)
\log p_{\bar{Z}^n}({\vc z}) {\rm d}{\vc z}
+\log{\rm e}
\nonumber\\
&=& {n}
\left\{\sum_{(w,t)\in {\cal D}} 
\int_{{\cal D}(w,t)}\hspace*{-4mm}
p_{Z^nW_nT_n}({\vc z},w,t){\rm d}{\vc z}
\right\}+\log{\rm e}
\nonumber\\
& & +\sum_{(w,t)\in {\cal D}}
\int_{{\cal D}(w,t)}\hspace*{-4mm}
p_{Z^nW_nT_n}({\vc z},w,t)
     \left\{\sum_{i=1}^n|z_i|\right\}{\rm d}{\vc z}\,.
\label{eqn:za0qz}
\eeqa
On the last term in (\ref{eqn:za0qz}), 
we have the following chain of inequalities:     
\beqa 
& &
\sum_{(w,t)\in {\cal D}}
\int_{{\cal D}(w,t)}\hspace*{-4mm}
p_{Z^nW_nT_n}({\vc z},w,t)
     \left\{\sum_{i=1}^n|z_i|\right\}{\rm d}{\vc z}
\nonumber\\
&\MLeq{a}& 
\sum_{(w,t)\in {\cal D}}
\left\{ 
\int_{{\cal D}(w,t)}\hspace*{-4mm}
p_{Z^nW_nT_n}({\vc z},w,t){\rm d}{\vc z}
\right\}^{\frac{1}{2}}
\nonumber\\
& &\times
\left\{ 
\int_{{\cal D}(w,t)}\hspace*{-4mm}
p_{Z^nW_nT_n}({\vc z},w,t)
\left\{\sum_{i=1}^n|z_i|\right\}^2{\rm d}{\vc z}
\right\}^{\frac{1}{2}}
\nonumber\\
&\MLeq{b}& 
\left\{
\sum_{(w,t)\in {\cal D}}
\int_{{\cal D}(w,t)}\hspace*{-4mm}
p_{Z^nW_nT_n}({\vc z},w,t)
     {\rm d}{\vc z}
\right\}^{\frac 1 2}
\nonumber\\
& &  \times
\left\{\sum_{(w,t)\in {\cal D}} 
\int_{{\cal D}(w,t)}\hspace*{-4mm}
p_{Z^nW_nT_n}({\vc z},w,t)
\left\{\sum_{i=1}^n|z_i|\right\}^2{\rm d}{\vc z}
\right\}^{\frac{1}{2}}
\nonumber\\
&\leq& 
\sqrt{e_{Z|US,i}^{(n)}}
\left\{ 
\int p_{Z^n}({\vc z})
\left\{\sum_{i=1}^n|z_i|\right\}^2{\rm d}{\vc z}
\right\}^{\frac{1}{2}}
\nonumber\\
&\MLeq{c}&\sqrt{e_{Z|US,i}^{(n)}}
\left\{
n\int p_{Z^n}({\vc z})
\left\{\sum_{i=1}^n|z_i|^2\right\}{\rm d}{\vc z}
\right\}^{\frac{1}{2}}
\nonumber\\
&=& \sqrt{e_{Z|US,i}^{(n)}}
\left\{n
\sum_{i=1}^n
\int z_i^2 p_{Z_i}({z}_i){\rm d}{z}_i
\right\}^{\frac{1}{2}}\,.
\label{eqn:xcc22}
\eeqa
Steps (a)-(c) follow from the Cauchy-Schwarz inequality.
On the other hand, we have
\beqa
\sum_{i=1}^n
\int z_i^2 p_{Z_i}({z}_i){\rm d}{z}_i
&=&\sum_{i=1}^n{\rm \bf E}
\left[\left|X_i+\xi_{2,i}\right|^2\right] 
\nonumber\\
&=&\sum_{i=1}^n
{\rm \bf E}\left[\left|X_i\right|^2\right]
+\sum_{i=1}^n
{\rm \bf E}\left[\left|\xi_{2,i}\right|^2\right] 
\nonumber\\
& \leq & n(P_1+N_2)\,.
\label{eqn:zaapp}
\eeqa
Combining (\ref{eqn:za002})-(\ref{eqn:aabbb2}) and 
(\ref{eqn:za0qz})-(\ref{eqn:zaapp}), we have 
\beqa
&      & \frac{1}{n}H(L_{n,i}|L_n^{(i-1)}Z^{nB})
\nonumber\\
& \geq &  r_1+r_2-I(X;Z|US)-2\epsilon-\frac{3+\log{\rm e}}{n}
\nonumber\\
&      & -r_2{e}_i^{(n)}
        -\left[e_{Z|US,i}^{(n)}+\sqrt{(P_1+N_2)e_{Z|US,i}^{(n)}}\right]
\nonumber\\
&      & -\left\{{\ts \frac{1}{2}\log(2\pi{\rm e}N_2)}\right\}
         e_{Z|XS,i}^{(n)}\,,
\nonumber
\eeqa
completing the proof.\hfill \QED

\subsection{
Proof of Lemma \ref{lm:conv1}
}

{\it Proof of Lemma \ref{lm:conv1}:} 
We first observe that we have the following 
chains of inequalities:  
\beqa
\log \left|{\cal M}_n\right|&=&H(M_n)
\nonumber\\
&=&I(M_n;Y^n) + H(M_n|Y^n)
\label{eqn:cvvv1}\\
&=&I(M_n;Z^n) + H(M_n|Z^n)\,,
\label{eqn:cvvv2}\\
\log \left|{\cal K}_n\right|&=&H(K_n)=H(K_n|M_n)
\nonumber\\
&=&I(K_n;Y^n|M_n)+ H(K_n|Y^nM_n)\,,
\label{eqn:cvvv3}\\
H(K_n|Z^n)&=&H(K_n|Z^nM_n)+I(K_n;M_n|Z^n)
\nonumber\\
          &=&H(K_n|M_n)-I(K_n;Z^n|M_n)
\nonumber\\
& &+I(K_n;M_n|Z^n)
\nonumber\\
          &=&I(K_n;Y^n|M_n)-I(K_n;Z^n|M_n)
\nonumber\\
& &       +H(K_n|Y^nM_n)+I(K_n;M_n|Z^n)
\nonumber\\
         &\leq& I(K_n;Y^n|M_n)-I(K_n;Z^n|M_n)
\nonumber\\
         & & +H(K_n|Y^nM_n)+H(M_n|Z^n)\,,
\label{eqn:cvvv4}
\\
 &\leq& \log |{\cal K}_n|-I(K_n;Z^n|M_n)
\nonumber\\
         & & +H(K_n|Y^nM_n)+H(M_n|Z^n)\,.
\label{eqn:cvvv4b}
\eeqa
Here, we suppose that $(R_0,R_1,R_{\rm e})\in {\cal R}_{\rm s}^*(\Ch)$. 
Set $\lambda^{(n)}\defeq$ $ 
\max\{\lambda_1^{(n)},$ $\lambda_2^{(n)}\}$. 
Then, by Fano's inequality we have
\beq
\left.
\ba{rcl}
H(M_n|Y^n)&\leq &\log \left|{\cal M}_n\right|\lambda^{(n)}+1
\\
H(M_n|Z^n)&\leq &\log \left|{\cal M}_n\right|\lambda^{(n)}+1
\\
H(K_n|Y^nM_n)&\leq &\log \left|{\cal K}_n\right|
                  \lambda^{(n)}+1\,.
\ea
\right\}
\label{eqn:convv2}
\eeq
Set 
$$
\left.
\ba{rcl}
\tau_{1,n}&\defeq& 
\frac{1}{n}\log \left|{\cal M}_n \right|\lambda^{(n)}+ \frac{1}{n}
\vSpa\\
\tau_{2,n}&\defeq& 
\frac{1}{n}\log \left|{\cal K}_n \right|\lambda^{(n)}+\frac{1}{n}\,.
\ea
\right\}
$$
\Irr{From (\ref{eqn:cvvv1})-(\ref{eqn:convv2})}, we have
\Irr{
\beq
\hspace*{-2mm}
\left.
\ba{rcl}
\frac{1}{n}\log \left|{\cal M}_n\right| 
&\leq &\frac{1}{n}\min\{I(M_n;Y^n),I(M_n;Z^n)\}+\tau_{1,n} 
\vSpa\\
\frac{1}{n}\log \left|{\cal K}_n\right| &\leq & 
\frac{1}{n}I(K_n;Y^n|M_n)+\tau_{2,n}  
\vSpa\\
\frac{1}{n}H(K_n|Z^n) 
 &\leq &\frac{1}{n}\log\left|{\cal K}_n\right|
           -\frac{1}{n}I(K_n;Z^n|M_n)
\vSpa\\
 &    &+{\tau}_{1,n}+{\tau}_{2,n} 
\vSpa\\
\frac{1}{n}H(K_n|Z^n) &\leq&
\frac{1}{n}I(K_n;Y^n|M_n)-\frac{1}{n}I(K_n;Z^n|M_n)
\vSpa\\
 &    &+{\tau}_{1,n}+{\tau}_{2,n}. 
\vSpa\\
\ea
\right\}
\label{eqn:convv1d}
\eeq
}
Set
\beq
\left.
\ba{rcl}
\delta_{1,n}
&\defeq&\tau_{1,n}
+\left[R_0-\frac{1}{n}\log \left|{\cal M}_n\right|\right]^{+}
\\
\delta_{2,n}
&\defeq&\tau_{2,n}
+\left[R_1-\frac{1}{n}\log \left|{\cal K}_n\right|\right]^{+}
\\
\delta_{3,n}&\defeq&\tau_{1,n}+ \tau_{2,n}
+\left[R_{\rm e}-\frac{1}{n}H(K_n|Z^n)\right]^{+}
\\
& &+\left[\frac{1}{n}\log \left|{\cal K}_n \right|-R_1\right]^{+}
\\
\delta_{4,n}&\defeq& \tau_{1,n}+ \tau_{2,n}
          +\left[R_{\rm e}-\frac{1}{n}H(K_n|Z^n)\right]^{+}\,.
\ea
\right\}
\label{eqn:convv1f}
\eeq
It is obvious that when $(R_0,R_1,$ $R_{\rm e})$ $\in {\cal R}_{\rm s}^{\Irr{*}}(\Ch)$, 
the above $\delta_{i,n},$ $i=1,2,3,4$ tend to zero as $n\to\infty$. 
From  (\ref{eqn:convv1d}) and (\ref{eqn:convv1f}), 
we have (\ref{eqn:convv1}) for 
$(R_0,R_1,$ $R_{\rm e})$ $\in {\cal R}_{\rm s}^{\Irr{*}}(\Ch)$. 
\hfill\QED

\subsection{
Proof of Lemma \ref{lm:conv2}
}

{\it Proof of Lemma \ref{lm:conv2}:} 
We first prove (\ref{eqn:cv1}) and (\ref{eqn:cv2}). 
We have the following chains of inequalities:
\beqa
& &I(M_n;Y^n)=H(Y^n)-H(Y^n|M_n)
\nonumber\\
&=&\sum_{i=1}^n
\left\{H(Y_i|Y^{i-1})-H(Y_i|Y^{i-1}M_n)\right\}
\nonumber\\
&\leq&
\sum_{i=1}^n
   \left\{H(Y_i)-H(Y_i|Y^{i-1}Z^{i-1}S_{i}M_n)
   \right\}
\nonumber\\
&=&\sum_{i=1}^n I(U_iS_i;Y_i)\,,
\nonumber\\
& &
I(M_n;Z^n)=H(M_n)-H(M_n|Z^{n})
\nonumber\\
&=&\sum_{i=1}^n\left\{ H(M_n|Z^{i-1})-H(M_n|Z^{i}) \right\}
\nonumber\\
&\MEq{a}&
\sum_{i=1}^n
\left\{H(M_n|Z^{i-1}S_i)-H(M_n|Z^i)\right\}
\nonumber\\
&\leq &
\sum_{i=1}^n
\left\{H(M_n|Z^{i-1}S_i)-H(M_n|Z^iS_i)\right\}
\nonumber\\
&=&\sum_{i=1}^n I(M_n;Z_i|Z^{i-1}S_i)
\nonumber\\
&=&
\sum_{i=1}^n
\left\{H(Z_i|Z^{i-1}S_i)-H(Z_i|Z^{i-1}S_iM_n)\right\}
\label{eqn:Conv}\\
&\leq&
\sum_{i=1}^n
\left\{H(Z_i|S_i)-H(Z_i|Y^{i-1}Z^{i-1}S_iM_n)\right\}
\nonumber\\
&=&\sum_{i=1}^n I(U_i;Z_i|S_i)\,.
\nonumber
\eeqa
Step $\mbox{\rm (a)}$ follows from 
$S_i\to M_n \to  Z^{i-1}$. 
Next, we prove (\ref{eqn:cv2xyz}). 
We have the following chain of inequalities:
\beqno
& &I(K_nM_n;Y^n) \MLeq{a}I(X^n;Y^n)
=\sum_{i=1}^nI(Y_i;X^n|Y^{i-1})
\nonumber\\
&=&\sum_{i=1}^n
\left\{H(Y_i|Y^{i-1})-H(Y_i|Y^{i-1}X^n)\right\}
\nonumber\\
&\leq&\sum_{i=1}^n
\left\{H(Y_i)-H(Y_i|Y^{i-1}X^nS_i)\right\}
\nonumber\\
&\MEq{b}&
\sum_{i=1}^n
\left\{H(Y_i)-H(Y_i|X_iS_i)\right\}
=\sum_{i=1}^nI(X_iS_i;Y_i)
\eeqno
Step ${\mbox{\rm (a)}}$ follows from 
$Y^n\to X^n\to K_nM_n$. 
Step ${\mbox{\rm (b)}}$ follows from
$Y_i\to X_iS_i \to Y^{i-1}X_{[i]}\,.$
Thirdly, we prove (\ref{eqn:cv3}). 
We have the following chain of inequalities:
\beqa
& &I(K_n;Y^n|M_n)\leq I(K_n;Y^nZ^n|M_n)
\nonumber\\
&=&I(K_nM_n;Y^nZ^n|M_n)\MLeq{a}I(X^n;Y^nZ^n|M_n)
\nonumber\\
&=&H(X^n|M_n)-H(X^n|Y^nZ^nM_n)
\nonumber\\
&=&\sum_{i=1}^n
\left\{H(X^n|Y^{i-1}Z^{i-1}M_n)
-H(X^n|Y^{i}Z^{i}M_n)\right\}
\nonumber\\
&{\MEq{b}}&
\sum_{i=1}^n 
\left\{H(X^n|Y^{i-1}Z^{i-1}M_nS_i)-H(X^n|Y^iZ^iM_n)\right\}
\nonumber\\
&\leq &
\sum_{i=1}^n 
\left\{H(X^n|Y^{i-1}Z^{i-1}M_nS_i)-H(X^n|Y^iZ^iM_nS_i)\right\}
\nonumber\\
&=&\sum_{i=1}^nI(X^n;Y_iZ_i|U_iS_i)
\nonumber\\
&=&\sum_{i=1}^n\left\{
H(Y_iZ_i|U_iS_i)-H(Y_iZ_i|U_iS_iX^n)
\right\}
\nonumber\\
&\MEq{c}&\sum_{i=1}^n \left\{H(Y_iZ_i|U_iS_i)
-H(Y_iZ_i|X_iS_i)\right\}
\nonumber\\
&=&\sum_{i=1}^n I(X_i;Y_iZ_i|U_iS_i){\,.}
\nonumber
\eeqa
Step {\rm (a)} follows from the Markov chain 
$Y^nZ^n\to X^n \to K_nM_n$. 
Step {\rm (b)} follows 
from $S_i\to Z^{i-1} $ $\to X^nY^{i-1}M_n$.
Step {\rm (c)} follows from 
$Y_iZ_i\to X_iS_i \to U_iX_{[i]}\,.$
Fourthly, we prove (\ref{eqn:cv4}).
We have the following chain of inequalities:
\beqa
& &I(K_n;Y^n|M_n)-I(K_n;Z^n|M_n)
\nonumber\\
&\leq&I(K_n;Y^nZ^n|M_n)-I(K_n;Z^n|M_n)
\nonumber\\
&=&I(K_n;Y^n|Z^nM_n)= I(K_nM_n;Y^n|Z^nM_n) 
\nonumber\\
&\MLeq{a}&I(X^n;Y^n|Z_nM_n)
\nonumber\\
&=&H(Y^n|Z^nM_n)-H(Y^n|Z^nX^nK_nM_n)
\nonumber\\
&=&\sum_{i=1}^n
\left\{H(Y_i|Y^{i-1}Z^{n}M_n)-H(Y_i|Y^{i-1}Z^{n}X^n)\right\}
\nonumber\\
&\leq&\sum_{i=1}^n
\left\{H(Y_i|Y^{i-1}Z^{i}M_n)-H(Y_i|Y^{i-1}Z^{n}S_iX^n)\right\}
\nonumber\\
&\MEq{b}&\sum_{i=1}^n
\left\{H(Y_i|Y^{i-1}Z^{i}S_iM_n)-H(Y_i|Y^{i-1}Z^{n}S_iX^n)\right\}
\nonumber\\
&=&\sum_{i=1}^n
\left\{H(Y_i|U_iS_iZ_i)-H(Y_i|U_iS_iZ^{n}S_iX^n)\right\}
\nonumber\\
&\MEq{c}&\sum_{i=1}^n
\left\{H(Y_i|U_iS_iZ_i)-H(Y_i|U_iS_iZ_iX_i)\right\}
\nonumber\\
&=& \sum_{i=1}^n I(X_i;Y_i|Z_iU_iS_i)\,.
\nonumber
\eeqa
Step {\rm (a)} follows from the Markov chain 
$Y^nZ^n\to X^n \to K_nM_n$. 
Step $\mbox{\rm (b)}$ follows from that 
$S_i=g_i(Z^{i-1})$ is a function of $Z^{i-1}$ 
in the case where $\{g_i\}_{i=1}^n$ is restricted 
to be deterministic. In the case where $\{g_i\}_{i=1}^n$ 
is allowed to be stochastic, if $\Ch$ belongs to the
\Cls, we have the following Markov chain: 
\beq
S_i\to Z^{i-1}\to Y^iZ^iK_nM_n\,.
\eeq  
Step {\rm (b)} follows from the above Markov chain. 
Step {\rm (c)} follows from 
$Y_i\to Z_iX_iS_i $ 
$\to Y^{i-1}Z_{[i]}X_{[i]}\,.$
Finally, we prove (\ref{eqn:cv5}).   
We have the following chain of inequalities:
\beqa
& &I(K_n;Z^n|M_n)= H(Z^n|M_n)-H(Z^n|K_nM_n)
\nonumber\\
&\MEq{a} &H(Z^n|M_n)-H(Z^n|X^n)
\nonumber\\
&=&\sum_{i=1}^n
\left\{H(Z_i|Z^{i-1}M_n)-H(Z_i|Z^{i-1}X^n)\right\}
\nonumber\\
&\MEq{b}&
\sum_{i=1}^n\left\{H(Z_i|Z^{i-1}S_iM_n)-H(Z_i|Z^{i-1}S_iX^n)\right\}
\nonumber\\
&\geq&\sum_{i=1}^n\left\{H(Z_i|U_iS_i)-H(Z_i|X_iS_i)\right\}
\nonumber\\
&\MEq{c}&\sum_{i=1}^n
\left\{H(Z_i|U_iS_i)-H(Z_i|X_iS_iU_i)\right\}
\nonumber\\
&=&   \sum_{i=1}^n I(X_i;Z_i|U_iS_i)\,.
\nonumber
\eeqa
Step {\rm (a)} follows from that 
$f_n$ is a one-to-one mapping. 
%
Step $\mbox{\rm (b)}$ follows from that 
$S_i=g_i(Z^{i-1})$ is a function of $Z^{i-1}$ 
in the case where $\{g_i\}_{i=1}^n$ is restricted 
to be deterministic. In the case where $\{g_i\}_{i=1}^n$ 
is allowed to be stochastic, if $\Ch$ belongs to the
\Cls, we have the following Markov chain: 
\beq
S_i\to Z^{i-1}\to Z_iM_nX^n\,.
\eeq  
Step {\rm (b)} follows from the above Markov chain. 
Step {\rm (c)} follows from 
$Z_i\to X_iS_i $$\to U_i\,.$
Thus, the proof of Lemma \ref{lm:conv2} is completed.
\hfill\QED

\subsection{
Proofs of Lemmas \ref{lm:conv3a} and \ref{lm:conv3b} 
}

In this appendix we prove Lemmas \ref{lm:conv3a} 
and \ref{lm:conv3b}.
We first present a lemma necessary to prove those lemmas.
\begin{lm}
\label{lm:conv3d}
\beqa
I(M_n;Y^n)&\leq& \sum_{i=1}^nI(Y_{i+1}^nZ^{i-1}S_iM_n;Y_i)\,,
\label{eqn:cvk1a}\\ 
I(M_n;Z^n)&\leq& \sum_{i=1}^nI(Y_{i+1}^nZ^{i-1}M_n;Z_i|S_i)\,,
\label{eqn:cvk2a}\\
I(K_nM_n;Y^n)
&\leq &\sum_{i=1}^n I(Y_{i+1}^nZ^{i-1}S_iK_nM_n;Y_i)\,,
\label{eqn:cvk3}\\
I(K_nM_n;Z^n)
&\leq&\sum_{i=1}^n I(Y_{i+1}^nZ^{i-1}K_nM_n;Z_i|S_i)\,,
\label{eqn:cvk4}
\eeqa
\beqa
& & I(Y^n;K_n|M_n)-I(Z^n;K_n|M_n)
\nonumber\\
&=  & \sum_{i=1}^n
        \left\{
        I(K_n;Y_i|Y_{i+1}^nZ^{i-1}M_nS_i)
       \right.
\nonumber\\
&     &\qquad\left. 
        -I(K_n;Z_i|Y_{i+1}^nZ^{i-1}M_nS_i)
        \right\}\,.
\label{eqn:cvk5}
\eeqa
\end{lm}

{\it Proof:}
We first prove (\ref{eqn:cvk1a}) and (\ref{eqn:cvk2a}). 
We have the following chains of inequalities:  
\beqa
I(M_n;Y^n)
&=&\sum_{i=1}^n
\left\{
H(Y_i|Y_{i+1}^n)-H(Y_i|Y_{i+1}^nM_n)
\right\}
\nonumber\\
&\leq&\sum_{i=1}^n
\left\{
H(Y_i)-H(Y_i|Y_{i+1}^nZ^{i-1}S_i M_n)
\right\}
\nonumber\\
&=&\sum_{i=1}^nI(Y_{i+1}^nZ^{i-1}S_iM_n;Y_i),
\nonumber\\
I(M_n;Z^n)
&\MLeq{a}&\sum_{i=1}^n
\left\{
H(Z_i|Z^{i-1}S_i)-H(Z_i|Z^{i-1}S_iM_n)
\right\}
\nonumber\\
&\leq &
\sum_{i=1}^n
\left\{
H(Z_i|S_i)-H(Z_i|Y_{i+1}^nZ^{i-1}S_iM_n)
\right\}
\nonumber\\
&=&\sum_{i=1}^n
I(Y_{i+1}^nZ^{i-1}M_n;Z_i|S_i)\,.
\nonumber
\eeqa
Step $\mbox{\rm (a)}$ follows from 
(\ref{eqn:Conv}). Next, we prove (\ref{eqn:cvk3}) 
and (\ref{eqn:cvk4}). 
We have the following chains of inequalities: 
\beqa
& & I(K_nM_n;Y^n)= H(Y^n)-H(Y^n|K_nM_n)
\nonumber\\
&=&\sum_{i=1}^n
\left\{
H(Y_i|Y_{i+1}^n)-H(Y_i|Y_{i+1}^nK_nM_n)
\right\}
\nonumber\\
&\leq&\sum_{i=1}^n
\left\{
H(Y_i)-H(Y_i|Y_{i+1}^nZ^{i-1}S_iK_nM_n)
\right\}
\nonumber\\
&=&\sum_{i=1}^n I(Y_{i+1}^nZ^{i-1}S_iK_nM_n;Y_i)\,,
\nonumber\\
& & I(K_nM_n;Z^n)
\nonumber\\
&=& H(K_nM_n|Z^n)-H(K_nM_n|Z^n)
\nonumber\\
&=&\sum_{i=1}^n
\left\{
H(K_nM_n|Z^{i-1})-H(K_nM_n|Z^{i})
\right\}
\nonumber\\
&\leq&\sum_{i=1}^n
\left\{
H(K_nM_n|Z^{i-1})-H(K_nM_n|Z^{i}S_i)
\right\}
\nonumber\\
&\MEq{a}&\sum_{i=1}^n
\left\{
H(K_nM_n|Z^{i-1}S_i)-H(K_nM_n|Z^{i}S_i)
\right\}
\nonumber\\
&=&\sum_{i=1}^nI(K_nM_n;Z_i|Z^{i-1}S_i)
\nonumber\\
&=&\sum_{i=1}^n
\left\{
H(Z_i|Z^{i-1}S_i)-H(Z_i|Z^{i-1}S_iK_nM_n)
\right\}
\nonumber\\
&\leq&\sum_{i=1}^n
\left\{
H(Z_i|S_i)-H(Z_i|Y_{i+1}^nZ^{i-1}S_iK_nM_n)
\right\}
\nonumber\\
&=&\sum_{i=1}^n I(Y_{i+1}^nZ^{i-1}K_nM_n;Z_i|S_i)\,.
\nonumber
\eeqa
Step $\mbox{\rm (a)}$ follows from  
$S_i\to Z^{i-1}\to K_nM_n$. 
Finally, we prove (\ref{eqn:cvk5}). 
We first observe 
the following two identities: 
\beqa
\hspace*{-7mm}& & H(Y^n|M_n) - H(Z^n|M_n)
\nonumber\\
\hspace*{-7mm}&=&\sum_{i=1}^n
\left\{ 
H(Y_i|Y_{i+1}^n Z^{i-1} M_n) - H(Z_i|Y_{i+1}^n Z^{i-1}M_n)
\right\},
\label{eqn:cvk61} 
\\
\hspace*{-7mm}& & H(Y^n|K_nM_n) - H(Z^n|K_nM_n)
\nonumber\\
\hspace*{-7mm}&=&\sum_{i=1}^n
\left\{ 
H(Y_i|Y_{i+1}^n Z^{i-1} K_nM_n)
\right.
\nonumber\\
\hspace*{-7mm}& &\qquad\left. 
-H(Z_i|Y_{i+1}^n Z^{i-1}K_nM_n)
\right\}.
\label{eqn:cvk70} 
\eeqa
Those identities follow from an elementary computation 
based on the chain rule of entropy. 
Subtracting (\ref{eqn:cvk70}) from (\ref{eqn:cvk61}), we have 
\beqno
& & I(Y^n;K_n|M_n)-I(Z^n;K_n|M_n)
\nonumber\\
&=  & \sum_{i=1}^n
        \left\{
        I(K_n;Y_i|Y_{i+1}^nZ^{i-1}M_n)
       \right.
\nonumber\\
&     &\qquad\left.  
        -I(K_n;Z_i|Y_{i+1}^nZ^{i-1}M_n)
        \right\}
\\
&=  & \sum_{i=1}^n
        \left\{
        -H(K_n|Y_{i}^nZ^{i-1}M_n)
        +H(K_n|Y_{i+1}^nZ^{i}M_n)
        \right\}
\\
&\leq & \sum_{i=1}^n
        \left\{
        -H(K_n|Y_{i}^nZ^{i-1}M_nS_i)
        +H(K_n|Y_{i+1}^nZ^{i}M_n)
        \right\}
\\
&\MEq{a}&\sum_{i=1}^n
         \left\{
         -H(K_n;|Y_{i}^nZ^{i-1}M_nS_i)
         +H(K_n|Y_{i+1}^nZ^{i}M_nS_i)
        \right\}
\\
&=  & \sum_{i=1}^n
        \left\{
        I(K_n;Y_i|Y_{i+1}^nZ^{i-1}M_nS_i)
       \right.
\nonumber\\
&     &\qquad\left.  
        -I(K_n;Z_i|Y_{i+1}^nZ^{i-1}M_nS_i)
        \right\}.
\eeqno
Step $\mbox{\rm (a)}$ follows from that 
$S_i=g_i(Z^{i-1})$ is a function of $Z^{i-1}$ 
in the case where $\{g_i\}_{i=1}^n$ is restricted 
to be deterministic. In the case where $\{g_i\}_{i=1}^n$ 
is allowed to be stochastic, if $\Ch$ belongs to the
\Cls, we have the following Markov chain: 
\beq
S_i\to Z^{i-1} \to Z_iY_{i+1}^{n}K_nM_n. 
\eeq  
Step {\rm (a)} follows from the above Markov chain. 
\hfill\QED 

Next, we present a lemma necessary to prove 
Lemma \ref{lm:conv3a}. 

\begin{lm}\label{lm:convlmD}\ \ 
For any sequence $\{U_i\}_{i=1}^{n}$ 
of random variables, we have 
\beqa 
& &I(K_nM_n;Y^n) \leq \sum_{i=1}^nI(X_iU_iS_i;Y_i)\,,
\label{eqn:convZ1}
\\
& &I(K_nM_n;Z^n) \leq \sum_{i=1}^nI(X_iU_i;Z_i|S_i)\,.
\label{eqn:convZ2}
\eeqa
\end{lm}

{\it Proof:} We first prove (\ref{eqn:convZ1}). We have the 
following chain of inequalities:   
\beqa
& &I(K_nM_n;Y^n) \MLeq{a} I(X^n;Y^n)
=H(Y^n)-H(Y^n|X^n)
\nonumber\\
&=    & \sum_{i=1}^n \left\{H(Y_i|Y^{i-1})-H(Y_i|Y^{i-1}X^n)\right\}
\nonumber\\
&\leq & \sum_{i=1}^n \left\{ H(Y_i)-H(Y_i|Y^{i-1}X^nS_i)\right\} 
\nonumber\\
&{\MEq{b}}& \sum_{i=1}^n \left\{H(Y_i)-H(Y_i|X_iS_i)\right\} 
\nonumber\\
&\leq&  \sum_{i=1}^n \left\{H(Y_i)-H(Y_i|X_iU_iS_i)\right\} 
=\sum_{i=1}^n I(X_iU_iS_i;Y_i)\,.
\nonumber
\eeqa
Step {\rm (a)} follows from the Markov chain 
$Y^n\to X^n \to K_nM_n$. 
Step {\rm (b)} follows from 
$Y_i\to X_iS_i \to Y^{i-1}X_{[i]}\,.$
Next, we prove (\ref{eqn:convZ2}). We have the following 
chain of inequalities:   
\beqa
& &I(K_nM_n;Z^n)\MLeq{a}I(X^n;Z^n)=H(X^n)-H(X^n|Z^n)
\nonumber\\
&=    & \sum_{i=1}^n \left\{H(X^n|Z^{i-1})-H(X^n|Z^{i})\right\}
\nonumber\\
&\leq & \sum_{i=1}^n \left\{H(X^n|Z^{i-1})-H(X^n|Z^{i}S_i)\right\}
\nonumber\\
& \MEq{b}& 
\sum_{i=1}^n \left\{H(X^n|Z^{i-1}S_i)-H(X^n|Z^{i}S_i)\right\}
\nonumber\\
&=&
\sum_{i=1}^n I(X^n;Z^i|Z^{i-1}S_i)
\nonumber\\
&=&\sum_{i=1}^n \left\{H(Z_i|Z^{i-1}S_i)-H(Z_i|Z^{i}X^nS_i)\right\}
\nonumber\\
& \MEq{c}& 
\sum_{i=1}^n \left\{ H(Z_i|S_i)-H(Z_i|X_iS_i)\right\}
\nonumber\\
&\leq & \sum_{i=1}^n 
\left\{H(Z_i|S_i)-H(Z_i|X_iU_iS_i)\right\}
=\sum_{i=1}^nI(X_iU_i;Z_i|S_i)\,.
\nonumber
\eeqa
Step {\rm (a)} follows from the Markov chain 
$Z^n\to X^n \to K_nM_n$. 
Step {\rm (b)} follows from $S_i\to Z^{i-1}\to X^n$. 
Step {\rm (c)} follows from 
$Z_i\to X_iS_i \to Z^{i-1}X_{[i]}\,.$
Thus, the proof of Lemma \ref{lm:convlmD} is completed.
\hfill\QED

{\it Proof of Lemma \ref{lm:conv3a}:} 
Set $U_i=Y_{i+1}^nZ^{i-1}M_n$. It can easily be verified that $U_i$, 
$X_iS_iZ_i$, $Y_i$ form a Markov chain 
$U_i \to X_iS_iZ_i \to Y_i $
in this order.  
From (\ref{eqn:cvk1a}), (\ref{eqn:cvk2a}), and (\ref{eqn:cvk5}) 
in Lemma \ref{lm:conv3d}, we obtain   
\newcommand{\wide}{\hspace*{-8mm}}
\beqno
I(M_n;Y^n)&\leq& \sum_{i=1}^nI(U_iS_i;Y_i),
\\ 
I(M_n;Z^n)&\leq& \sum_{i=1}^nI(U_i;Z_i|S_i), 
\eeqno
and 
\beqa
&     &I(Y^n;K_n|M_n)-I(Z^n;K_n|M_n)
\nonumber\\
&\leq & 
      \sum_{i=1}^n\{I(K_n;Y_i|U_iS_i)
                   -I(K_n;Z_i|U_iS_i)\}\,,
\label{eqn:conv13}
\eeqa
respectively. 
From (\ref{eqn:convZ1}), (\ref{eqn:convZ2}) 
in Lemma \ref{lm:convlmD}, we obtain   
\beqno
I(K_nM_n;Y^n)&\leq& \sum_{i=1}^nI(X_iU_iS_i;Y_i),
\\ 
I(K_nM_n;Z^n)&\leq& \sum_{i=1}^nI(X_iU_i;Z_i|S_i), 
\eeqno
respectively. 
It remains to evaluate an upper bound of 
$$
I(K_n;Y_i|U_iS_i)-I(K_n;Z_i|U_iS_i)\,.
$$
We have the following chain of inequalities:
\beqa
& &  I(K_n;Y_i|U_iS_i)-I(K_n;Z_i|U_iS_i)
\nonumber\\
&=& H(Y_i|U_iS_i)-H(Y_i|K_nM_nU_iS_i)
\nonumber\\
& &-H(Z_i|U_iS_i)+H(Z_i|K_nM_nU_iS_i)
\nonumber\\
&{\MEq{a}}& H(Y_i|U_iS_i)-H(Y_i|X^nU_iS_i)
\nonumber\\
&  &-H(Z_i|U_iS_i)+H(Z_i|X^nU_iS_i)
\nonumber\\
&= & H(Y_i|U_iS_i)
\nonumber\\
&  &-H(Y_i|Z_iX^nU_iS_i)-I(Y_i;Z_i|X^nU_iS_i)
\nonumber\\
&  &-H(Z_i|U_iS_i)
\nonumber\\
&  &+H(Z_i|Y_iX^nU_iS_i)+I(Y_i;Z_i|X^nU_iS_i)
\nonumber\\
&= & H(Y_i|U_iS_i)-H(Y_i|Z_iX^nU_iS_i)
\nonumber\\
&  &-H(Z_i|U_iS_i)+H(Z_i|Y_iX^nU_iS_i)
\nonumber\\
&\MEq{b}& 
H(Y_i|U_iS_i)-H(Y_i|Z_iX_iS_i)
\nonumber\\
&  &-H(Z_i|U_iS_i)+H(Z_i|Y_iX^nU_iS_i)
\nonumber\\
&\leq & H(Y_i|U_iS_i)-H(Y_i|Z_iX_iU_iS_i)
\nonumber\\
&     &-H(Z_i|U_iS_i)+H(Z_i|Y_iX_iU_iS_i)
\nonumber\\
&=& I(Y_i;Z_iX_i|U_iS_i)-I(Z_i;Y_iX_i|U_iS_i)
\nonumber\\
&=& I(X_i;Y_i|U_iS_i)-I(X_i;Z_i|U_iS_i)\,.
\nonumber
\eeqa
Step {\rm (a)} follows from $X^n=f_n(K_n,M_n)$ and $f_n$ 
is a one-to-one mapping. Step {\rm (b)} follows from 
$Y_i\to Z_iX_iS_i \to U_iX_{[i]}\,.$
Finally, we prove (\ref{eqn:cva4ssx}).   
We have the following chain of inequalities:
\beqa
& &I(K_n;Z^n|M_n)= H(Z^n|M_n)-H(Z^n|K_nM_n)
\nonumber\\
&\MEq{a} &H(Z^n|M_n)-H(Z^n|X^n)
\nonumber\\
&=&\sum_{i=1}^n
\left\{H(Z_i|Z^{i-1}M_n)-H(Z_i|Z^{i-1}X^n)\right\}
\nonumber\\
&\MEq{b}&
\sum_{i=1}^n\left\{H(Z_i|Z^{i-1}S_iM_n)-H(Z_i|Z^{i-1}S_iX^n)\right\}
\nonumber\\
&\geq&\sum_{i=1}^n\left\{H(Z_i|U_iS_i)-H(Z_i|X_iS_i)\right\}
\nonumber\\
&=  &\sum_{i=1}^n
\left\{
I(X_i;Z_i|U_iS_i)-I(U_i;Z_i|X_iS_i)
\right\}\,.
\nonumber
\eeqa
Step {\rm (a)} follows from that 
$f_n$ is a one-to-one mapping.
Step $\mbox{\rm (b)}$ follows from that 
$S_i=g_i(Z^{i-1})$ is a function of $Z^{i-1}$ 
in the case where $\{g_i\}_{i=1}^n$ is restricted 
to be deterministic. In the case where $\{g_i\}_{i=1}^n$ 
is allowed to be stochastic, if $\Ch$ belongs to the
\Cls, we have the following Markov chain: 
\beq
S_i\to Z^{i-1} \to Z_iM_nX^n.
\eeq  
Step {\rm (b)} follows from the above Markov chain. 
\hfill\QED  

{\it Proof of Lemma \ref{lm:conv3b}:}
This lemma immediately follows from Lemma \ref{lm:conv3d}.
\hfill\QED  

\subsection{
Proof of Lemma \ref{lm:conv3aa} 
}

In this appendix we prove Lemma \ref{lm:conv3aa}.

{\it Proof of Lemma \ref{lm:conv3aa}:} \ \  
Set $U_i\defeq Y^{i-1}Z_{i+1}^{n}M_n$. 
It can easily be verified that $U_i$, 
$X_iS_iZ_i$, $Y_i$ form a Markov chain 
$U_i \to X_iS_iZ_i \to Y_i$ in this order. 
In a manner similar to the proof 
of Lemma \ref{lm:conv3d}, we obtain 
the following chains of inequalities:  
\beqa
I(M_n;Y^n)
&=&\sum_{i=1}^n
\left\{
H(Y_i|Y^{i-1})-H(Y_i|Y^{i-1}M_n)
\right\}
\nonumber\\
&\leq&\sum_{i=1}^n
\left\{
H(Y_i)-H(Y_i|Y^{i-1}Z_{i+1}^{n}M_n)
\right\}
\nonumber\\
&=&\sum_{i=1}^nI(Y^{i-1}Z_{i+1}^nM_n;Y_i),
\nonumber\\
I(M_n;Z^n)&=&H(Z^n)-H(Z^n|M_n)
\nonumber\\
&=&\sum_{i=1}^nH(Z_i|Z^{i-1})-\sum_{i=1}^nH(Z_i|Z_{i+1}^nM_n)
\nonumber\\
&\MLeq{a}&
\sum_{i=1}^n
\left\{
H(Z_i|S_i)-H(Z_i|Z_{i+1}^nM_n)
\right\}
\nonumber\\
&\leq &
\sum_{i=1}^n
\left\{
H(Z_i|S_i)-H(Z_i|Y^{i-1}Z_{i+1}^{n}S_iM_n)
\right\}
\nonumber\\
&=&\sum_{i=1}^n
I(Y^{i-1}Z_{i+1}^nM_n;Z_i|S_i)\,.
\nonumber
\eeqa
Step $\mbox{\rm (a)}$ follows from that 
$S_i=g_i(Z^{i-1})$ is a function of $Z^{i-1}$ 
in the case where $\{g_i\}_{i=1}^n$ is restricted 
to be deterministic. In the case where $\{g_i\}_{i=1}^n$ 
is allowed to be stochastic, if $\Ch$ belongs to the
\Cls, we have the following Markov chain: 
\beq
S_i\to Z^{i-1} \to Z_iM_nX^n.
\eeq  
Step (a) follows from the above Markov chain. 
Hence, we have 
\beqno
I(M_n;Y^n)&\leq& \sum_{i=1}^nI(U_i;Y_i),
\\ 
I(M_n;Z^n)&\leq& \sum_{i=1}^nI(U_i;Z_i|S_i). 
\eeqno
Furthermore, by taking $\{U_i\}_{i=1}^n$ be constant 
in (\ref{eqn:convZ1}), (\ref{eqn:convZ2}) 
in Lemma \ref{lm:convlmD}, we obtain   
\beqno
I(K_nM_n;Y^n)&\leq& \sum_{i=1}^nI(X_iS_i;Y_i),
\\ 
I(K_nM_n;Z^n)&\leq& \sum_{i=1}^nI(X_i;Z_i|S_i), 
\eeqno
respectively. It remains to evaluate an upper bound of 
$$
I(K_n;Y^n|M_n)-I(K_n;Z^n|M_n)\,.
$$
Since $f_n$ is deterministic, we have
\beqa
& &I(K_n;Y^n|M_n)-I(K_n;Z^n|M_n)\,
\nonumber\\
&=&H(Y^n|M_n)-H(Z^n|M_n)
   -H(Y^n|X^n)
\nonumber\\
& &+H(Z^n|X^n)\,. 
\label{eqn:conv777} 
\eeqa
We separately evaluate the following two quantities:
\beqno
H(Y^n|M_n)-H(Z^n|M_n),
H(Y^n|X^n)-H(Z^n|X^n).
\eeqno
We observe the following two identities: 
\beqa
\hspace*{-9mm}& & H(Y^n|M_n) - H(Z^n|M_n)
\nonumber\\
\hspace*{-9mm}&=&\sum_{i=1}^n
\left\{ 
H(Y_i|Y^{i-1} Z_{i+1}^n M_n)-H(Z_i|Y^{i-1}Z_{i+1}^nM_n)
\right\},\label{eqn:cvk61b} 
\\
\hspace*{-9mm}& & -H(Y^n|X^n) + H(Z^n|X^n)
\nonumber\\
\hspace*{-9mm}&=&\sum_{i=1}^n
\left\{ 
-H(Y_i|Y_{i+1}^n Z^{i-1} X^n)
+H(Z_i|Y_{i+1}^n Z^{i-1}X^n)
\right\}. 
\label{eqn:cvk70b}
\eeqa
Those identities follow from an elementary computation based 
on the chain rule of entropy. From (\ref{eqn:cvk61b}), we 
have  
\beqa
& &
H(Y^n|M_n)-H(Z^n|M_n)
\nonumber\\
&=&\sum_{i=1}^n \left\{H(Y_i|U_i)-H(Z_i|U_i)\right\}\,.
\label{eqn:cvk62b} 
\eeqa
Next, we evaluate an upper bound of     
$$
-H(Y_i|Y_{i+1}^n Z^{i-1}X^n)
+H(Z_i|Y_{i+1}^n Z^{i-1}X^n)\,.
$$
Set 
$
\tilde{U}_i\defeq Y_{i+1}^n Z^{i-1}X_{[i]}\,.
$
We have the following chain of inequalities:  
\beqa 
& & -H(Y_i|Y_{i+1}^n Z^{i-1}X^n)
    +H(Z_i|Y_{i+1}^n Z^{i-1}X^n)
\nonumber\\
&=&-H(Y_i|X_i\tilde{U}_i)+H(Z_i|X_i\tilde{U}_i)
\nonumber\\
&\leq &-H(Y_i|X_iS_i\tilde{U}_i)+H(Z_i|X_i\tilde{U}_i)
\\
&{\MEq{a}} &-H(Y_i|X_iS_i\tilde{U}_i)+H(Z_i|X_iS_i\tilde{U}_i)
\nonumber\\
&= &-H(Y_i| Z_iX_iS_i\tilde{U_i})
    +I(Y_i;Z_i|X_iS_i\tilde{U_i})
\nonumber\\
&  &+H(Z_i| Y_iX_iS_i\tilde{U_i})
    -I(Y_i;Z_i|X_iS_i\tilde{U_i})
\nonumber\\
&= &-H(Y_i|Z_iX_iS_i\tilde{U_i})+ H(Z_i|Y_iX_iS_i\tilde{U_i})
\nonumber\\
&\MEq{b}
&-H(Y_i|Z_iX_iS_i) +H(Z_i|Y_iX_iS_i\tilde{U}_i)
\nonumber\\
&\leq &-H(Y_i|Z_iX_iS_i) +H(Z_i|Y_iX_iS_i)
\nonumber\\
& =   &   -H(Y_i|X_iS_i)+I(Y_i;Z_i|X_iS_i)
\nonumber\\
&     &   +H(Z_i|X_iS_i)-I(Y_i;Z_i|X_iS_i)
\nonumber\\
&=    &   -H(Y_i|X_iS_i) +H(Z_i|X_iS_i){\,.}
\label{eqn:conv23b}
\eeqa
Step $\mbox{\rm (a)}$ follows from that 
$S_i=g_i(Z^{i-1})$ is a function of $Z^{i-1}$ 
in the case where $\{g_i\}_{i=1}^n$ is restricted 
to be deterministic. In the case where $\{g_i\}_{i=1}^n$ 
is allowed to be stochastic, if $\Ch$ belongs to the
\Cls, we have the following Markov chain: 
\beq
S_i\to Z^{i-1} \to Z_iY_{i+1}^{n}X^n. 
\eeq  
Step (a) follows from the above Markov chain. 
Step {\rm (b)} follows from 
$Y_i\to$ $Z_iX_iS_i$ $\to \tilde{U}_i\,.$
Combining 
(\ref{eqn:conv777}), 
(\ref{eqn:cvk70b}), 
(\ref{eqn:cvk62b}), 
and (\ref{eqn:conv23b}), 
we obtain
\beqno
& &I(K_n;Y^n|M_n)-I(K_n;Z^n|M_n)
\nonumber\\
&\leq &\sum_{i=1}^n \{H(Y_i|U_i) -H(Z_i|U_i)
\nonumber\\
& &\qquad      -H(Y_i|X_iS_i)+H(Z_i|X_iS_i)\}
\nonumber\\
&\leq &\sum_{i=1}^n \{H(Y_i|U_i) -H(Z_i|U_i)
\nonumber\\
& &\qquad        -H(Y_i|X_iS_iU_i)+H(Z_i|X_iS_i)\}
\nonumber\\
&= &\sum_{i=1}^n\{I(X_iS_i;Y_i|U_i)-I(X_iS_i;Z_i|U_i)
\nonumber\\
& &\qquad+I(U_i;Z_i|X_iS_i)\}
\nonumber\\
&= &\sum_{i=1}^n\{I(X_i;Y_i|U_iS_i)-I(X_i;Z_i|U_iS_i)
\nonumber\\
& &\qquad+\ZeTaI+I(U_i;Z_i|X_iS_i)\}\,.
\eeqno
Finally, we prove (\ref{eqn:cva4r}).   
We have the following chain of inequalities:
\beqa
& &I(K_n;Z^n|M_n)= H(Z^n|M_n)-H(Z^n|K_nM_n)
\nonumber\\
&\MEq{a} &H(Z^n|M_n)-H(Z^n|X^n)
\nonumber\\
&=&\sum_{i=1}^n
\left\{H(Z_i|Z_{i+1}^{n}M_n)-H(Z_i|Z^{i-1}X^n)\right\}
\nonumber\\
&\MEq{b}&
\sum_{i=1}^n\left\{H(Z_i|Z_{i+1}^{n}M_n)-H(Z_i|Z^{i-1}S_iX^n)\right\}
\nonumber\\
&\geq&\sum_{i=1}^n\left\{H(Z_i|U_iS_i)-H(Z_i|X_iS_i)\right\}
\nonumber\\
&=&\sum_{i=1}^n
\left\{I(X_i;Z_i|U_iS_i)-I(U_i;Z_i|X_iS_i)\right\}\,.
\nonumber
\eeqa
Step {\rm (a)} follows from that 
$f_n$ is a one-to-one mapping. 
Step $\mbox{\rm (b)}$ follows from that 
$S_i=g_i(Z^{i-1})$ is a function of $Z^{i-1}$ 
in the case where $\{g_i\}_{i=1}^n$ is restricted 
to be deterministic. In the case where $\{g_i\}_{i=1}^n$ 
is allowed to be stochastic, if $\Ch$ belongs to the
\Cls, we have the following Markov chain: 
\beq
S_i\to Z^{i-1} \to Z_iX^n. 
\eeq  
Step (b) follows from the above Markov chain. 
Thus, the proof of Lemma \ref{lm:conv3aa} is completed.
\hfill\QED

\subsection{
Proof of Lemma \ref{lm:Rndau} 
}

We first observe that by the Cauchy-Schwarz inequality we have 
\beqno
{\bf E}_S\left({\bf E}_{X(S)}X(S)\right)^2
&\leq &{\bf E}_S
\left[
\left(
\sqrt{{\bf E}_{X(S)} X^2(S) }\sqrt{ {\bf E}_{X(S)} 1} 
\right)^2
\right]\\
&= & {\bf E}_S{\bf E}_{X(S)}X^2(S)\leq P_1\,.
\eeqno
Then, there exits $\alpha \in [0,1]$ such that
\beqno
&&{\bf E}_S\left({\bf E}_{X(S)}X(S)\right)^2=\bar{\alpha} P_1\,.
\eeqno
We derive an upper bound of $h(Y)$. We have the following 
chain of inequalities:   
\beqa
& &
h(Y)=h(X+S+\xi_1)
\nonumber\\
&\leq &{\ts \frac{1}{2}} 
\log
\left\{
(2\pi{\rm e})\left({\bf E}_{XS}|X+S|^2+ N_1\right)
\right\}
\nonumber\\
&=&{\ts \frac{1}{2}} 
   \log
   \left\{
   (2\pi{\rm e})
   \left({\bf E}_{X}{X^2}
       +2{\bf E}_{XS}XS 
        +{\bf E}_{S}S^2 +N_1 \right)
   \right\}
\nonumber\\
&\leq& {\ts \frac{1}{2}} 
   \log
   \left\{
   (2\pi{\rm e})
   \left(P_1+P_2+2{\bf E}_{XS}XS 
        +N_1 \right)
   \right\}\,.
\label{eqn:ppp0} 
\eeqa
By the Cauchy-Schwarz inequality we have 
\beqa
&    &{\bf E}_{XS}XS
 =  {\bf E}_S\left[S{\bf E}_{X(S)}X(S)\right]
\nonumber\\
&\leq&\sqrt{{\bf E}_S S^2 }
 \sqrt{{\bf E}_{S}\left({\bf E}_{X(S)}X(S)\right)^2}
=\sqrt{P_2}\sqrt{\bar{\alpha}P_1}\,.
\label{eqn:ppp} 
\eeqa
From (\ref{eqn:ppp0}) and (\ref{eqn:ppp}),
we have 
$$
h(Y)\leq\ts 
{\ts \frac{1}{2}} \log
\left\{ (2 \pi{\rm e})\left(P_1+P_2+\sqrt{\bar{\alpha}P_1 P_2}+N_1\right)
\right\}\,.
$$
Next, we estimate an upper bound of $h(Y|S)$. 
We have the following chain of inequalities:
\beqa
& & h(Y|S)={\bf E}_S\left[h(X(S)+\xi_1)\right]
\nonumber\\ 
      &\leq&{\bf E}_S
\left[\ts \frac{1}{2}
\log\left\{(2\pi{\rm e})
\left({\bf V}_{X(S)}\left[X(S)\right]+ N_1 \right)
     \right\}
\right]
\nonumber\\
&=&{\bf E}_S
\Bigl[\ts \frac{1}{2}
\log
 \Bigl\{(2\pi{\rm e})
 \Bigl(
       {\bf E}_{X(S)}[X^2(S)]
 \Bigr.\Bigr.\Bigr.
\nonumber\\
& &
\left.
    \left.
     \left.
   -\left({\bf E}_{X(S)}X(S)\right)^2+ N_1 
     \right)
   \right\}
  \right]
\nonumber\\
&\leq & \ts \frac{1}{2}
\log
 \Bigl\{(2\pi{\rm e})
 \Bigl(
       {\bf E}_S{\bf E}_{X(S)}[ X^2(S)]
 \Bigr.\Bigr.
\nonumber\\
& &\left.
   \left. 
   -{\bf E}_S\left({\bf E}_{X(S)}X(S)\right)^2+ N_1 
  \right)
 \right\}
\nonumber\\
&\leq & \ts \frac{1}{2}\log
   \left\{(2\pi{\rm e})\left(\alpha P_1+ N_1 \right)\right\}\,.
\nonumber
\eeqa
Similarly, we obtain 
\beqa
h(Z|S)&\leq& \ts \frac{1}{2}\log
   \left\{(2\pi{\rm e})\left(\alpha P_1+ N_2 \right)\right\}\,,
\nonumber\\
h(\tilde{Y}|S)&\leq& \ts \frac{1}{2}\log
   \left\{(2\pi{\rm e})\left(\alpha P_1+ \tilde{N}_1\right)\right\}\,.
\label{eqn:zzz0}
\eeqa
Since 
\beqno
& & h(\tilde{Y}|S) \geq h(\tilde{Y}|XS)
=\ts \frac{1}{2}\log \left\{(2\pi{\rm e})\tilde{N}_1\right\}
\eeqno
and (\ref{eqn:zzz0}), there exists $\beta\in [0,1]$ such that
$$
h(\tilde{Y}|US)=
\ts \frac{1}{2}\log
\left\{(2\pi{\rm e})\left(\beta\alpha P_1+ \tilde{N}_1\right)\right\}\,.
$$
Finally, we derive lower bounds of $h(Y|US)$ and $h(Z|US)$. 
{
Let $ \tilde{Y}(u,s)$ be a random variable with 
a conditional distribution of $\tilde{Y}$ for given $(U,S)$$=(u,s)$. 
Similar notations are used for $Y$ and $Z$. 
From the relation (\ref{eqn:eqqq}) between $X,S,$ $Y,Z,$ and 
$\tilde{Y}$, we have 
\beqa
Y(u,s)&=&\tilde{Y}(u,s)+\bar{a}(s+\tilde{\xi}_2)\,,
\label{eqn:zzoo}\\
Z(u,s)&=&\tilde{Y}(u,s)-{a}(s+\tilde{\xi}_2)\,.
\label{eqn:zzooo}
\eeqa
Note that $\tilde{Y}(u,s)$ is independent of $\tilde{\xi}_2$.
Applying entropy power inequality to (\ref{eqn:zzoo}) 
and (\ref{eqn:zzooo}), 
we have}
{
\beqno  
\ts \frac{1}{2 \pi{\rm e}} 2^{2h(Y(u,s))}
&\geq& \ts \frac{1}{2\pi{\rm e}} 2^{2h(\tilde{Y}(u,s))}
          +\frac{1}{2\pi{\rm e}} 2^{2h \left(\bar{a}(s+{\tilde{\xi}_2})\right)} 
\nonumber\\
&=& \ts \frac{1}{2\pi{\rm e}} 2^{2h(\tilde{Y}(u,s))}+ \bar{a}^2\tilde{N}_2\,,
\nonumber\\
\ts \frac{1}{2 \pi{\rm e}} 2^{2h(Z(u,s))}
&\geq& \ts \frac{1}{2\pi{\rm e}} 2^{2h(\tilde{Y}(u,s))}
          +\frac{1}{2\pi{\rm e}} 2^{2h\left({a}(s+{\tilde{\xi}_2})\right)} 
\nonumber\\
&=& \ts \frac{1}{2\pi{\rm e}} 2^{2h(\tilde{Y}(u,s))}+ {a}^2\tilde{N}_2\,,
\eeqno
from which we have 
\beqa
h(Y(u,s))&\geq &F_1\left(h(\tilde{Y}(u,s))\right)\,,
\label{eqn:asdd}\\
h(Z(u,s))&\geq &F_2\left(h(\tilde{Y}(u,s))\right)\,,
\label{eqn:asdd2}
\eeqa
where
\beqno
F_1(\gamma) &\defeq &\frac{1}{2}
\log\left(2^{2\gamma}+(2\pi{\rm e})\bar{a}^2\tilde{N}_2\right)\,,
\\
F_2(\gamma) &\defeq &\frac{1}{2}
\log\left(2^{2\gamma}+(2\pi{\rm e}){a}^2\tilde{N}_2\right)\,.
\eeqno
By a simple computation we can show that $F_i(\gamma),i=1,2$ are 
monotone increasing and convex functions of $\gamma$. 
Taking the expectation of both sides of (\ref{eqn:asdd}) with respect to 
$(U,S)$, we have  
\beqa
      & &h(Y|US)
\nonumber\\
      &=&{{\bf E}}_{US}\left[h({Y}(U,S))\right]
       \geq  {\bf E}_{US}\left[
                          F_1\left( h(\tilde{Y}(U,S))\right)
                          \right]
\nonumber\\
      & \MGeq{a}& F_1\left({\bf E}_{US}\left[
                            h(\tilde{Y}(U,S))\right]
                          \right)
       = F_1\left(h(\tilde{Y}|US)\right)
      \nonumber\\
      &=& \ts\frac{1}{2}\log
\left\{(2\pi{\rm e})
    \left(
         \beta{\alpha}P_1+\tilde{N}_1 + \bar{a}^2\tilde{N}_2
    \right)
\right\}
\nonumber\\
      &=& \ts \frac{1}{2}\log
         \left\{(2\pi{\rm e})
            \left(\beta{\alpha}P_1
            +\ts \frac{(1-\rho^2)N_1N_2}{N_1+N_2-2\rho\sqrt{N_1N_2}}
            \right.
         \right.
\nonumber\\
& & \qquad\quad 
     \left.\left.
     +\ts \frac{N_1^2+\rho^2N_1N_2-2\rho N_1\sqrt{N_1N_2}}
     {N_1+N_2-2\rho\sqrt{N_1N_2}}
\right)
\right\}
\nonumber\\
&=&\ts \frac{1}{2}
\log \left\{ 
(2\pi{\rm e})\left(\beta{\alpha}P_1+ N_1\right)
\right\}
\,.
\nonumber
\eeqa
Step {\rm (a)} follows from the convexity of $F_1(\gamma)$ 
and Jensen's inequality. 
Taking the expectation of both sides of (\ref{eqn:asdd2}) 
with respect to $(U,S)$, we have  
\beqa
  & &h(Z|US)
\nonumber\\
&=&{\bf E}_{US}\left[h({Z}(U,S))\right]
       \geq  {\bf E}_{US}
               \left[F_2\left( h(\tilde{Y}(U,S))\right)\right]
\nonumber\\
      & \MGeq{a}& F_2\left({\bf E}_{US}\left[h(\tilde{Y}(U,S))\right] \right)
      = F_2\left(h(\tilde{Y}|US)\right)
\nonumber\\
      &=& \ts \frac{1}{2}\log
\left\{(2\pi{\rm e})
    \left(
         \beta{\alpha}P_1+\tilde{N}_1 + {a}^2\tilde{N}_2
    \right)
\right\}
\nonumber\\
      &=& \ts \frac{1}{2}\log
         \left\{(2\pi{\rm e})
            \left(\beta{\alpha}P_1
            +\ts \frac{(1-\rho^2)N_1N_2}{N_1+N_2-2\rho\sqrt{N_1N_2}}
            \right.
         \right.
\nonumber\\
& & \qquad\quad 
     \left.\left.
     +\ts \frac{N_2^2+\rho^2N_1N_2-2\rho N_2\sqrt{N_1N_2}}
               {N_1+N_2-2\rho\sqrt{N_1N_2}}
\right)
\right\}
\nonumber\\
&=&\ts \frac{1}{2}
\log 
\left\{ 
(2\pi{\rm e})\left(\beta{\alpha}P_1+ N_2\right)
\right\}
\,.
\nonumber
\eeqa
Step {\rm (a)} follows from the convexity of $F_2(\gamma)$ 
and Jensen's inequality.} Thus, the proof of Lemma \ref{lm:Rndau} 
is completed. 
\hfill\QED


\begin{thebibliography}{99}

\bibitem{sh}C. E. Shannon, ``Communication theory of secrecy systems,''
{\em Bell Sys. Tech. Journal}, vol. 28, pp. 656-715, 1949.

\bibitem{yama1} H. Yamamoto,
``Coding theorems for Shannon cipher  system with correlated 
source outputs and common information,''
{\em IEEE Trans. Inform. Theory}, vol. 40 
pp. 85-95, 
1994. 

\bibitem{yama2} \underline{$\qquad$},
``Rate-distortion theory for the Shannon cipher system,'' 
{\em IEEE Trans. Inform. Theory}, vol. 43 
pp. 827-835, 
1997. 

\bibitem{Wyn1}A.~D.~Wyner, ``The wire-tap channel,'' 
{\em Bell Sys. Tech. Journal}, vol. 54, pp. 1355-1387, 1975.

\bibitem{CsiKor1}I.~Csisz{\' a}r and J.~K{\" o}rner, ``Broadcast
channels with confidential messages,'' 
{\em IEEE Trans. Inform. Theory}, vol. IT-24, pp. 339-348, 1978.

\bibitem{yama3} H. Yamamoto,
``A source-coding problem for  sources with 
additional outputs to keep secret from the receiver 
or wiretappers,''
{\em IEEE Trans. Inform. Theory}, vol. 29, 
pp. 918-923, 
1983. 

\bibitem{yama4} \underline{$\qquad$},
``On secret sharing  communication-systems with 2 or 3 channels,''
{\em IEEE Trans. Inform. Theory}, vol. 32, 
pp. 387-393, 
1986. 

\bibitem{yama5} \underline{$\qquad$},
``A rate-distortion problem for a communication-system with 
a secondary decoder to be hindered,''
{\em IEEE Trans. Inform. Theory}, vol. 34, 
pp. 835-842, 
1988. 
\bibitem{yama6} \underline{$\qquad$}, 
``Coding theorem for secret  sharing communication-systems 
  with 2 noisy channels,''
{\em IEEE Trans. Inform. Theory}, vol. 35, 
pp. 572-578, 
1989. 

\bibitem{yama7} \underline{$\qquad$},
``A coding theorem for secret sharing communication-systems 
  with  2 Gaussian wiretap channels,''
{\em IEEE Trans. Inform. Theory}, vol. 37, 
pp. 634-638, 
1991. 

\bibitem{Maurer}U. M. Maurer, 
``Secret key agreement by public discussion 
from common information,'' 
{\em IEEE Trans. Inform. Theory}, vol. 39, pp. 733-742, 1993.

\bibitem{ac1}R. Ahlswede and I.~Csisz{\' a}r, ``Common randomness in 
information theory and cryptography -Part I: Secret sharing,'' 
{\em IEEE Trans. Inform. Theory}, vol. 39, pp. 1121-1132, 1993.


\bibitem{cn}I.~Csisz{\' a}r and P. Narayan, ``Common randomness and secret
key generation with {a} helper,''    
{\em IEEE Trans. Inform. Theory}, vol. 46, pp. 344-366, 2000.

{
\bibitem{cn2}\underline{$\qquad$}, 
``Secrecy capacity for multiple terminals,''    
{\em IEEE Trans. Inform. Theory}, vol. 50, pp. 3047-3061, 2004.
}
\bibitem{cn3}\underline{$\qquad$}, 
``Secrecy capacities for multiterminal channel models,''    
{\em IEEE Trans. Inform. Theory}, 
vol. 54, 
pp. 2437-2452, 
2008.


\bibitem{ohrcc}Y.~Oohama, 
``Coding for relay channels with confidential messages,'' 
{\em in Proceedings of the IEEE Information 
Theory Workshop(ITW)}, Cairns, Australia, pp. 87-89, 2001.

\bibitem{ohrcc07}\underline{$\qquad$}, 
``Capacity Theorems for relay channels with confidential 
messages,'' 
{\em in Proceedings of the IEEE International Symposium of Information Theory(ISIT),}
Nice, France, pp. 926-930, 2007.

\bibitem{heye}X. He and A. Yener, ``On the equivocation region 
of relay channels with orthogonal components,'' 
{\em in Proceedings of the Asilomar Conference 
on Signals, Systems and Computers}, Pacific Grove, CA, USA, pp. 883-887, 2007.


\bibitem{heye3}\underline{$\qquad$}, 
``Cooperation with an untrusted relay: A secrecy perspective,'' 
{\em IEEE Trans. Inform. Theory}, vol. 56, 
pp. 3807-3827, 
2010.


\bibitem{lp}Y. Liang and H.V. Poor, 
``Multiple access channels with confidential messages,"
{\em IEEE Trans. Inform. Theory}, 
vol. 54, pp. 976-1002, 
2008.
%

\bibitem{liu} R. Liu, I. Mari\'c, P. Spasojevi\'c, and R. D. Yates,
``Discrete memoryless interference and broadcast channels with 
confidential messages: Secrecy rate regions,"
{\em IEEE Trans. Inform. Theory}, 
vol. 54, pp. 2493-2507, 
2008.

\bibitem{ty} E. Tekin and A. Yener, 
``The general Gaussian multiple-access and two-way wiretap channels: 
Achievable rates and cooperative jamming,"
{\em IEEE Trans. Inform. Theory}, 
vol. 54, pp. 2735-2751, 
2008.

\bibitem{lg}L. Lai and H. El Gamal, 
``The relay-eavesdropper channel: 
Cooperation for secrecy,'' 
{\em IEEE Trans. Inform. Theory}, 
vol. 54, pp. 4005-4019, 
2008.
%
\bibitem{cg}T.~M.~Cover and A.~El ~Gamal, ``Capacity theorems for the
relay channel,'' {\em IEEE Trans. Inform. Theory}, 
vol. IT-25, pp. 572-584, 1979.

\bibitem{ckB}I.~Csisz{\' a}r and J.~K{\" o}rner, 
{\it Information Theory: Coding Theorems for 
Discrete Memoryless Systems.} Academic Press, New York, 1981.

\bibitem{han}
T. S. Han, {\it Information-Spectrum Methods in Information
Theory. }Springer-Verlag, Berlin, New York, 2002. The Japanese 
edition was published by Baifukan-publisher, Tokyo, 1998.

\bibitem{lv}Y. Liang and V. V. Veeravalli, 
``Cooperative relay broadcast channels,"
{\em IEEE Trans. Inform. Theory},
vol. 53, pp. 900-928, 2007.

{
\bibitem{lk}Y. Liang and G. Kramer, 
``Rate regions for relay broadcast channels,"
{\em IEEE Trans. Inform. Theory},  
vol. 53, pp. 3517-3535, 2007.
}

%

\bibitem{lh}S. K. Leung-Yan-Cheong and M. Hellman,
``The Gaussian wire-tap channel," 
{\em IEEE Trans. Inform. Theory}, 
vol. IT-24, pp. 451-456, 1978.

\bibitem{CovTh}T. M. Cover and J. A. Thomas, 
{\it Elements of Information Theory}, Wiley, New York, 1991.  

\RevP{
\bibitem{tdcmj}A. Thangaraj, S. Dihidar, A. R. Calderbank,
S. W. McLaughlin, and J.-M. Merolla,
``Applications of LDPC codes to the wire-tap channel,"  
{\em IEEE Trans. Inform. Theory}, 
vol. 53, pp. 2933-2945, 2007.
}
\end{thebibliography}
\end{document}